\documentclass[10pt,aps,prd,twocolumn,showpacs,superscriptaddress,eqsecnum,longbibliography,nofootinbib]{revtex4-2}
\usepackage{mathrsfs,amsmath,amsthm,latexsym,amssymb,amsfonts,epsfig,cancel,enumerate,graphicx,subcaption,txfonts,overpic}
\usepackage{multirow}
\usepackage{xcolor}
\definecolor{navy}{RGB}{0,0,150}
\usepackage[colorlinks, linkcolor=blue, citecolor=blue, urlcolor=blue, plainpages=false, pdfstartview=FitH]{hyperref}
\usepackage{appendix}
\allowdisplaybreaks
\begin{document}
	
	\title{Observational Signatures of Accretion Disks around a Schwarzschild Black Hole in a Hernquist Dark Matter Halo}
	
	\author{Zhenglong Ban}
	\email{zlban123@163.com}
	\affiliation{College of Physics Science and Technology, Hebei University, Baoding 071002, China}
	
	\author{Jing-Ya Zhao}
	\email{Zhaojingya2002@163.com}
	\affiliation{College of Physics Science and Technology, Hebei University, Baoding 071002, China}
	
	\author{Tian-You Ren}
	\email{zlban123@163.com}
	\affiliation{College of Physics Science and Technology, Hebei University, Baoding 071002, China}
	
	\author{Yaobin Hua}
	\email{zlban123@163.com}
	\affiliation{College of Physics Science and Technology, Hebei University, Baoding 071002, China}
	
	\author{Rong-Jia Yang \footnote{Corresponding author}}
	\email{yangrongjia@tsinghua.org.cn}
	\affiliation{College of Physics Science and Technology, Hebei University, Baoding 071002, China}
	\affiliation{Hebei Key Lab of Optic-Electronic Information and Materials, Hebei University, Baoding 071002, China}
	\affiliation{National-Local Joint Engineering Laboratory of New Energy Photoelectric Devices, Hebei University, Baoding 071002, China}
	\affiliation{Key Laboratory of High-pricision Computation and Application of Quantum Field Theory of Hebei Province, Hebei University, Baoding 071002, China}
	
	\begin{abstract}
		We investigate how a Hernquist-type dark matter (DM) halo, parametrized by its core radius $r_{\mathrm{s}}$ and central density $\rho_{\mathrm{s}}$, influences both the gravitational wave (GW) emission from timelike periodic orbits and the electromagnetic appearance of a thin accretion disk around a Schwarzschild black hole (BH). By analyzing the effective potential for timelike geodesics, we show that the DM halo shifts the marginally bound orbit (MBO) and the innermost stable circular orbit (ISCO) outward, reflecting its modification of the spacetime geometry and the energy–angular momentum structure of particle motion. Employing a semi-analytical method, we compute orbital trajectories and the associated GW waveforms, revealing that the DM halo alters the characteristic zoom-whirl dynamics and induces measurable changes in waveform morphology. Furthermore, we generate direct and secondary images of the accretion disk across various observer inclinations and find that increasing $r_{\mathrm{s}}$ or $\rho_{\mathrm{s}}$ results in cooler, dimmer disks with modified flux distributions. Our results demonstrate that the presence of a DM halo imprints distinct signatures in both gravitational and electromagnetic observables, offering a multi-messenger pathway to probe DM environments near BHs.
	\end{abstract}
	
	\maketitle
	\section{Introduction}
	BHs stand among the most profound predictions of Einstein’s General Relativity (GR), originally emerging as vacuum solutions to the field equations yet now recognized as ubiquitous astrophysical entities residing in rich and dynamic environments. As cosmic laboratories where gravity operates in its strongest regime, BHs provide unparalleled opportunities to probe fundamental physics under extreme conditions. Over the past decade, BH astrophysics has entered a precision era: the Event Horizon Telescope (EHT) has delivered horizon-scale images of the supermassive BHs M87$^{*}$ and Sgr A$^{*}$ \cite{EventHorizonTelescope:2019dse,EventHorizonTelescope:2021bee,EventHorizonTelescope:2022wkp}, LIGO/Virgo collaborations have detected GWs from binary BH mergers \cite{LIGOScientific:2016aoc,KAGRA:2021vkt}, and the gravity instrument has observed relativistic infrared flares orbiting near our Galactic center \cite{GRAVITY:2020lpa}. Together, these breakthroughs offer direct observational access to the strong-gravity regime, moving beyond pure theoretical speculation toward empirical validation. Nevertheless, despite GR's remarkable empirical success, it remains incomplete: it cannot resolve the nature of BH singularities nor account for the enigmatic components of the universe: DM and dark energy. This underscores the need to study BHs not in isolation, but as embedded objects interacting with their surrounding astrophysical media, particularly DM halos, which may leave subtle yet detectable imprints on both orbital dynamics and electromagnetic signatures.
	
	Theoretical and observational studies consistently support the existence of extended DM halos enveloping galaxies. This conclusion is robustly inferred from the flatness of galactic rotation curves, strong and weak gravitational lensing measurements, and the anisotropies in the cosmic microwave background (CMB). According to the standard $\Lambda$CDM cosmological model, DM constitutes approximately $27\%$ of the total energy density of the universe and plays a pivotal role in the formation of structures and galactic evolution. Given that BHs typically reside at galactic centers, they are expected to be deeply embedded within such DM halos, implying that their local spacetime geometry and the dynamics of nearby matter are inevitably influenced by the gravitational potential of the surrounding DM. To describe the spatial distribution of these halos, various empirical and semi-analytical density profiles have been proposed. Among the most widely used are the Navarro–Frenk–White (NFW) \cite{Navarro:1995iw}, Burkert \cite{Burkert:1995yz}, Moore \cite{Moore:1994yx}, and Dehnen \cite{Dehnen:1993uh} profiles—each characterized by distinct central behaviors (e.g., cuspy versus cored) and asymptotic fall-offs, leading to different modifications of the gravitational field near BHs. Of particular relevance is the Hernquist profile, which, owing to its mathematical simplicity, physical motivation (originally derived for spherical stellar systems), and excellent fit to the mass distributions of elliptical galaxies and simulated dark halos, has become a standard choice in both astrophysical and relativistic studies. In this work, we adopt a Hernquist-type DM halo to systematically investigate its impact on the structure of periodic timelike orbits and the observational signatures of thin accretion disks around a Schwarzschild BH.
	
	BHs and GWs, two cornerstone predictions of GR, have now been robustly confirmed through landmark observational campaigns \cite{LIGOScientific:2016aoc,LIGOScientific:2016sjg,EventHorizonTelescope:2019dse,EventHorizonTelescope:2019uob}. Among the diverse GW sources, extreme-mass-ratio inspirals (EMRIs), in which a stellar-mass compact object spirals into a massive BH, offer a uniquely rich probe of strong-field gravity. The emitted waveforms encode detailed information about the spacetime geometry in the immediate vicinity of the event horizon, rendering EMRIs among the most promising avenues for testing the nature of BHs. Crucially, the structure of these waveforms is intimately tied to the orbital dynamics of the inspiraling body \cite{Glampedakis:2005hs}. In this context, special classes of orbits—particularly periodic geodesics—serve as valuable templates for identifying characteristic features in EMRI signals. Such periodic orbits have been rigorously shown to exist in both Schwarzschild and Kerr spacetimes \cite{Levin:2008mq,Levin:2008ci,Grossman:2008yk}. The dynamics of test particles in curved spacetime provide a powerful diagnostic tool for exploring the structure and strength of gravitational fields. In particular, bound orbits—ranging from stable circular trajectories to complex precessing paths—serve as essential probes in classical tests of gravity, assessments of long-term orbital stability in strong-field regimes, and modeling of GW sources \cite{Poisson:1993vp,Cutler:1993vq,Apostolatos:1993nu,Dai:2023cft,GRAVITY:2020gka,Cardoso:2014sna,Cunha:2022gde,Guo:2024cts,Mummery:2022ana}. Among these, periodic orbits represent a special class of motion in which a particle returns exactly to its initial position and momentum after a finite proper time. A striking feature of many such orbits near compact objects is zoom-whirl behavior: the particle alternates between rapid revolutions (``whirls'') close to the BH and outward excursions (``zooms'') to apastron. These orbits can be uniquely characterized by a triplet of integers $(z,w,v)$: the zoom number $z$ counts the number of radial oscillations (or ``leaves'') in one full cycle; the whirl number $w$ denotes how many additional azimuthal loops the particle completes near periastron before zooming out again; and the vertex number $v$ labels the sequence of successive apastron passages relative to the initial one $(v=0)$. Owing to their sensitivity to spacetime geometry, zoom-whirl periodic orbits have been extensively studied across a range of gravitational frameworks—including GR and its modifications, offering valuable insights into the nature of compact objects and the underlying theory of gravity \cite{Grossman:2008yk,Fujita:2009bp,Healy:2009zm,Wei:2019zdf,Azreg-Ainou:2020bfl,Deng:2020yfm,Wang:2022tfo,QiQi:2024dwc,Alloqulov:2025ucf,Wang:2025hla}.
	
	Astrophysical processes, particularly those occurring in the strong-field regime near compact objects, provide a crucial testing ground for gravitational theories—both within GR and beyond. Among these, the electromagnetic emission from thin accretion disks serves as a direct observational probe of spacetime geometry. To model and interpret such emissions, modern studies rely on two primary imaging methodologies: the semi-analytic approach and full numerical ray-tracing combined with radiative transfer. In a seminal contribution, Luminet employed the semi-analytic method to compute the direct and secondary images of a thin accretion disk around a Schwarzschild BH, deriving the observed brightness from an analytical expression for the relativistic radiation flux \cite{Luminet:1979nyg}. Since then, the optical appearance and physical properties of thin disks have been extensively investigated across a variety of background spacetimes, including those of BHs, wormholes, and other exotic compact objects \cite{Hou:2022eev,Zhang:2024lsf,Gyulchev:2019tvk,Shaikh:2019hbm,Bambi:2019tjh,Johannsen:2016uoh,Gates:2020sdh,Okyay:2021nnh,Liu:2024brf,Feng:2025ljz,Yin:2025coq,Feng:2024iqj,He:2022lrc}. Parallel efforts have extended these analyses into alternative theories of gravity, exploring how modifications to Einstein's theory, such as in scalar-tensor models, regular BH spacetimes, or solutions inspired by quantum corrections, affect the structure of BH shadows and the morphology of accretion disks \cite{Cai:2025pan,Cai:2025rst,Huang:2023ilm,Guo:2023grt,Liu:2021lvk,Guo:2022rql}. Collectively, these works demonstrate that disk imaging not only reveals the behavior of matter under extreme gravity but also encodes subtle imprints of the underlying gravitational framework.
	
	The remainder of this paper is organized as follows. In Sec.~\ref{section2}, we present the spacetime metric of a Schwarzschild BH embedded in a Hernquist DM halo and derive the equations governing timelike geodesics, with particular attention to the dynamics of test particles, including the locations of the ISCOs and MBOs. Section~\ref{section3} focuses on periodic orbits, where we characterize their topology via the rational number $q$ and investigate how the halo parameters $\rho_{\mathrm{s}}$ and $r_{\mathrm{s}}$ influence the energy and orbital angular momentum of these trajectories. The gravitational waveforms emitted by an extreme mass-ratio inspiral (EMRI) system driven by such periodic orbits are computed and analyzed in Sec.~\ref{section4}. Moving to the photon sector, Sec.~\ref{section5} examines null geodesics in the equatorial plane, detailing the structure of critical photon orbits. In Sec.~\ref{section6}, we construct the deflection angle $\varphi(b)$ diagram to model the imaging of a thin accretion disk, generating both direct and secondary images and analyzing the apparent radiation flux distribution as observed from various inclination angles. Finally, we summarize our key findings and discuss their implications in Sec.~\ref{section7}.
	
	\section{Schwarzschild Black Hole in a Hernquist Dark Matter Halo and the timelike geodesics}
	\label{section2}
	The spacetime metric for the Schwarzschild–Hernquist system takes the form \cite{Jha:2025xjf}
	\begin{equation}
		ds^{2}=-f(r)\mathrm{d}t^{2}+\frac{\mathrm{d}r^{2}}{f(r)}+r^{2}\left(\mathrm{d}\theta^{2}+\sin^{2}\theta \mathrm{d} \phi^{2}\right),\label{xianyuan}
	\end{equation}
	where 
	\begin{equation}
		f(r)=1-\frac{2M}{r}-\frac{4\pi\rho_{\mathrm{s}}r_{\mathrm{s}}^{3}}{r+r_{\mathrm{s}}}.\label{fr}
	\end{equation}
	where $\rho_{\mathrm{s}}$ and $r_{\mathrm{s}}$ represent the central DM density and the core radius, respectively. In the limiting case where either $\rho_{\mathrm{s}}=0$ or $r_{\mathrm{s}}=0$, the above metric reduces to the standard Schwarzschild solution.
	
\begin{figure*}[htbp]
		\centering
		\begin{subfigure}{0.32\textwidth}
			\centering
			\includegraphics[width=\linewidth]{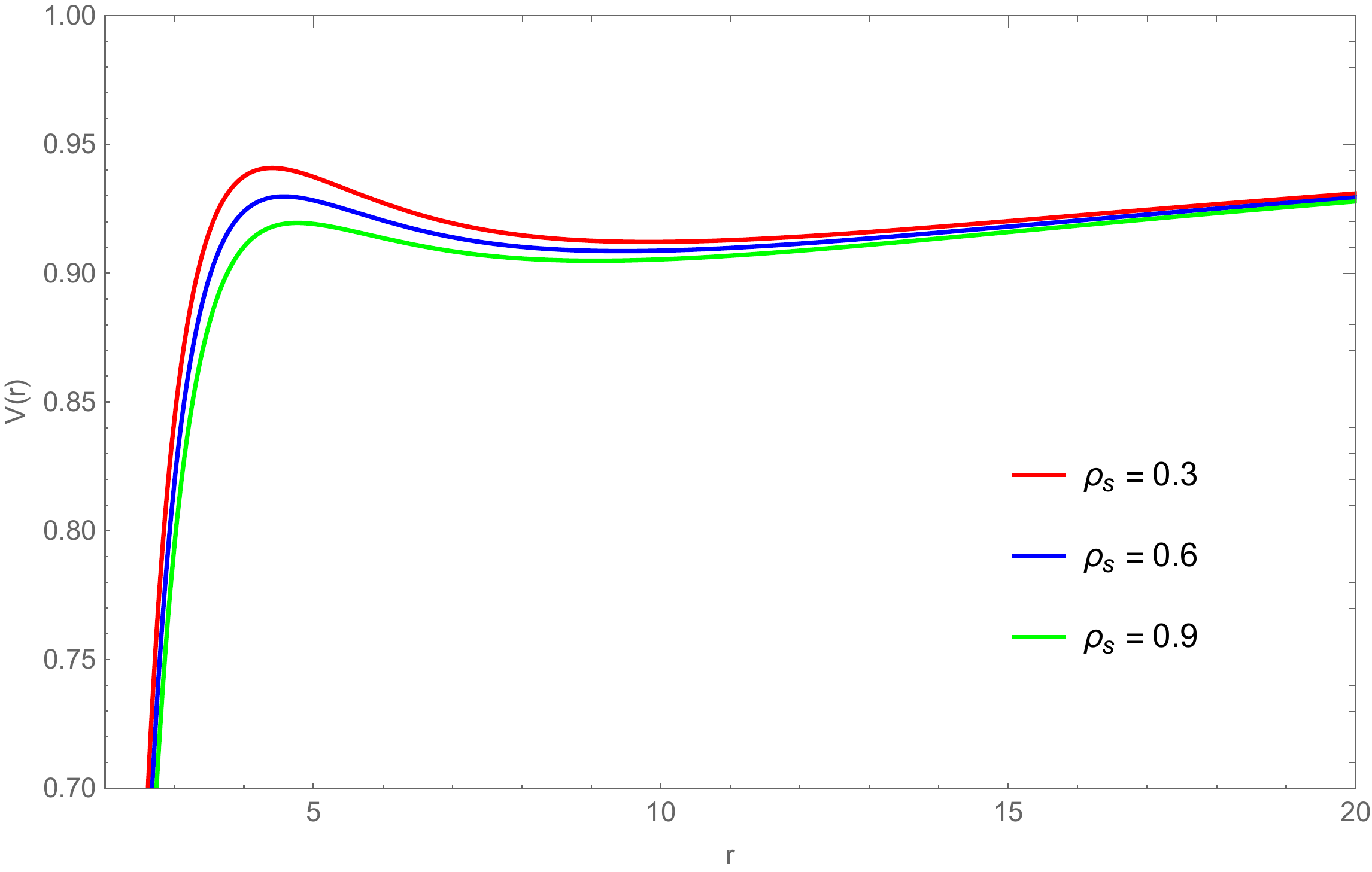}
		\end{subfigure}%
		\hfill
		\begin{subfigure}{0.32\textwidth}
			\centering
			\includegraphics[width=\linewidth]{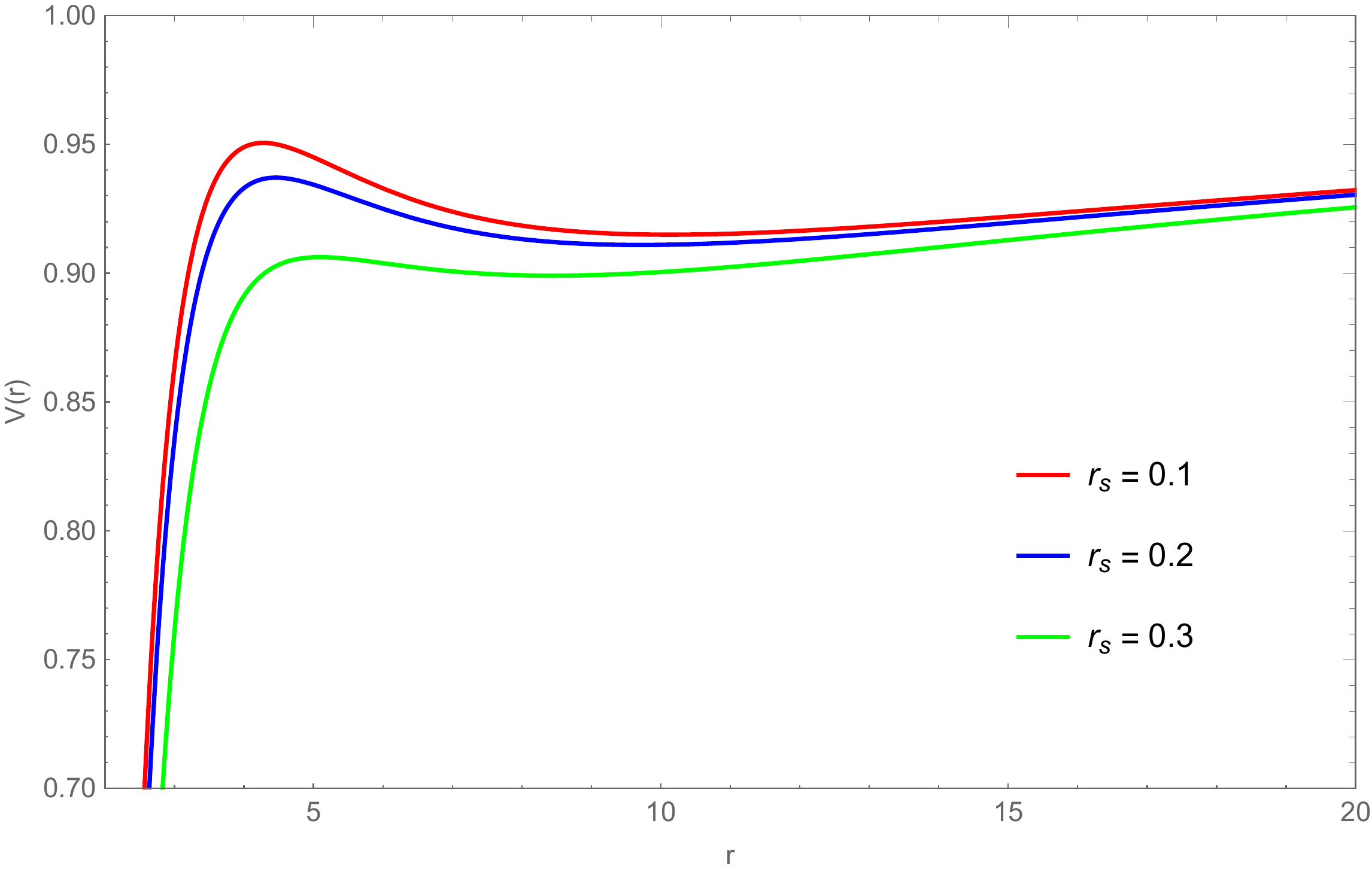}
		\end{subfigure}%
		\hfill
		\begin{subfigure}{0.32\textwidth}
			\centering
			\includegraphics[width=\linewidth]{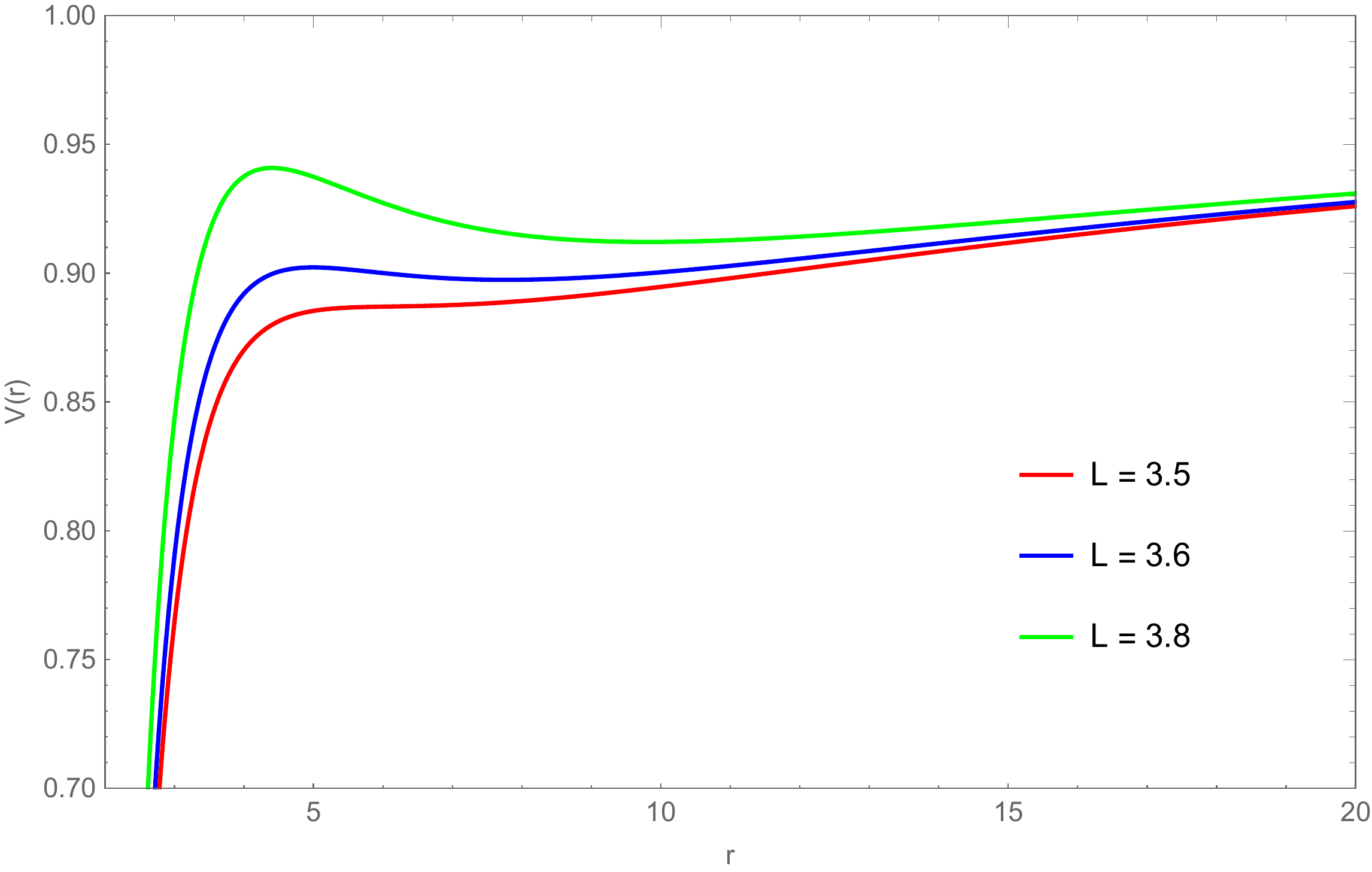}
		\end{subfigure}
		\caption{The effective potential $V_{\mathrm{eff}}$ for different values of the Hernquist halo parameter and angular momentum $L$. Left: $r_{\mathrm{s}}=0.2$, $L=3.8$; Middle: $\rho_{\mathrm{s}}=0.4$, $L=3.8$; Right: $r_{\mathrm{s}}=0.3$, $\rho_{\mathrm{s}}=0.4$.}
		\label{epotential}
	\end{figure*}

\begin{figure*}[htbp]
		\centering
		\begin{subfigure}{0.38\textwidth}
			\includegraphics[width=\linewidth]{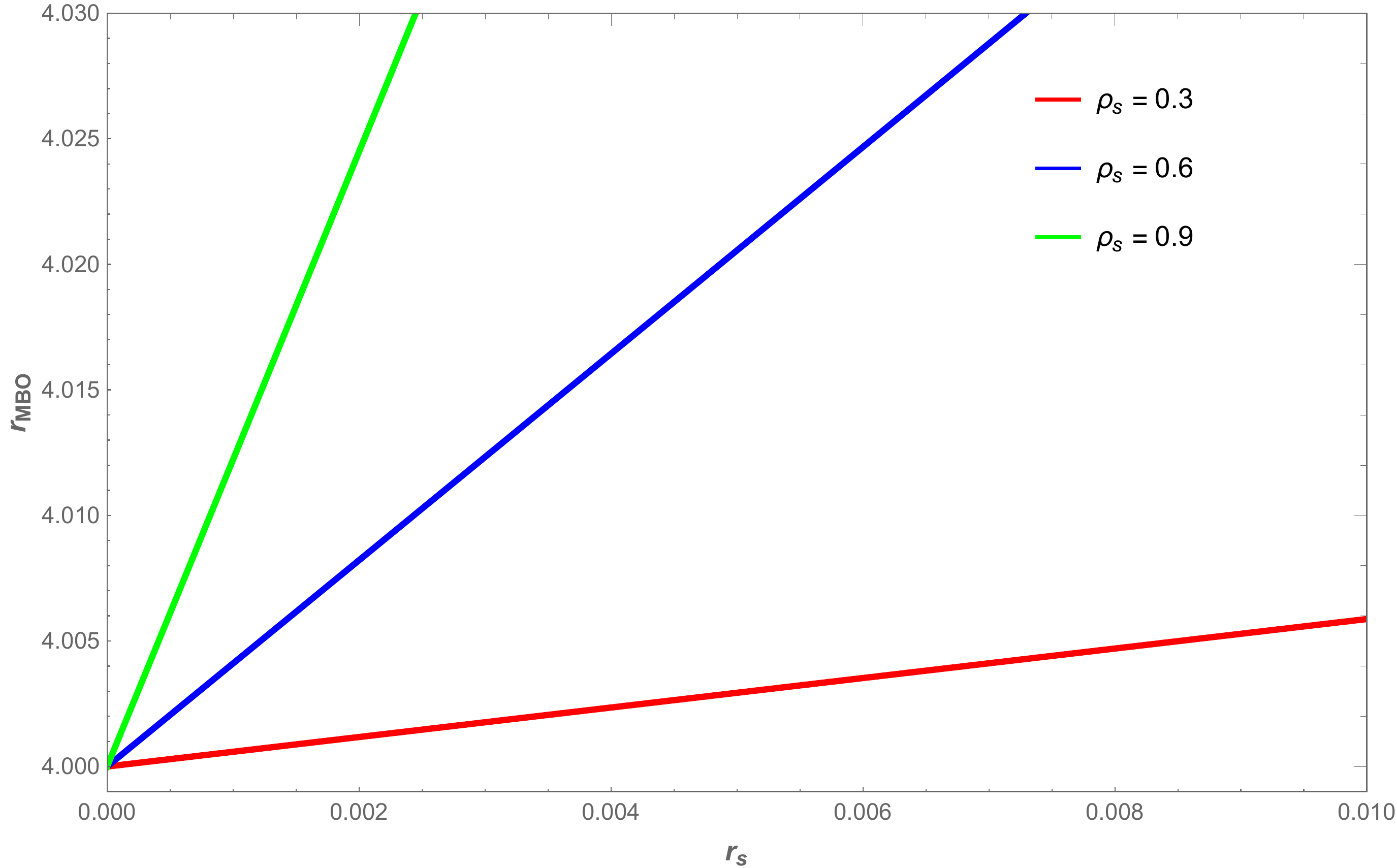}
		\end{subfigure}%
		\hspace{0.13\textwidth}% 
		\begin{subfigure}{0.38\textwidth}
			\includegraphics[width=\linewidth]{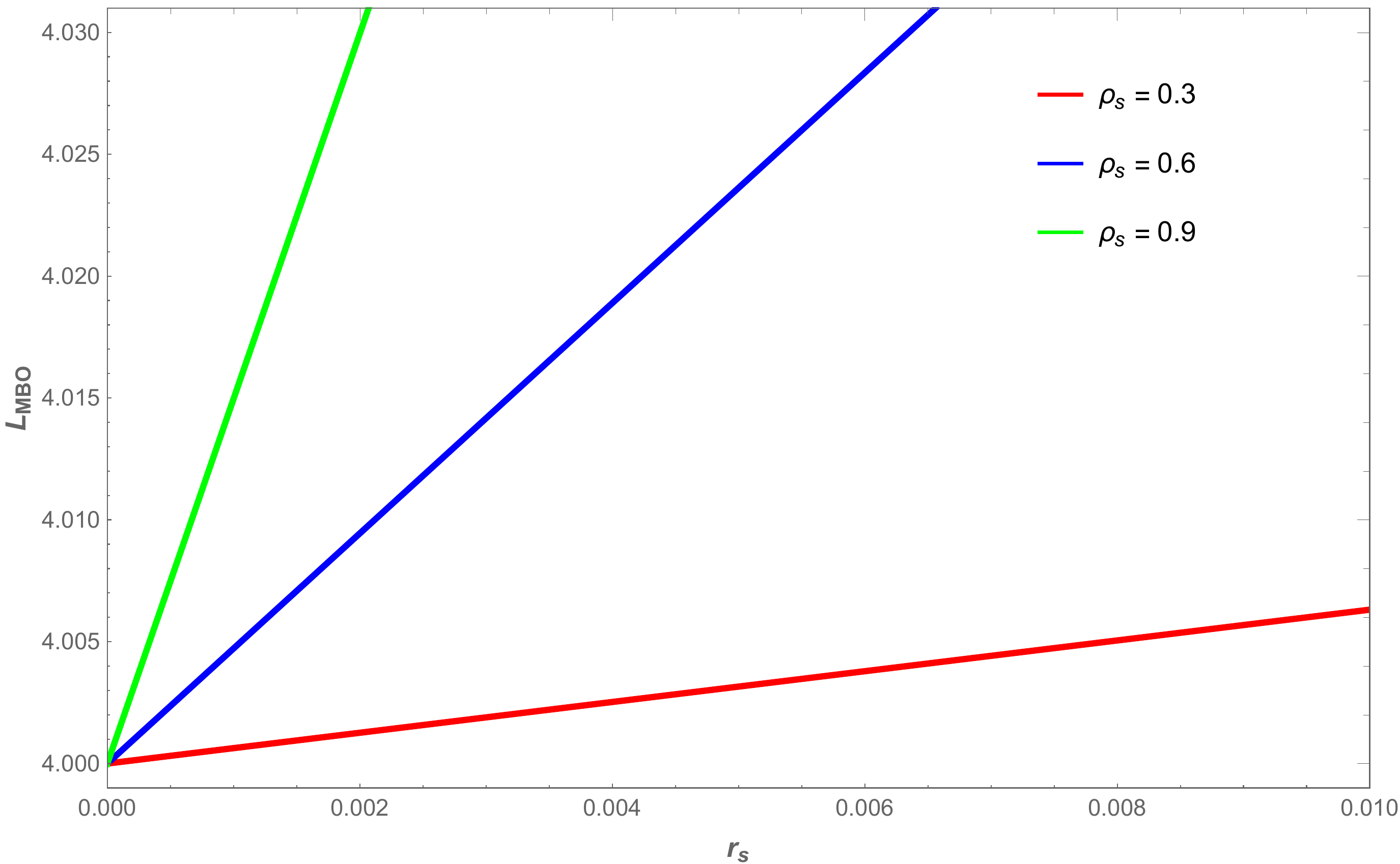}
		\end{subfigure}
		\caption{Left: The radius of the marginally bound orbit ($r_{\mathrm{MBO}}$) as a function of the dark matter halo core radius $r_{\mathrm{s}}$ for several values of the central density $\rho_{\mathrm{s}}$. Right: The corresponding angular momentum ($L_{\mathrm{MBO}}$) versus $r_{\mathrm{s}}$, with the same set of $\rho_{\mathrm{s}}$ values.}
		\label{rlmbo}
	\end{figure*}

	Due to the spherical symmetry of the Schwarzschild–Hernquist spacetime, the geodesic motion of a massive test particle can be restricted to the equatorial plane $(\theta=\pi/2)$ without loss of generality. For a neutral, non-spinning test particle moving along a timelike geodesic in this effective spacetime, the dynamics are governed by the Lagrangian 
	\begin{equation}
		\mathcal{L}=\frac{m}{2}g_{\mu\nu}\frac{\mathrm{d}x^{\mu}}{\mathrm{d}\lambda}\frac{\mathrm{d}x^{\nu}}{\mathrm{d}\lambda},\label{La}
	\end{equation}
	here, with $m$ set to 1, $g_{\mu\nu}$ is the metric tensor encoding the geometry of the spacetime, while $\lambda$ serves as an affine parameter parametrizing the particle's worldline. From the Lagrangian, one can define the covariant components of the generalized momentum as
	\begin{equation}
		p_{\mu}=\frac{\partial\mathcal{L}}
		{\partial\dot{x}^{\mu}}.\label{pmu}
	\end{equation}
	The equations of motion for the test particle then take the following form:
	\begin{equation}
		\begin{split}
			p_{t} &= g_{tt}\dot{t}=-E,\\
			p_{\phi} &= g_{\phi\phi}\dot{\phi}=L,\\
			p_{r} &= g_{rr}\dot{r},\\
			p_{\theta} &= g_{\theta\theta}\dot{\theta}=0.\label{meq}
		\end{split}    
	\end{equation}
	where $E$ and $L$ denote the conserved energy and orbital angular momentum of the test particle, respectively. For timelike geodesics confined to the equatorial plane, the normalization condition $g_{\mu\nu}\dot{x}^{\mu}\dot{x}^{\nu}=-1$ leads directly to the effective potential:
	\begin{equation}
		V_{\mathrm{eff}}=f(r)\left(1+\frac{L^{2}}{r^{2}}\right).\label{Veff}
	\end{equation}
	Figure~\ref{epotential} shows that the effective potential depends sensitively on both the DM halo parameters $(\rho_{\mathrm{s}},r_{\mathrm{s}})$, and the particle's angular momentum $L$: increasing the halo mass deepens the potential well, while increasing $L$ raises the potential.
	The asymptotic behavior of the effective potential (Eq. \ref{Veff}) is such that $V_{\mathrm{eff}} \rightarrow1$ as $r\rightarrow\infty$. As a result, the critical energy $E=1$ is established for particles moving along bound or unbound orbits, with particles in bound orbits required to satisfy $E\leq1$. General bound orbits exist only for specific combinations of energy $E$ and angular momentum $L$, which must lie within the following ranges:
	\begin{equation}
		L_{\mathrm{ISCO}} \leq L \quad \text{and} \quad E_{\mathrm{ISCO}} \leq E \leq E_{\mathrm{MBO}} = 1. \label{range}
	\end{equation}
	where $E_{\mathrm{ISCO}}$ and $L_{\mathrm{ISCO}}$ are the energy and angular momentum of the ISCO, and $E_{\mathrm{MBO}}$ is the energy of the MBO, defined by
	\begin{equation}
		V_{\mathrm{eff}} = 1,\ \frac{\mathrm{d}V_{\mathrm{eff}}}{\mathrm{d}r} = 0.\label{mbo}
	\end{equation}
	The solution to Eq. (\ref{mbo}) defines the radius and angular momentum of the MBO, denoted respectively by $r_{\mathrm{MBO}}$ 
	and $L_{\mathrm{MBO}}$. Figure~\ref{rlmbo} shows their dependence on the central DM density $\rho_{\mathrm{s}}$ for a Schwarzschild BH embedded in a Hernquist halo. Both $r_{\mathrm{MBO}}$ 
	and $L_{\mathrm{MBO}}$ increase monotonically with $\rho_{\mathrm{s}}$, reflecting the growing gravitational influence of the DM distribution.
	
\begin{figure*}[htbp]
		\centering
		\begin{subfigure}{0.32\textwidth}
			\centering
			\includegraphics[width=\linewidth]{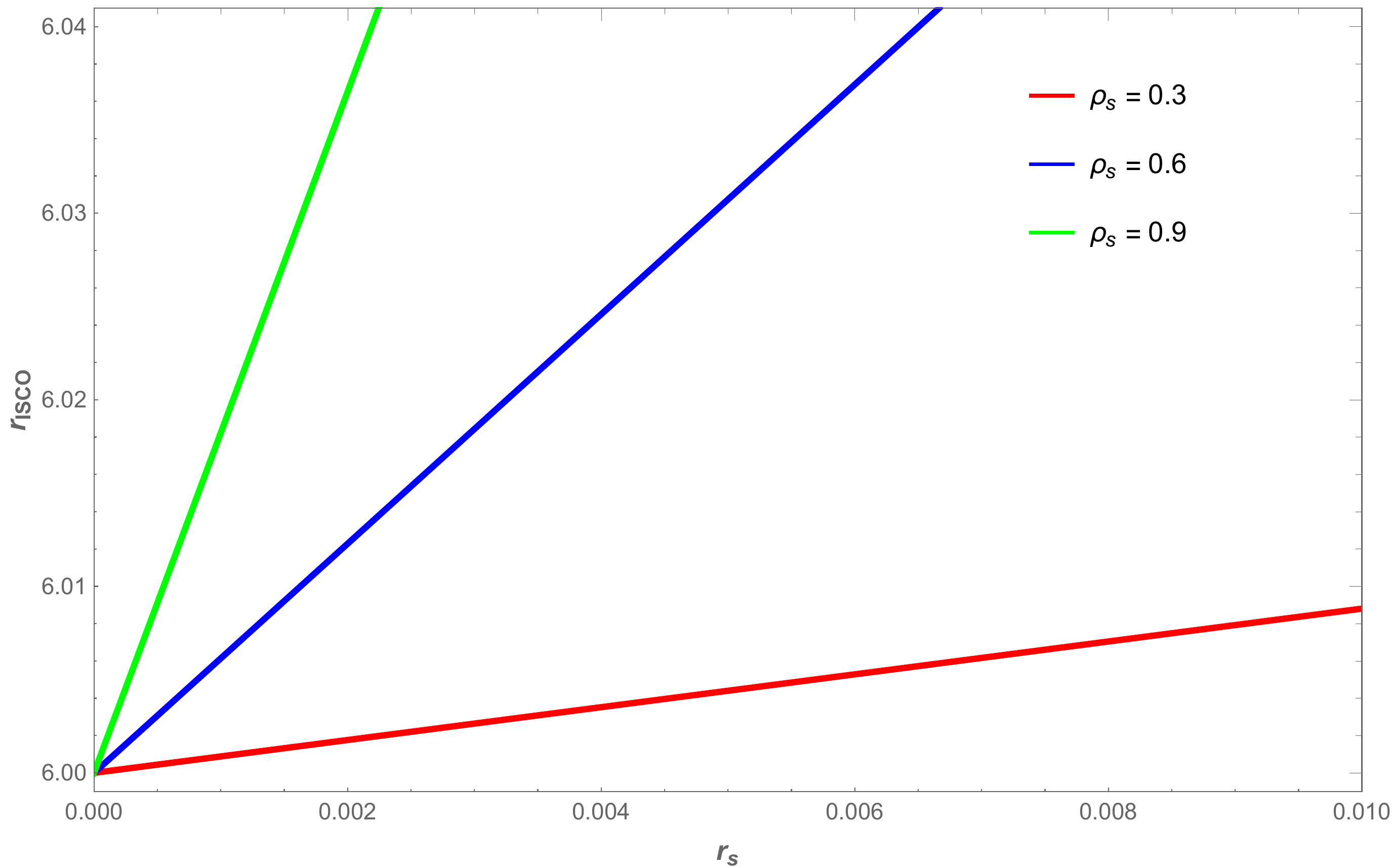}
		\end{subfigure}%
		\hfill
		\begin{subfigure}{0.32\textwidth}
			\centering
			\includegraphics[width=\linewidth]{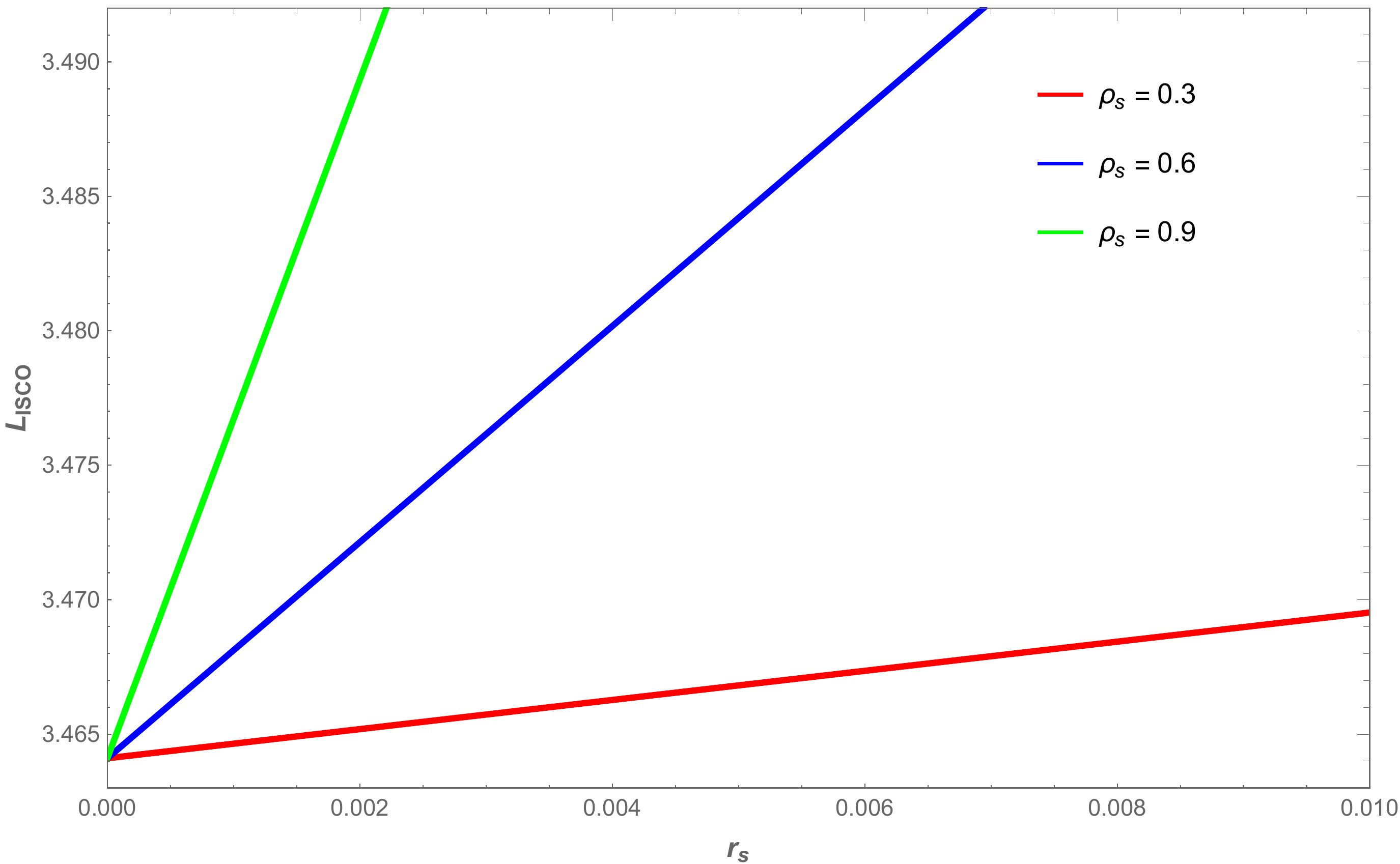}
		\end{subfigure}%
		\hfill
		\begin{subfigure}{0.32\textwidth}
			\centering
			\includegraphics[width=\linewidth]{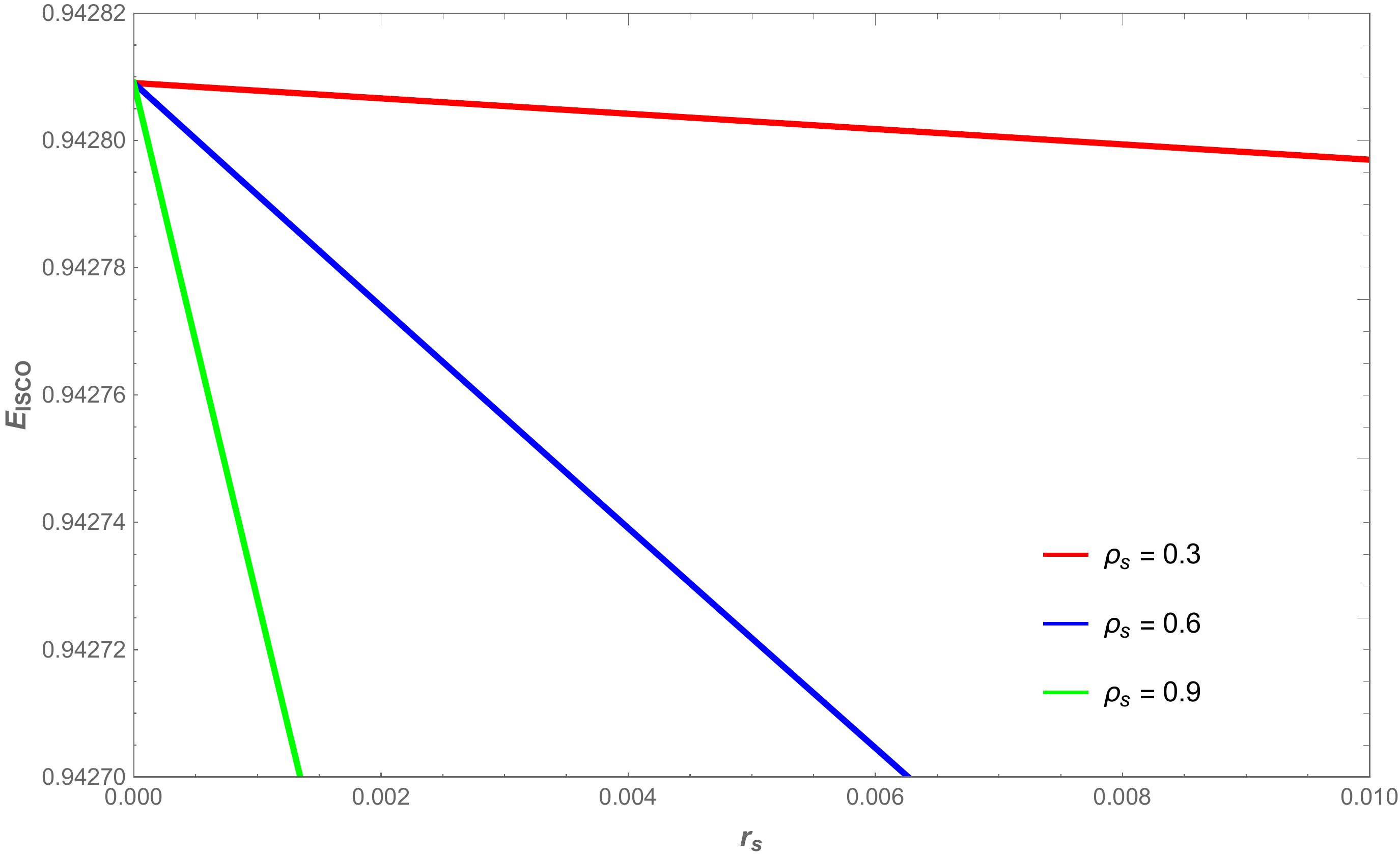}
		\end{subfigure}
		
		\caption{The ISCO radius $r_{\mathrm{ISCO}}$, the angular momentum $L_{\mathrm{ISCO}}$, and the energy $E_{\mathrm{ISCO}}$ as functions of the Hernquist halo scale radius $r_{\mathrm{s}}$, for different values of the central density $\rho_{\mathrm{s}}$.}
		\label{rleisco}
	\end{figure*}

\begin{figure*}[htbp]
		\centering
		\begin{subfigure}{0.38\textwidth}
			\includegraphics[width=\linewidth]{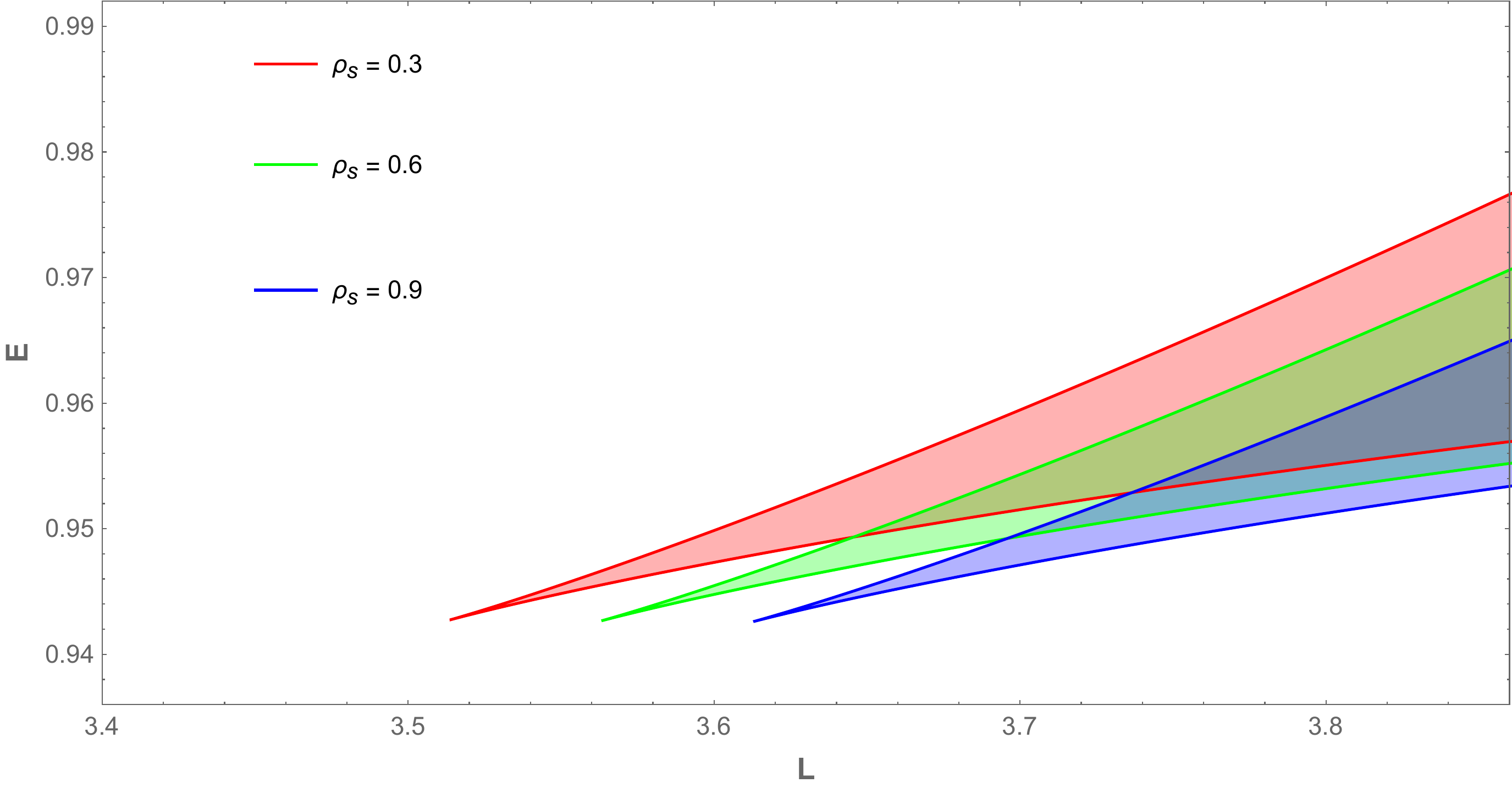}
		\end{subfigure}%
		\hspace{0.13\textwidth}% 控制间距
		\begin{subfigure}{0.38\textwidth}
			\includegraphics[width=\linewidth]{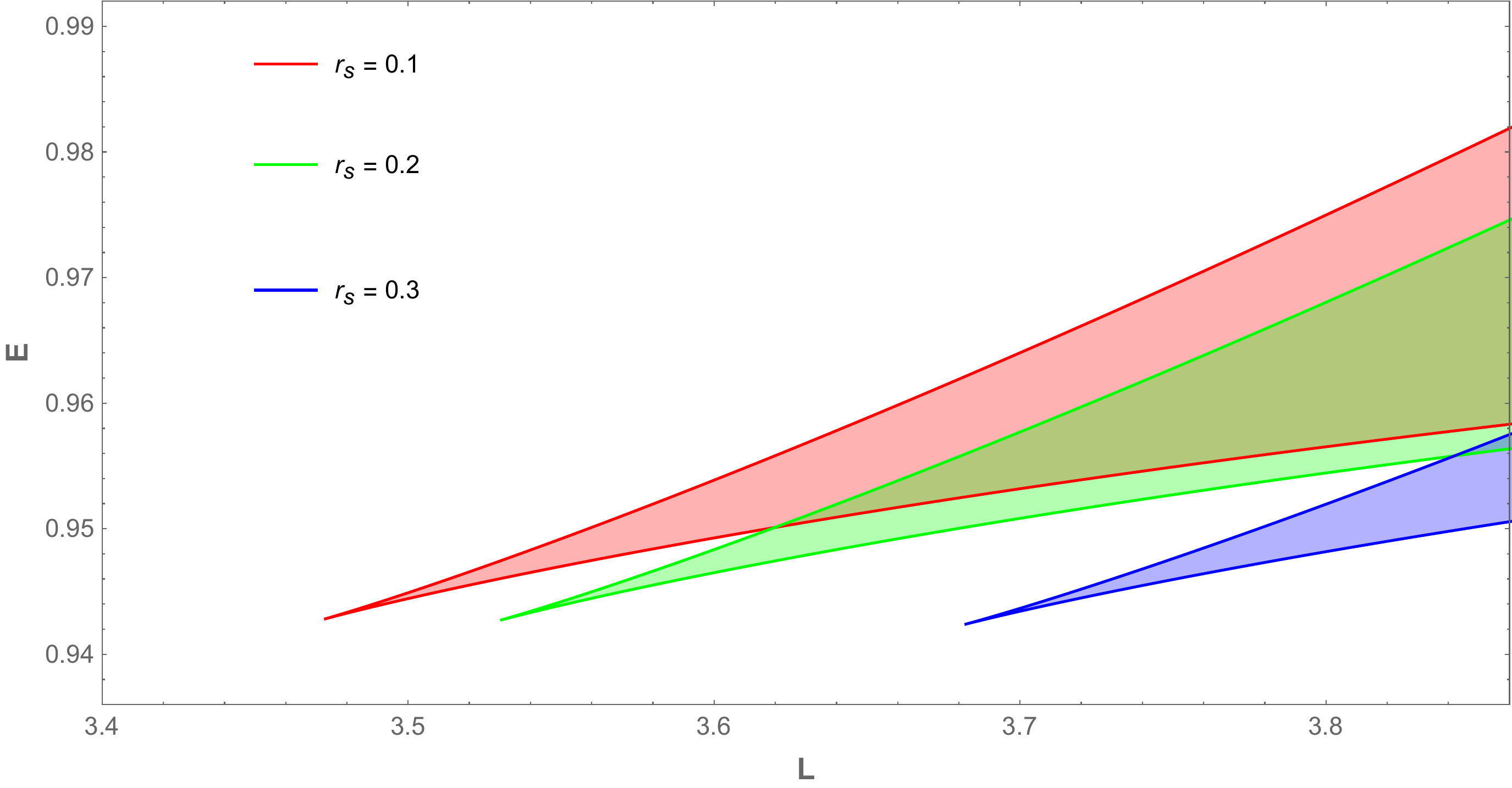}
		\end{subfigure}
		
		\caption{The allowed parameter space of the energy and the orbital angular momentum for the bound orbits around the Schwarzschild BH embedded in a Hernquist dark matter halo with different values of the parameters. Left: $r_{\mathrm{s}}=0.2$; Right: $\rho_{\mathrm{s}}=0.4$.}
		\label{el}
	\end{figure*}

	The ISCO is another significant class of bound orbits, determined by the following set of conditions:
	\begin{equation}
		\dot{r} = 0,\quad \frac{\mathrm{d}V_{\mathrm{eff}}}{\mathrm{d}r} = 0 \quad \text{and} \quad \frac{\mathrm{d}^{2}V_{\mathrm{eff}}}{\mathrm{d}r^{2}} = 0.
	\end{equation}
	Figure~\ref{rleisco} displays the radius, the specific orbital angular momentum, and the energy of the ISCO as functions of the DM halo's scale radius $r_{\mathrm{s}}$, for several values of the central density $\rho_{\mathrm{s}}$. Both $r_{\mathrm{ISCO}}$ and $L_{\mathrm{ISCO}}$ increase with larger $r_{\mathrm{s}}$ or $\rho_{\mathrm{s}}$, reflecting the enhanced gravitational pull of the surrounding DM distribution. Correspondingly, the ISCO energy $E_{\mathrm{ISCO}}$ decreases with increasing $r_{\mathrm{s}}$ and $\rho_{\mathrm{s}}$, a consequence of the outward migration of the ISCO away from the BH in response to the stronger halo gravity.
	To further characterize bound orbits, Figure~\ref{el} displays the allowed region in the $(E,L)$ parameter space for a Schwarzschild BH surrounded by a Hernquist DM halo, obtained from Eq. (\ref{range}). An increase in the halo parameters results in a downward shift of this region.
	
	\section{The periodic orbits around the Schwarzschild black hole embedded in a Hernquist dark matter halo}
	\label{section3}
	
\begin{figure*}[htbp]
		\centering
		\begin{subfigure}{0.45\textwidth}
			\includegraphics[width=3in, height=5.5in, keepaspectratio]{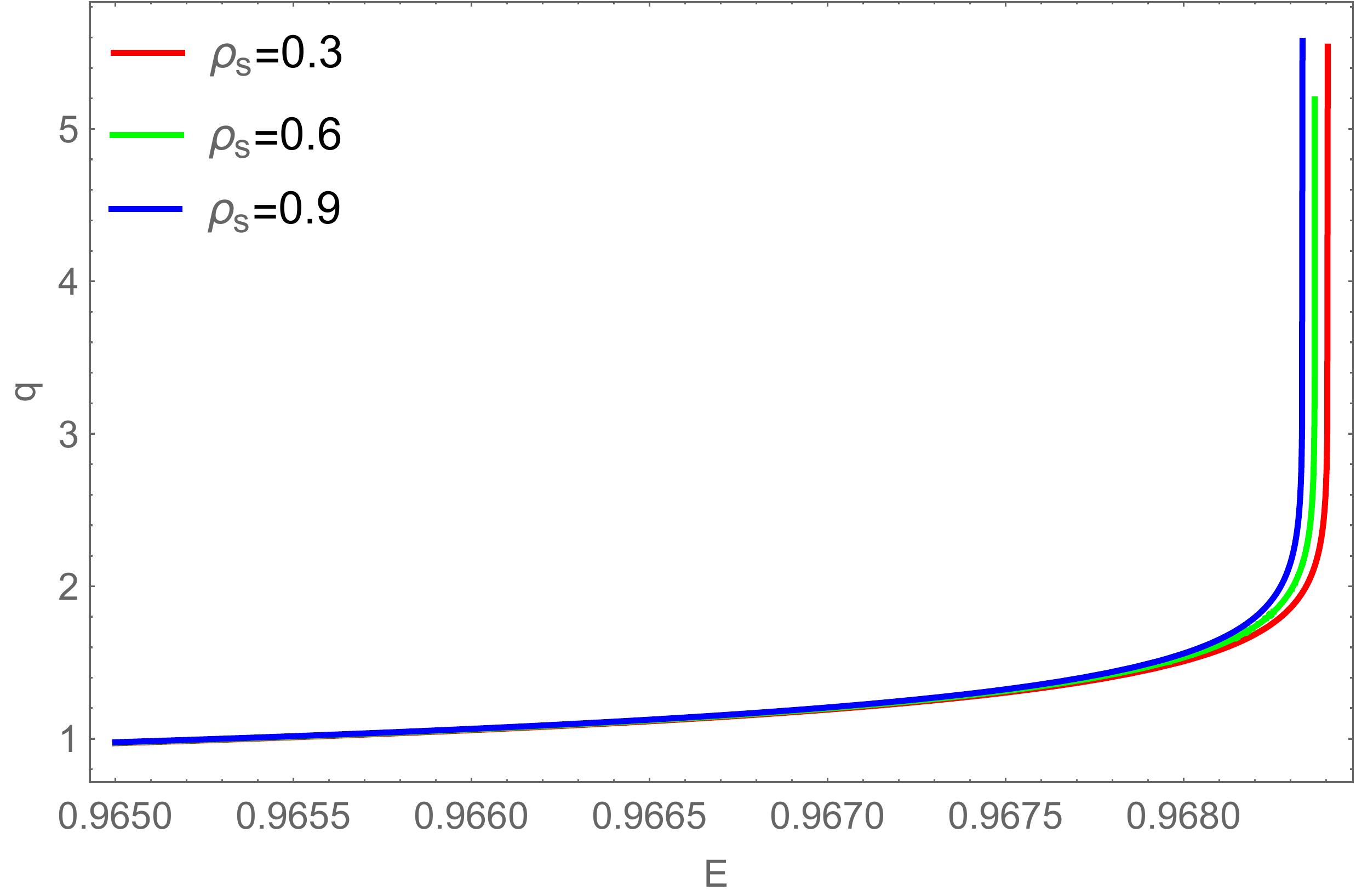}
		\end{subfigure}
		\hfill
		\begin{subfigure}{0.45\textwidth}
			\includegraphics[width=3in, height=5.5in,keepaspectratio]{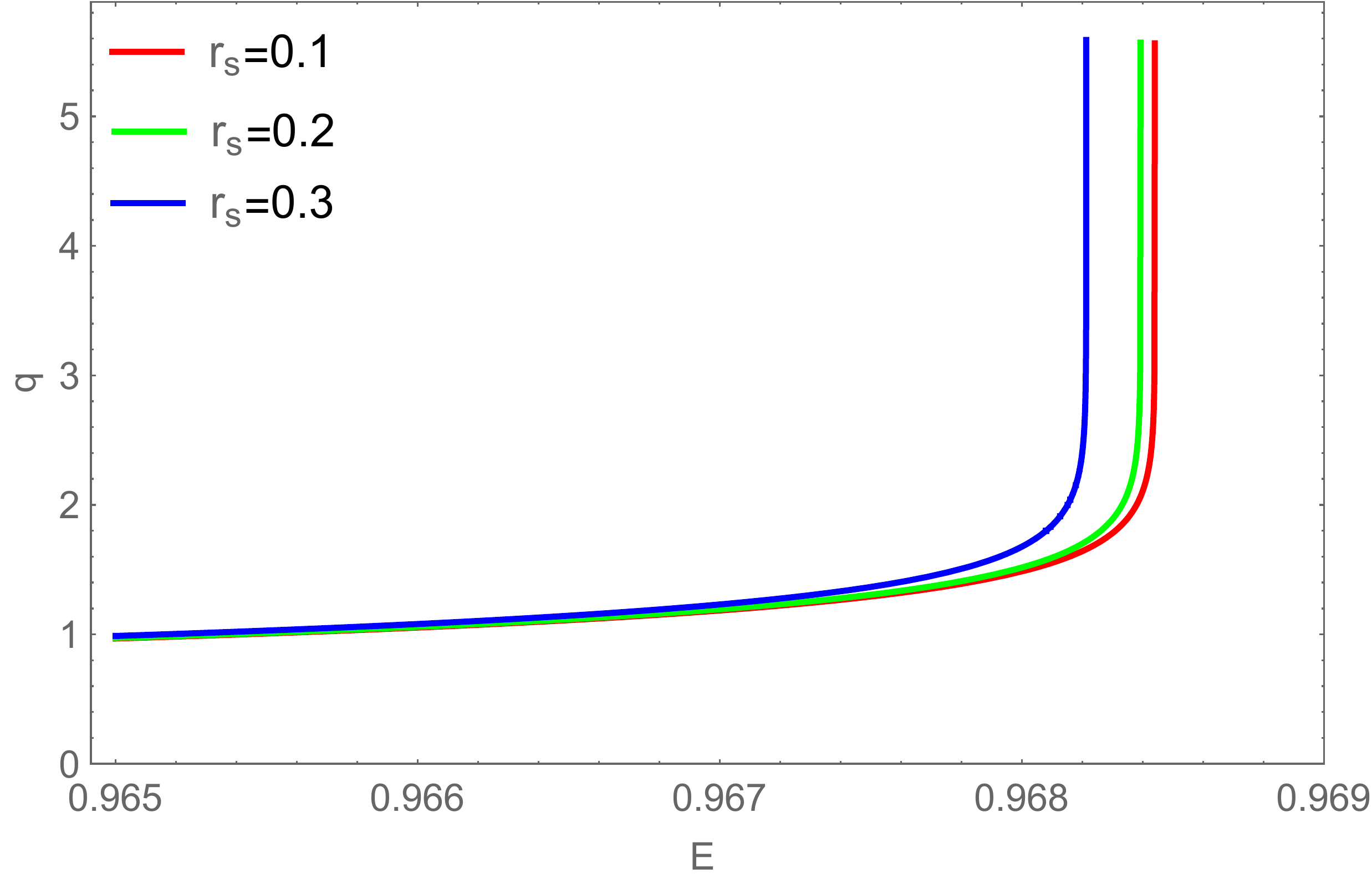}
		\end{subfigure}
		\begin{subfigure}{0.45\textwidth}
			\includegraphics[width=3in, height=5.5in, keepaspectratio]{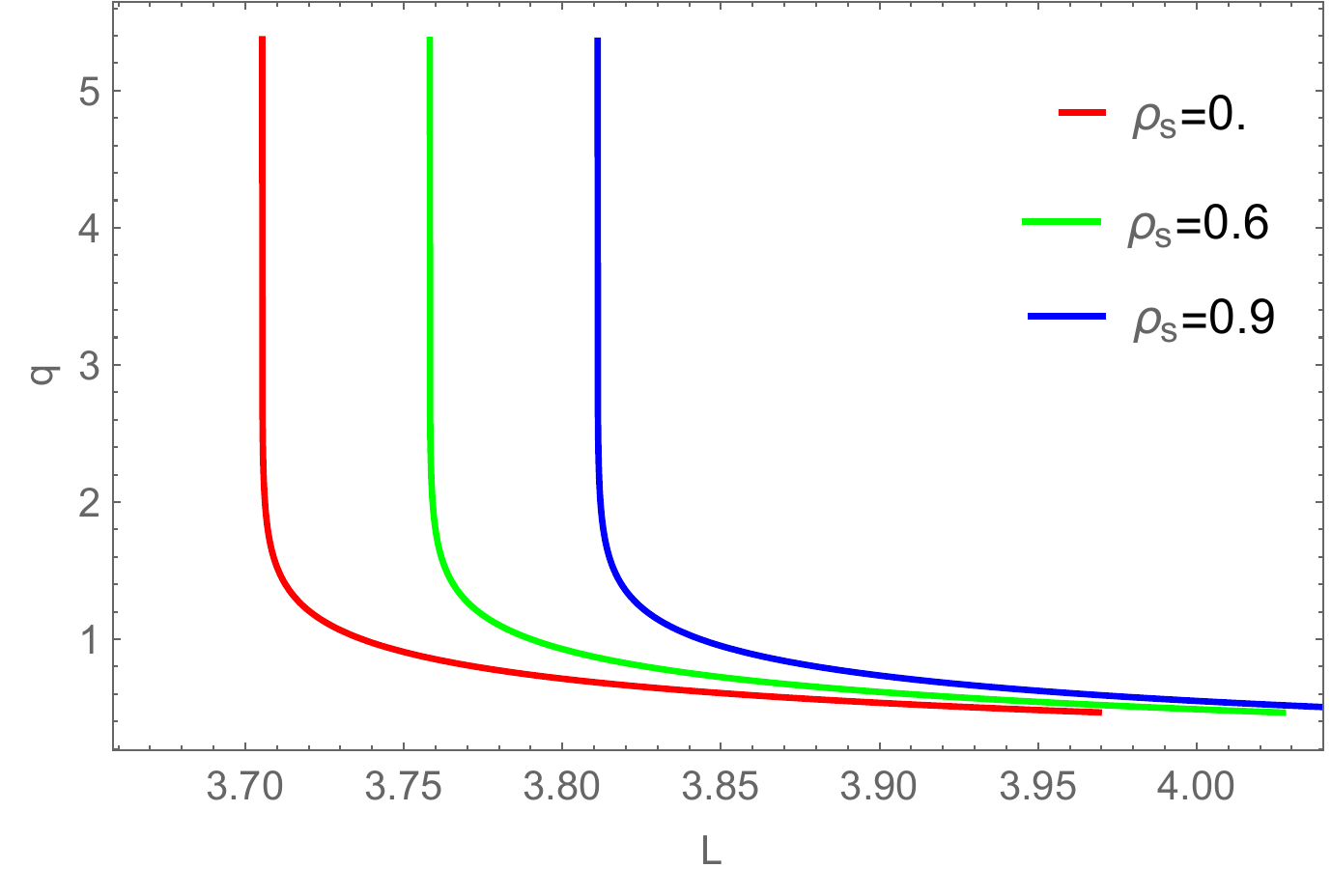}
		\end{subfigure}
		\hfill
		\begin{subfigure}{0.45\textwidth}
			\includegraphics[width=3in, height=5.5in, keepaspectratio]{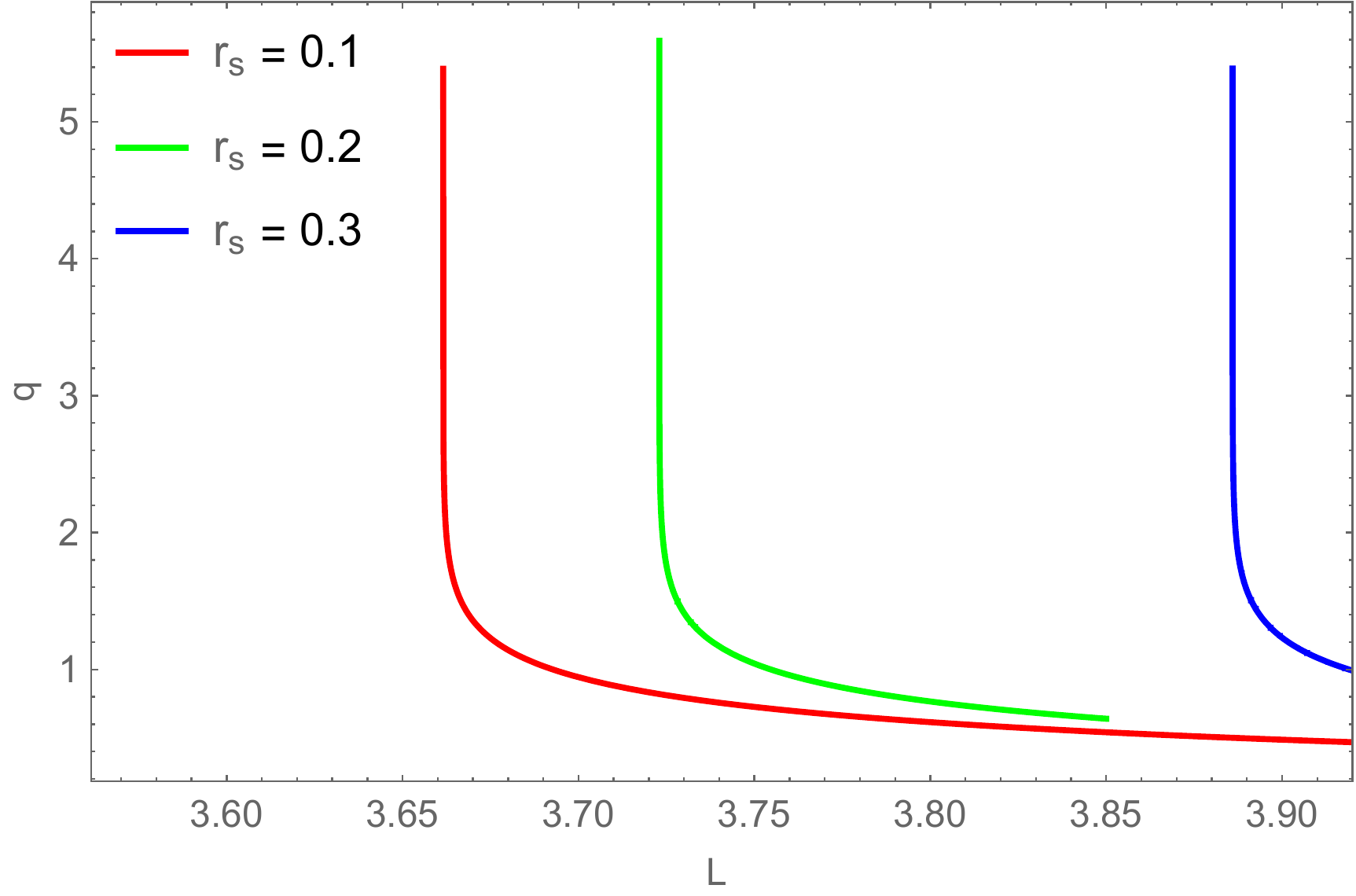}
		\end{subfigure}
		\caption{Top panel: The dependence of the rational number $q$ on the energy $E$ for periodic orbits around a Schwarzschild black hole embedded in a Hernquist dark matter halo. The left panel shows the results for varying central density $\rho_{\mathrm{s}}$ (with fixed scale radius $r_{\mathrm{s}} = 0.2$), while the right panel displays the dependence on the scale radius $r_{\mathrm{s}}$ (with fixed central density $\rho_{\mathrm{s}} = 0.4$). In both cases, the orbital angular momentum is set to $L = \frac{1}{2}(L_{\rm MBO} + L_{\rm ISCO})$. Bottom panel: The rational number $q$ as a function of the orbital angular momentum $L$, with particle energy fixed as $E = 0.96$. Different curves correspond to distinct combinations of the halo parameters $\rho_{\mathrm{s}}$ and $r_{\mathrm{s}}$.}
		\label{qel}
	\end{figure*}

	In order to explore the impact of the Hernquist DM halo on particle motion, we investigate the corresponding periodic orbits around the Schwarzschild BH embedded in such a halo. Among all bound orbits, periodic orbits play a particularly special role, as their fundamental orbital frequencies maintain a rational relationship. Each periodic orbit can be uniquely described by a triplet of integers $(z,w,v)$, referred to as the zoom, whirl, and vertex numbers, respectively. The rational relationship among the orbital frequencies can be expressed in the form given by \cite{Levin:2008mq}.
	\begin{equation}
		q=\frac{\omega_{\phi}}{\omega_{r}}-1=w+\frac{v}{z},\label{qwvz}  
	\end{equation}
	where $\omega_{\phi}$ and $\omega_{r}$ represent the azimuthal and radial orbital frequencies, respectively. Based on the equations of motion, the rational number  is given by \cite{Yang:2024lmj,Shabbir:2025kqh,Jiang:2024cpe}:
	\begin{equation}
		q=\frac{1}{\pi}\int_{r_{1}}^{r_{2}}\frac{\dot{\phi}}{\dot{r}}-1=\frac{1}{\pi}\int_{r_{1}}^{r_{2}}\frac{1}{r^{2}\sqrt{E^{2}-V_{\mathrm{eff}}}}.\label{qint}  
	\end{equation}
	For a given periodic orbit, $r_{1}$ and $r_{2}$ represent the minimum and maximum orbital radii, i.e., the periapsis and apoapsis, respectively. Figure~\ref{qel} presents the behavior of the rational number $q$ for bound periodic orbits around a Schwarzschild BH immersed in a Hernquist DM halo, plotted as a function of the particle's specific energy $E$ (top panel) and specific angular momentum $L$ (bottom panel), for several choices of the halo parameters $\rho_{\mathrm{s}}$ and $r_{\mathrm{s}}$. In the upper panel, $q$ grows monotonically with $E$ over most of the allowed range, followed by a steep upturn as $E$ approaches the upper bound of the bound-orbit domain—signaling the onset of orbital instability near the separatrix. Notably, increasing either $\rho_{\mathrm{s}}$ or $r_{\mathrm{s}}$ shifts the entire $q(E)$ curve toward lower energies, reflecting the deepening of the effective potential well due to the enhanced DM contribution. The lower panel reveals that $q$ diverges sharply as $L$ decreases toward its minimum admissible value, while at larger $L$, the rational number $q$ declines smoothly, indicating a transition toward more circular-like motion.
	
\begin{figure*}[htbp]
		\centering
		
		\begin{subfigure}{0.3\textwidth}
			\centering
			\includegraphics[width=\linewidth, keepaspectratio]{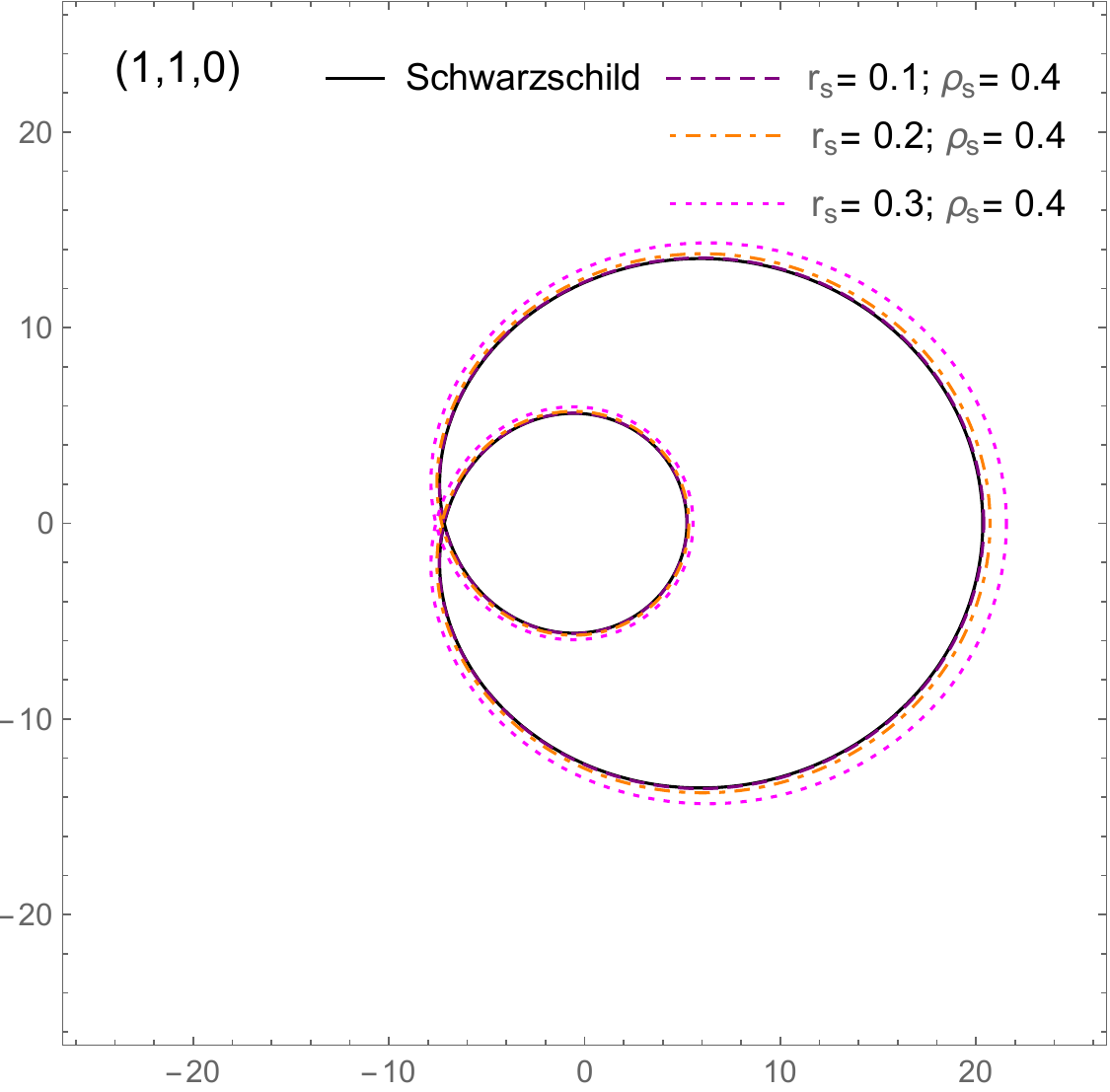}
		\end{subfigure}
		\hfill
		\begin{subfigure}{0.3\textwidth}
			\centering
			\includegraphics[width=\linewidth, keepaspectratio]{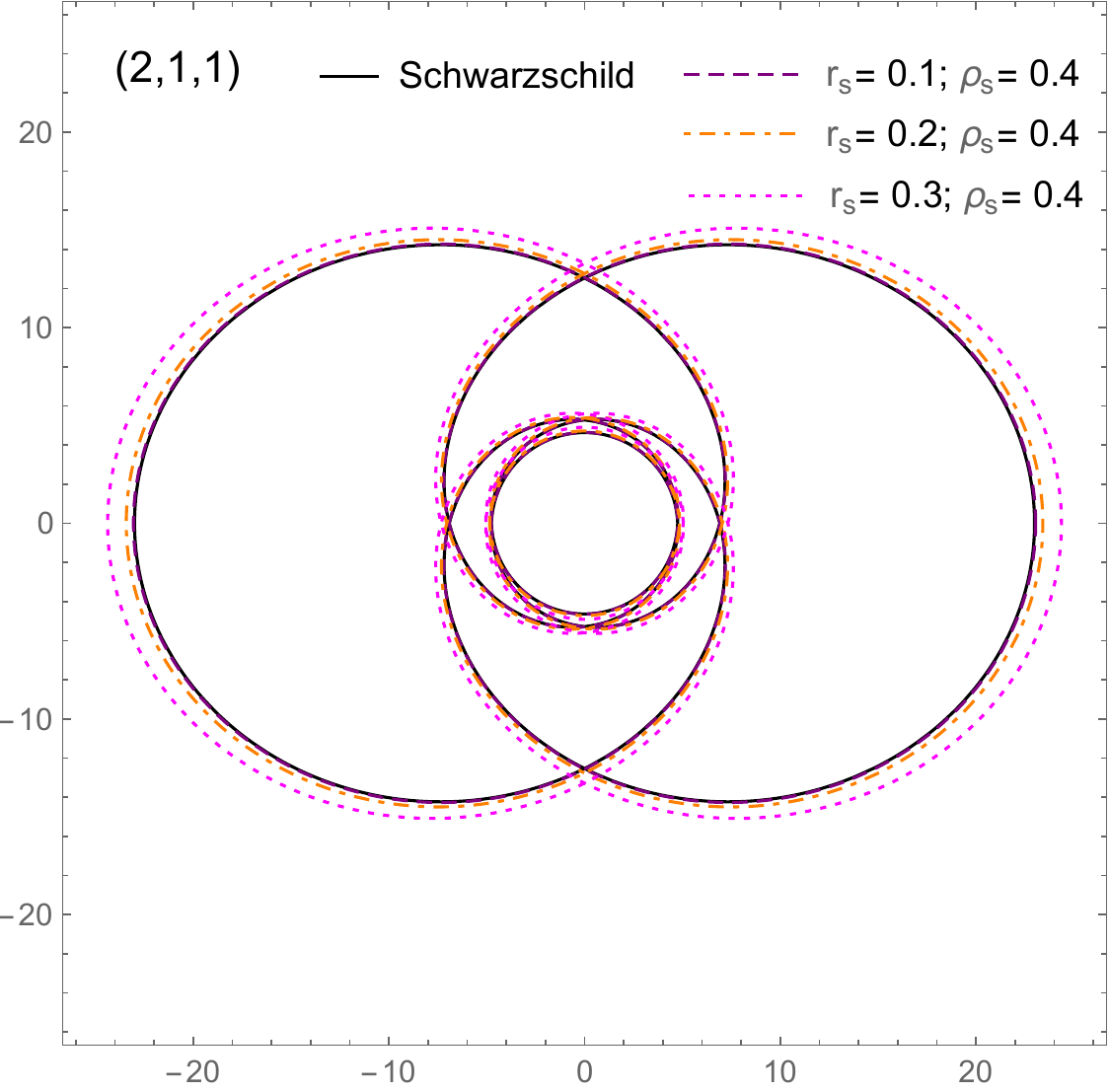}
		\end{subfigure}
		\hfill
		\begin{subfigure}{0.3\textwidth}
			\centering
			\includegraphics[width=\linewidth, keepaspectratio]{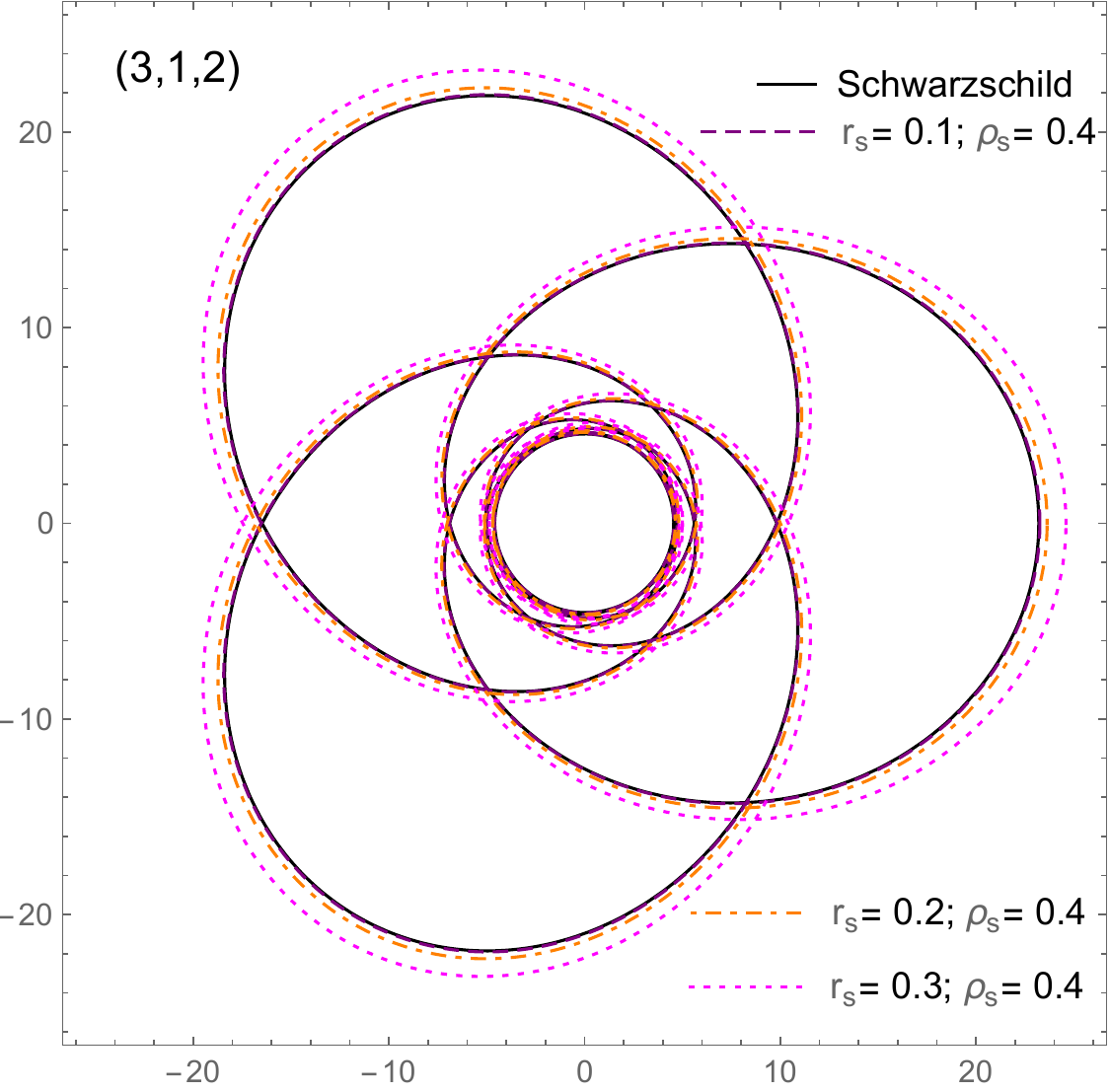}
		\end{subfigure}
		
		\vspace{0.5cm} 
		
		\begin{subfigure}{0.3\textwidth}
			\centering
			\includegraphics[width=\linewidth, keepaspectratio]{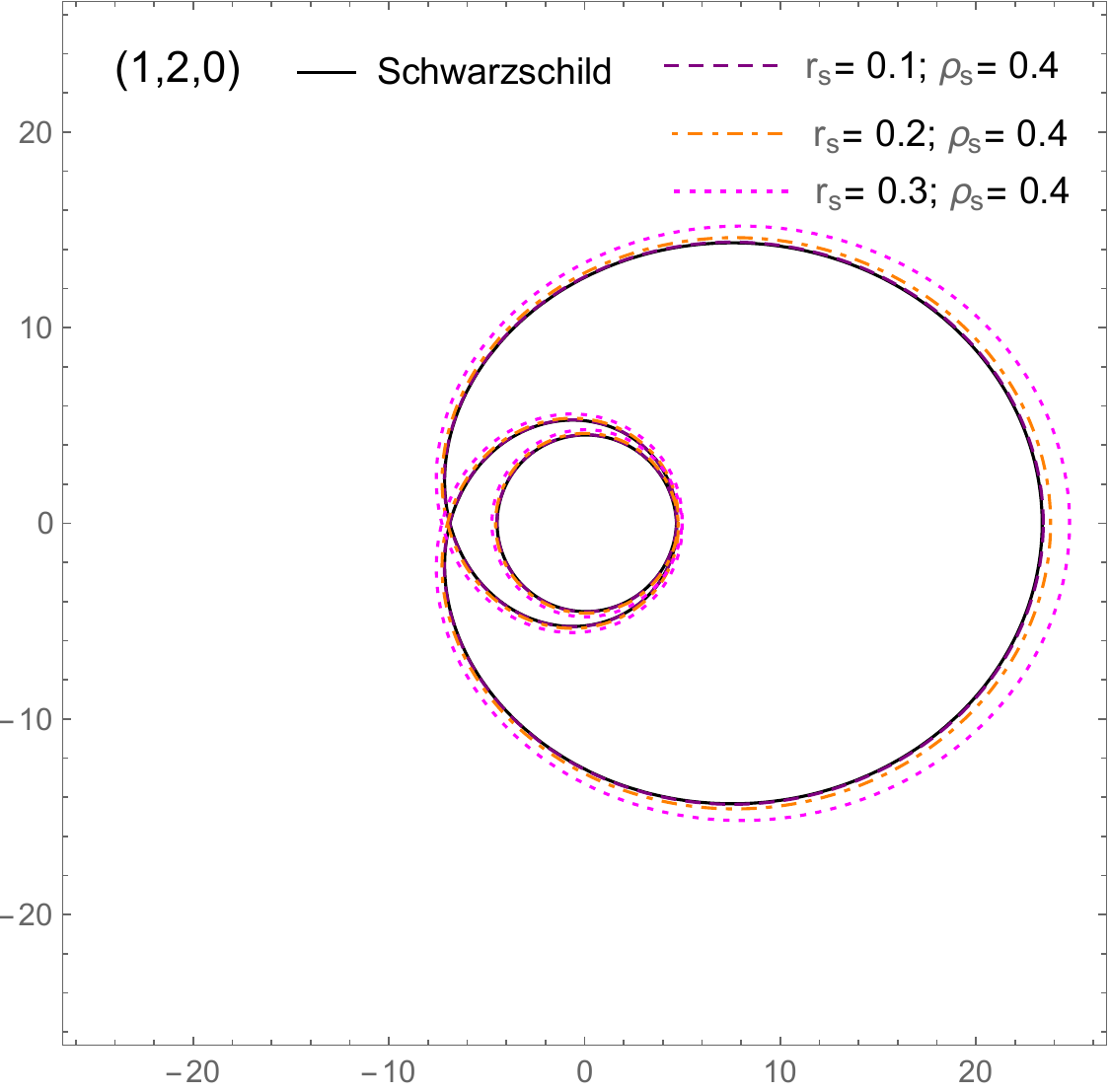}
		\end{subfigure}
		\hfill
		\begin{subfigure}{0.3\textwidth}
			\centering
			\includegraphics[width=\linewidth, keepaspectratio]{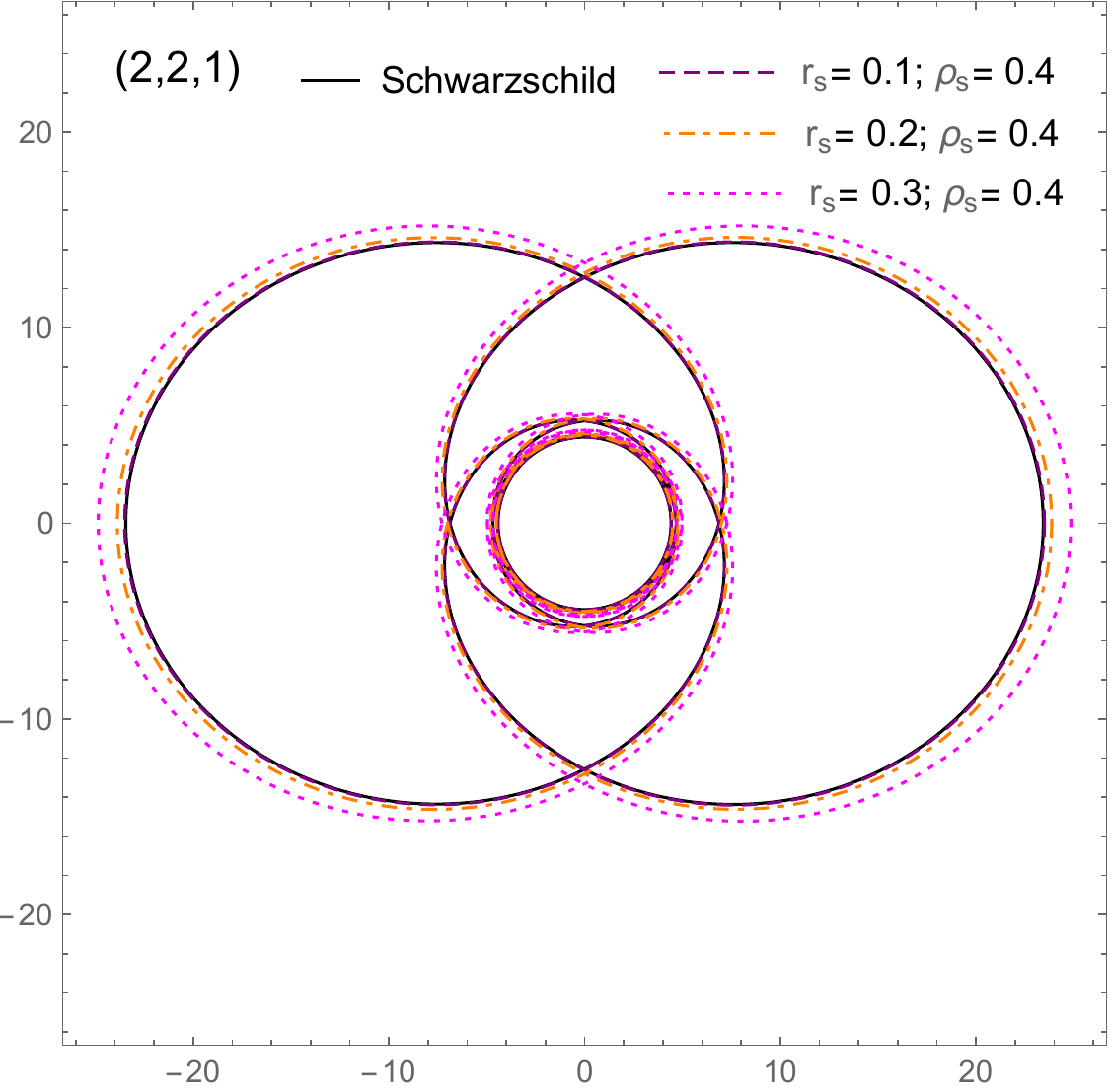}
		\end{subfigure}
		\hfill
		\begin{subfigure}{0.3\textwidth}
			\centering
			\includegraphics[width=\linewidth, keepaspectratio]{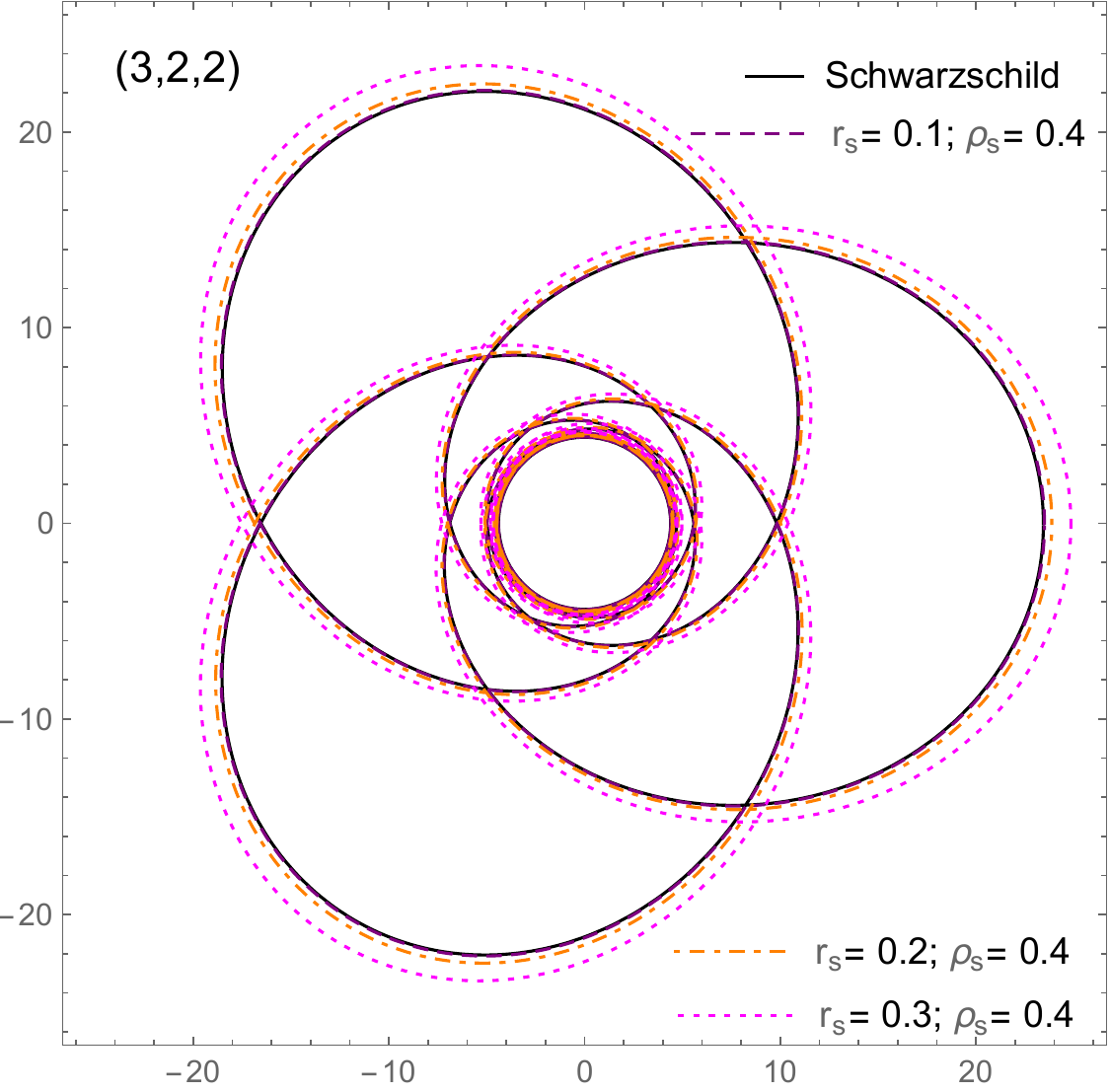}
		\end{subfigure}
		
		\caption{The periodic orbits labeled by $(z,w,v)$ around a Schwarzschild black hole embedded in a Hernquist dark matter halo, shown for various values of the scale radius $r_{\mathrm{s}}$, with fixed parameters $L=\frac{1}{2}(L_{\mathrm{MBO}}+L_{\mathrm{ISCO}})$ and $\rho_{\mathrm{s}}=0.4$.}
		\label{orbitL}
	\end{figure*}

\begin{figure*}[htbp]
		\centering
		
		\begin{subfigure}{0.3\textwidth}
			\centering
			\includegraphics[width=\linewidth, keepaspectratio]{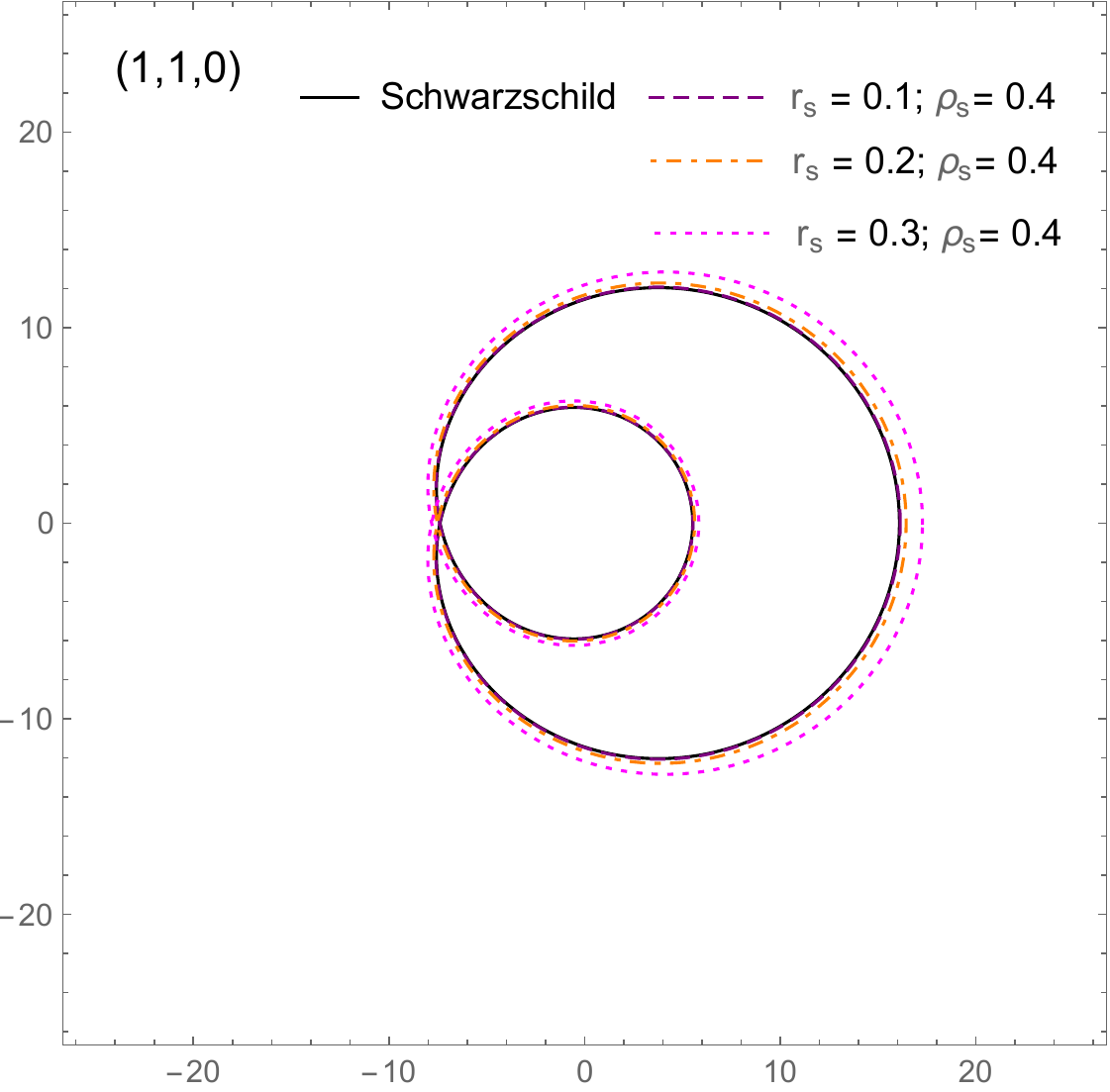}
		\end{subfigure}
		\hfill
		\begin{subfigure}{0.3\textwidth}
			\centering
			\includegraphics[width=\linewidth, keepaspectratio]{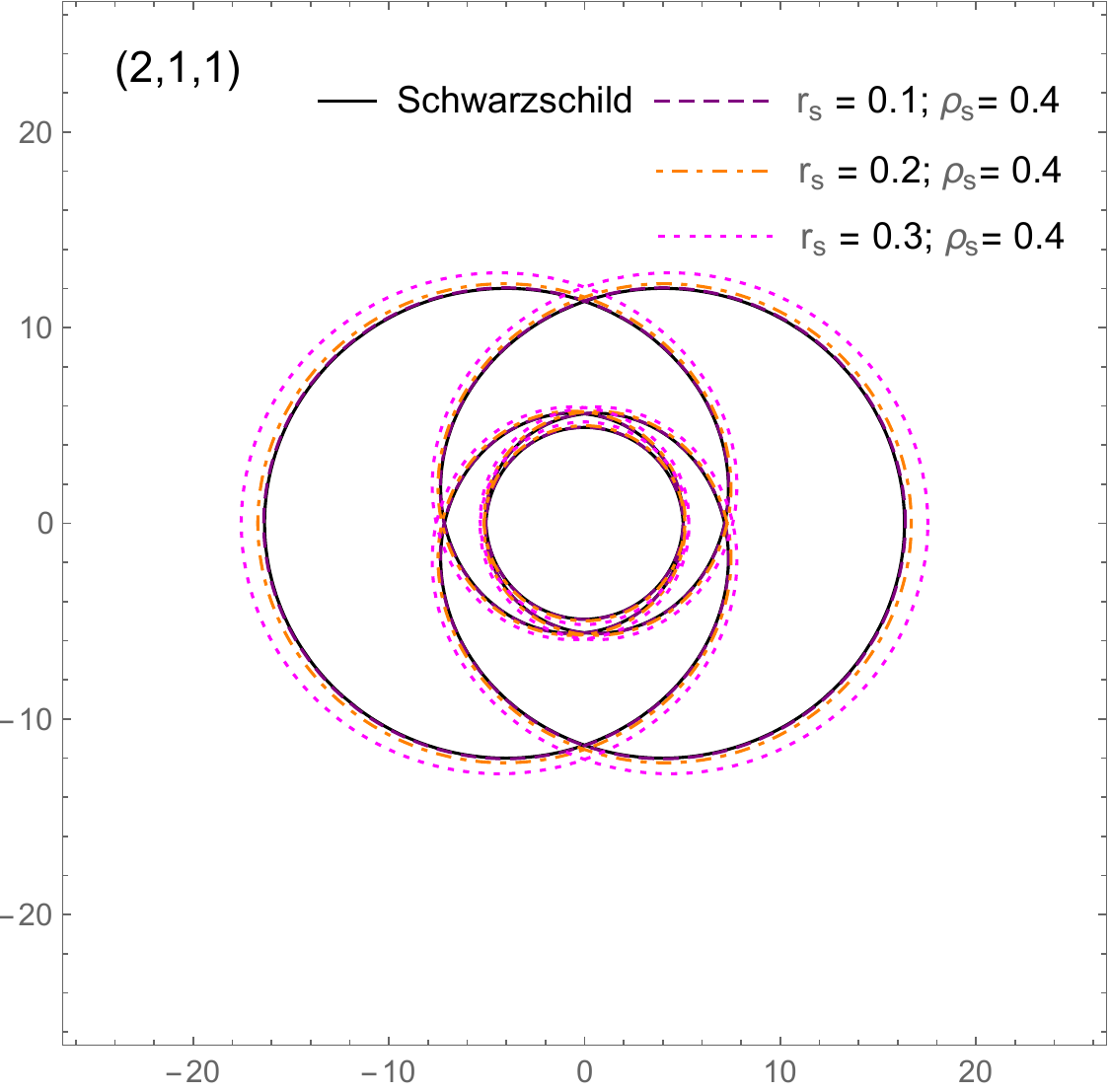}
		\end{subfigure}
		\hfill
		\begin{subfigure}{0.3\textwidth}
			\centering
			\includegraphics[width=\linewidth, keepaspectratio]{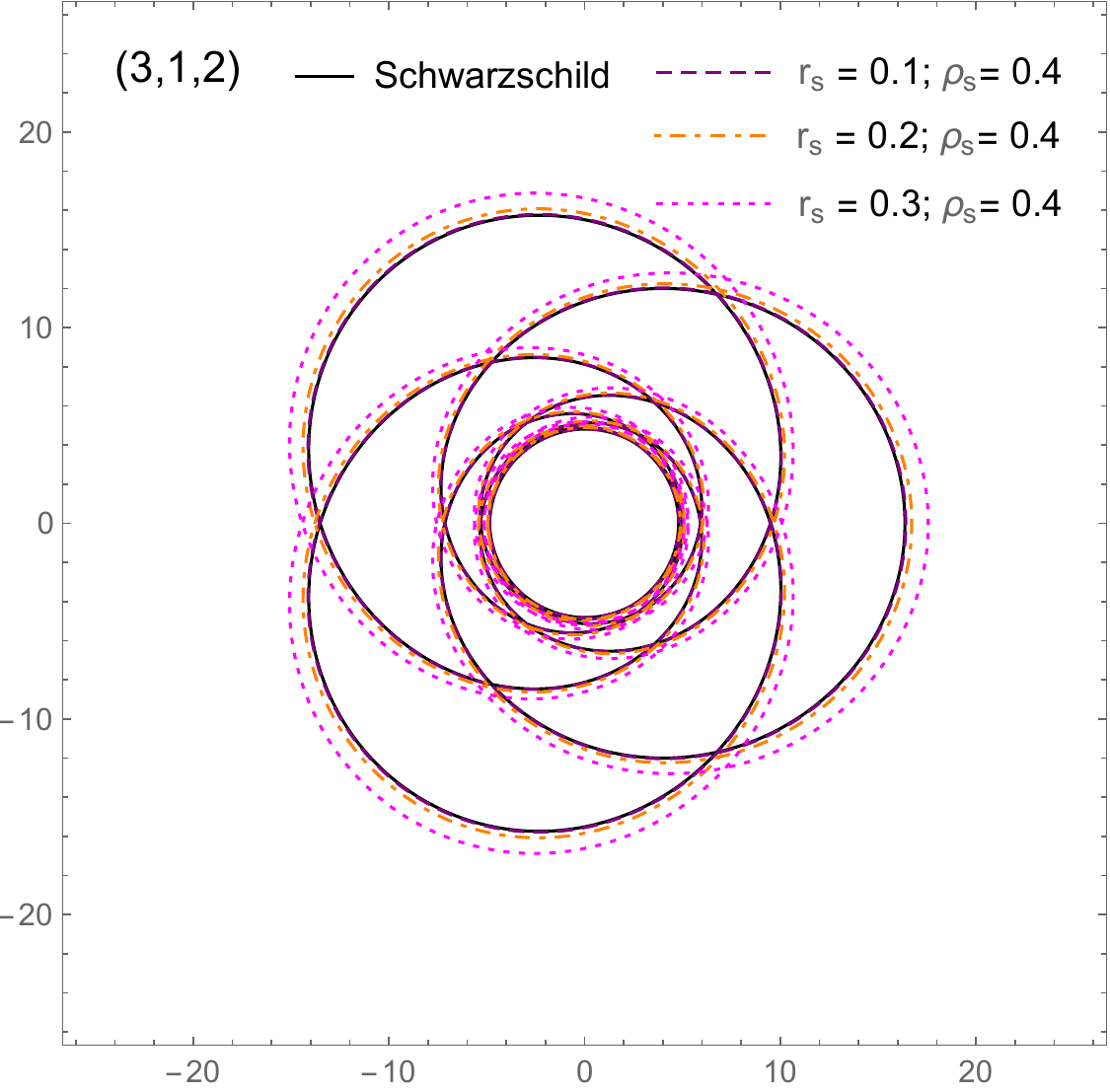}
		\end{subfigure}
		
		\vspace{0.5cm}

		\begin{subfigure}{0.3\textwidth}
			\centering
			\includegraphics[width=\linewidth, keepaspectratio]{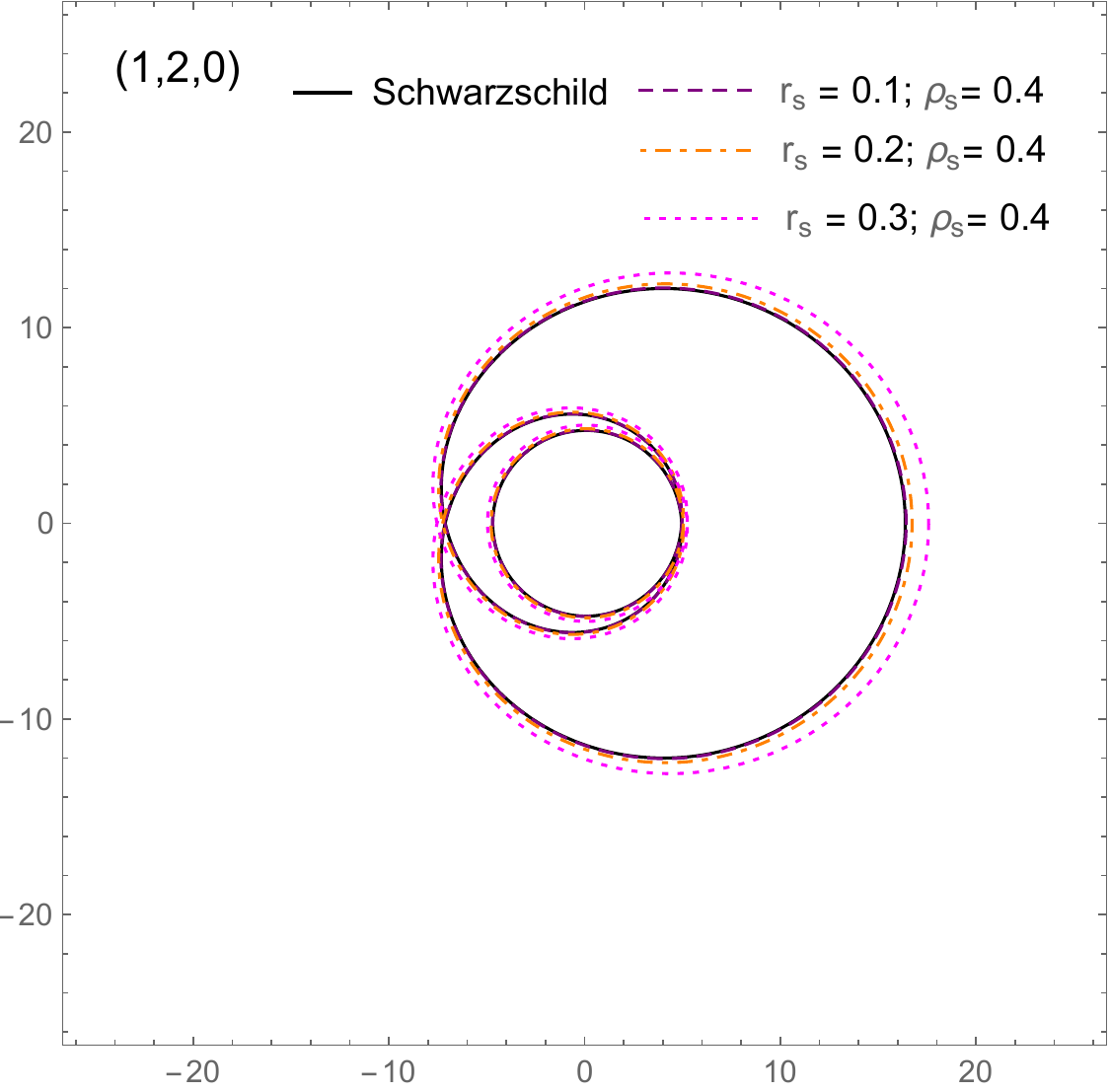}
		\end{subfigure}
		\hfill
		\begin{subfigure}{0.3\textwidth}
			\centering
			\includegraphics[width=\linewidth, keepaspectratio]{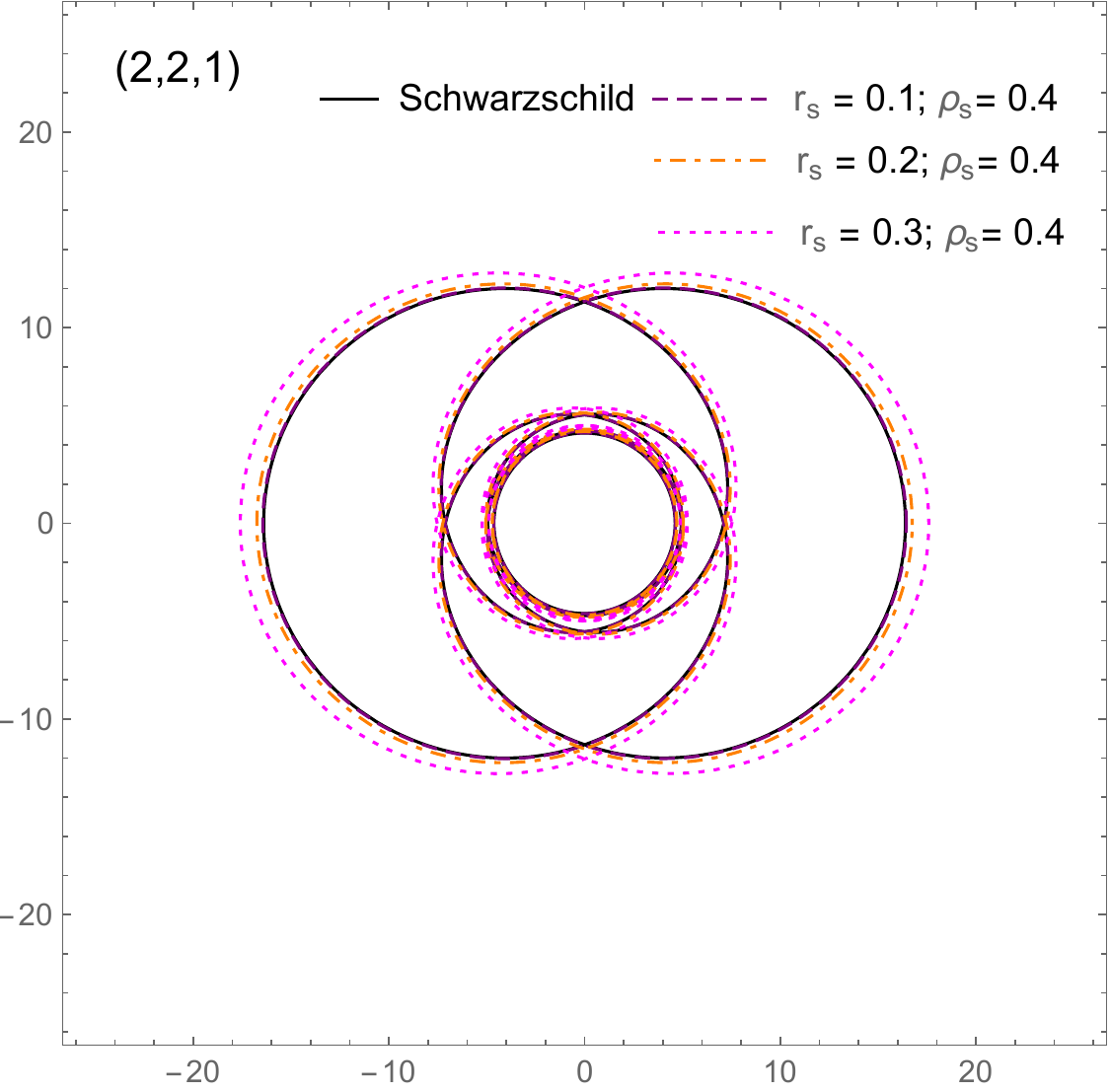}
		\end{subfigure}
		\hfill
		\begin{subfigure}{0.3\textwidth}
			\centering
			\includegraphics[width=\linewidth, keepaspectratio]{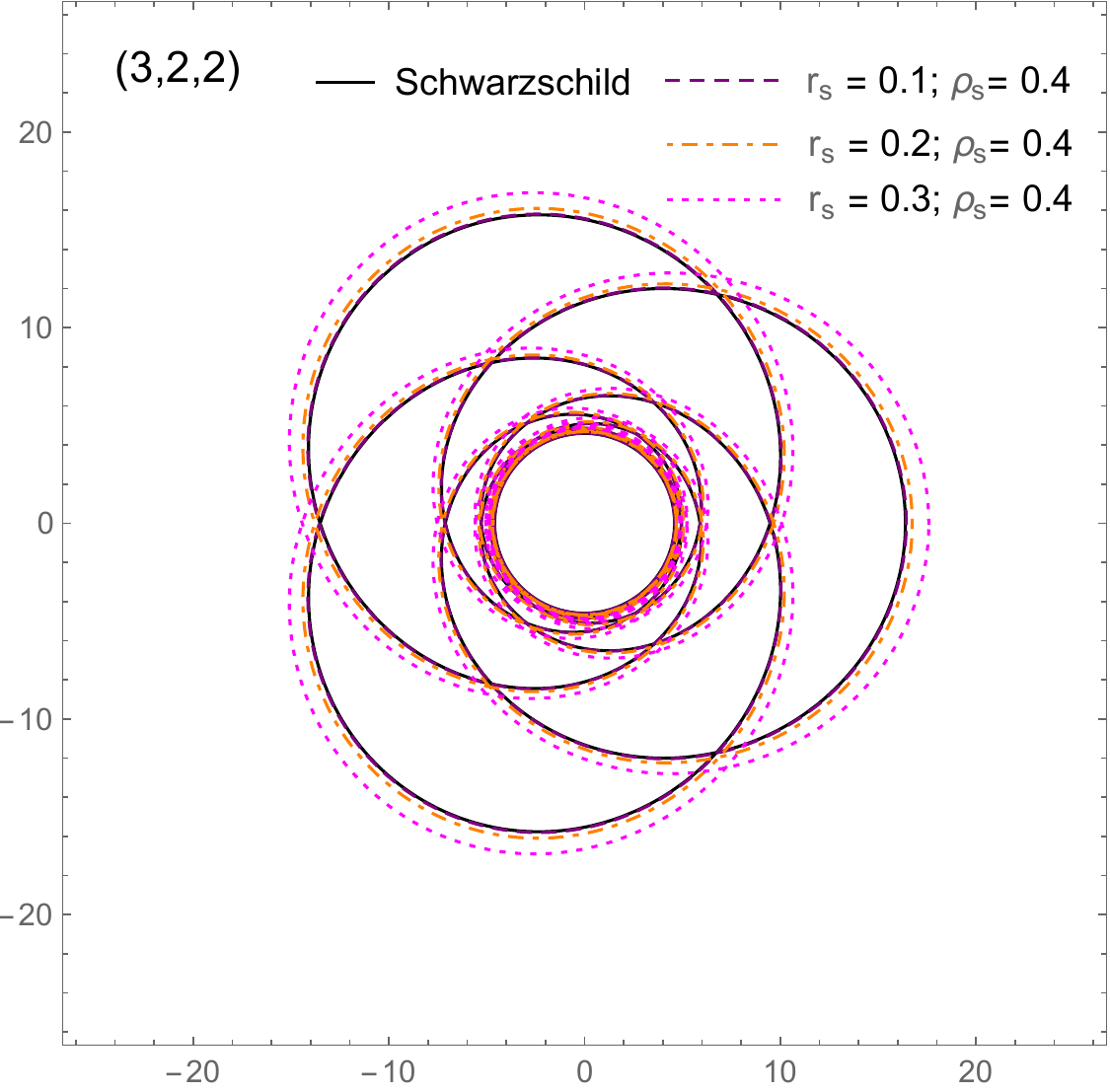}
		\end{subfigure}
		
		\caption{The periodic orbits labeled by $(z,w,v)$ around a Schwarzschild black hole embedded in a Hernquist dark matter halo, shown for various values of the scale radius $r_{\mathrm{s}}$, with fixed parameters $E=0.96$ and $\rho_{\mathrm{s}}=0.4$.}
		\label{orbitE}
	\end{figure*}

	The energy $E$ of periodic orbits corresponding to various integer triplets $(z,w,v)$ is computed numerically for a fixed orbital angular momentum $L=\frac{1}{2}(L_{\mathrm{MBO}}+L_{\mathrm{ISCO}})$; the resulting values are listed in Table~\ref{tabE}. Based on these data, the associated periodic trajectories around a Schwarzschild BH embedded in a Hernquist DM halo are plotted in Fig.~\ref{orbitL} for selected $(z,w,v)$ combinations. In parallel, Table~\ref{tabL} presents the orbital angular momentum $L$ of periodic orbits with the same set of $(z,w,v)$ labels, obtained under the fixed conditions $E=0.96$ and $r_{\mathrm{s}}=0.2$. Using the entries of Table~\ref{tabL}, we further illustrate the corresponding periodic orbits in Fig.~\ref{orbitE}, offering a complementary view of how the orbit morphology depends on the resonance triplet in the presence of the Hernquist halo.
	\begin{table*}[!htbp]  % 使用 table* 环境
		\setlength{\abovecaptionskip}{0.2cm}
		\setlength{\belowcaptionskip}{0.2cm}
		\centering
		\caption{The energy $E$ corresponding to the periodic orbits characterized by $(z,w,v)$ is listed for $L=\frac{1}{2}(L_{\mathrm{MBO}}+L_{\mathrm{ISCO}})$ and a Hernquist halo scale radius $r_{\mathrm{s}}=0.2$.}
		\label{tabE}
		\resizebox{\textwidth}{!}{%  % 改为 columnwidth
			\begin{tabular}{lcccccccc}
				\hline 
				$\mathbf{BHs}$ & $\rho_{\mathrm{s}}$ & $L$ & $E_{(1,1,0)}$ & $E_{(1,2,0)}$ & $E_{(2,1,1)}$ & $E_{(2,2,1)}$ & $E_{(3,1,2)}$ & $E_{(3,2,2)}$ \\
				\hline$\mathbf{Schwarzschild}$ & $ - $ & $3.732051$ & $0.965425$ & $0.968383$ & $0.968026$ & $0.968434$ & $0.968225$ & $0.968438$\\
				\hline
				$ $ & $0.3$ & $3.785544$ & $0.965381$ & $0.968344$ & $0.967987$ & $0.968396$ & $0.968186$ & $0.968400$ \\
				$\mathbf{DM}$ & $0.6$ & $3.839110$ & $0.965339$ & $0.968308$ & $0.967950$ & $0.968359$ & $0.968149$ & $0.968364$ \\
				$ $ & $0.9$ & $3.892746$ & $0.965299$ & $0.968273$ & $0.967914$ & $0.968325$ & $0.968114$ & $0.968329$ \\
				\hline 
			\end{tabular}%
		}
	\end{table*}
	
	\begin{table*}[!htbp]  % 使用 table* 环境
		\setlength{\abovecaptionskip}{0.2cm}
		\setlength{\belowcaptionskip}{0.2cm}
		\centering
		\caption{The orbital angular momenta $L$ for the periodic orbits labeled by $(z,w,v)$ in a Schwarzschild–Hernquist spacetime, computed at $E=0.96$ and $r_{\mathrm{s}}=0.2$.}
		\label{tabL}
		\resizebox{\textwidth}{!}{%  % 改为 columnwidth
			\begin{tabular}{lccccccc}
				\hline 
				$\mathbf{BHs}$ & $\rho_{\mathrm{s}}$ & $L_{(1,1,0)}$ & $L_{(1,2,0)}$ & $L_{(2,1,1)}$ & $L_{(2,2,1)}$ & $L_{(3,1,2)}$ & $L_{(3,2,2)}$ \\
				\hline
				$\mathbf{Schwarzschild}$ & $ - $  & $3.683588$ & $3.653406$ & $3.657596$ & $3.652701$ & $3.655335$ & $3.652636$ \\
				\hline
				$$ & $0.3$ & $3.736827$ & $3.706210$ & $3.710459$ & $3.705496$ & $3.708166$ & $3.705431$ \\
				$\mathbf{DM}$ & $0.6$ & $3.790127$ & $3.759074$ & $3.763382$ & $3.758351$ & $3.761057$ & $3.758284$ \\
				$ $ & $0.9$ & $3.843486$ & $3.811996$ & $3.816363$ & $3.811263$ & $3.814006$ & $3.811196$ \\
				\hline 
			\end{tabular}%
		}
	\end{table*}
	
	\section{Gravitational waveforms from periodic orbits}
	\label{section4}
	Using the family of periodic geodesics obtained in the preceding section, we model the associated gravitational radiation via the kludge waveform formalism introduced in \cite{Babak:2006uv}, applied to test-particle motion around a Schwarzschild BH surrounded by a Hernquist DM halo. In the context of EMRIs, the spacetime metric perturbation—including contributions up to second order—is provided by \cite{Maselli:2021men,Liang:2022gdk}.
	\begin{equation}
		h_{ij}=\frac{4G{\eta}M}{c^{4}D_{\mathrm{L}}}\left(v_{i}v_{j}-\frac{Gm}{r}n_{i}n_{j}\right).\label{hij}
	\end{equation}
	The system is parametrized by the BH mass $M$, the test particle mass $m$, and their symmetric mass ratio $\eta = mM/(m + M)^2$, with $D_{\rm L}$ denoting the luminosity distance to the source. The particle's motion is described by its spatial velocity $v_i$ and the radial unit vector $n_i$. When the emitted gravitational radiation is projected onto the detector's sky-fixed coordinates, the resulting plus ($h_+$) and cross ($h_\times$) polarization waveforms for periodic orbits are expressed as \cite{Liang:2022gdk,Will:2016sgx}.
	\begin{equation}
		h_{+}=-\frac{2\eta}{c^{4}D_{\mathrm{L}}}\frac{(GM)^{2}}{r}(1+\cos^{2}\iota)\cos(2\phi+2\omega),\label{hz}
	\end{equation}
	\begin{equation}
		h_{\times}=-\frac{4\eta}{c^{4}D_{\mathrm{L}}}\frac{(GM)^{2}}{r}\cos\iota\sin(2\phi+2\omega).\label{hc}
	\end{equation}
	Here, $\iota$ denotes the angle between the direction of the test particle's orbital angular momentum and the line of sight to the observer (i.e., the orbital inclination), while $\omega$ is the longitude of pericenter, specifying the orientation of the orbit's closest approach.
	
\begin{figure*}[htbp]
		\centering
		\begin{subfigure}{\textwidth}
			\centering
			\includegraphics[width=0.6\linewidth]{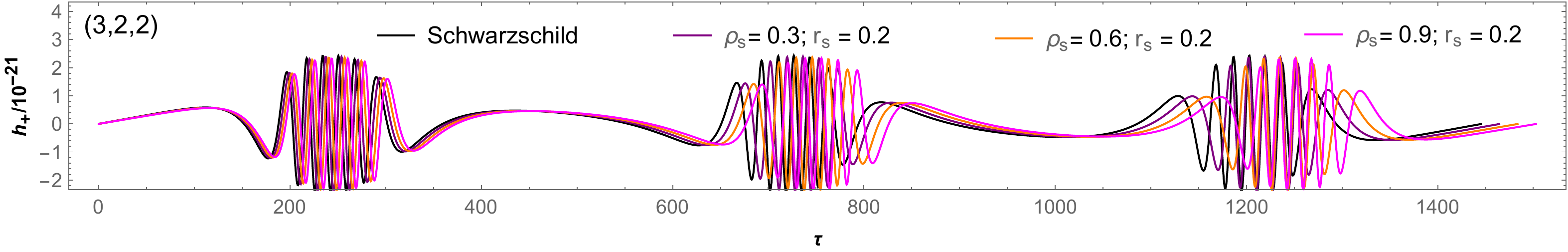}
			\caption*{$r_{\mathrm{s}}=0.2$}
		\end{subfigure}
		
		\vspace{0.5cm} % 调整上下间距
		
		\begin{subfigure}{\textwidth}
			\centering
			\includegraphics[width=0.6\linewidth]{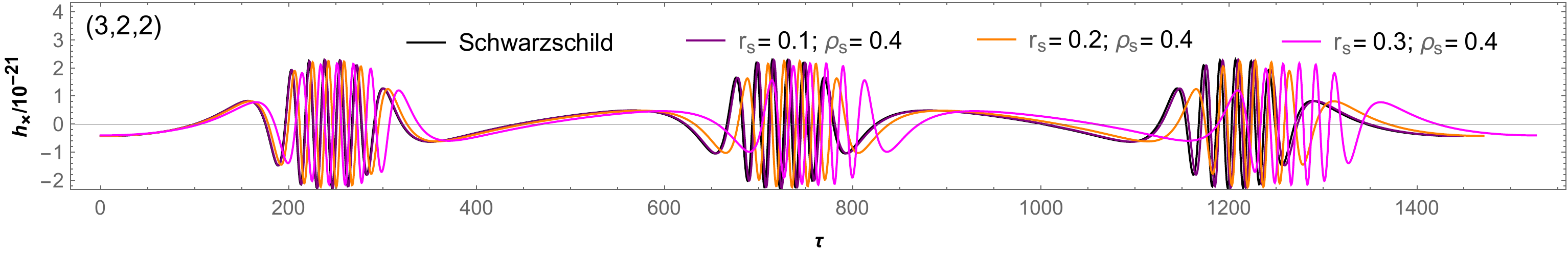}
			\caption*{$\rho_{\mathrm{s}}=0.4$}
		\end{subfigure}
		
		\caption{The gravitational waveforms emitted by a test particle of mass $m = 10\,M_\odot$ moving along $(3,2,2)$ the periodic orbits around a Schwarzschild black hole of mass $M = 10^7\,M_\odot$, embedded in a Hernquist dark matter halo. The orbital angular momentum is fixed at $L = \tfrac{1}{2}(L_{\rm MBO} + L_{\rm ISCO})$, and the energy is correspondingly determined by the orbit parameters. Top panel: waveforms for varying central density $\rho_{\mathrm{s}}$ with the scale radius held fixed at $r_{\mathrm{s}} = 0.2$. Bottom panel: the waveforms for varying scale radius $r_{\mathrm{s}}$ with the central density fixed at $\rho_{\mathrm{s}} = 0.4$.}
		\label{GW1}
	\end{figure*}

\begin{figure*}[htbp]
		\centering
		\begin{subfigure}{\textwidth}
			\centering
			\includegraphics[width=0.6\linewidth]{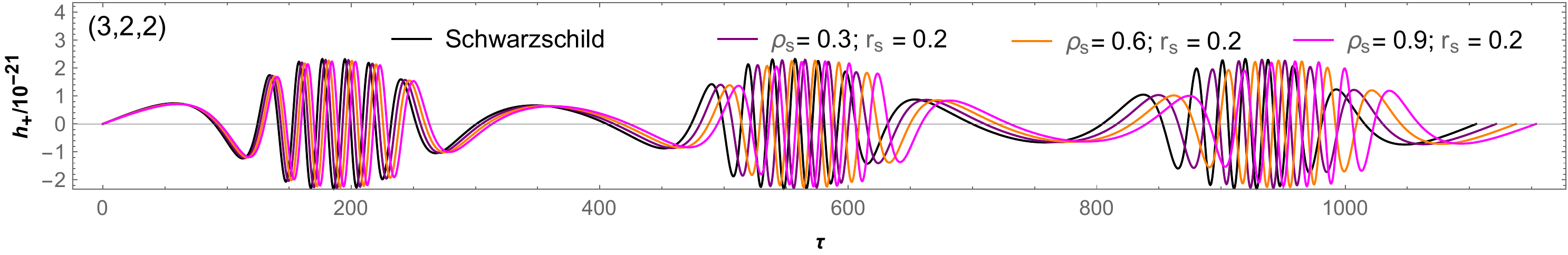}
			\caption*{$r_{\mathrm{s}}=0.2$}
		\end{subfigure}
		
		\vspace{0.5cm} % 调整上下间距
		
		\begin{subfigure}{\textwidth}
			\centering
			\includegraphics[width=0.6\linewidth]{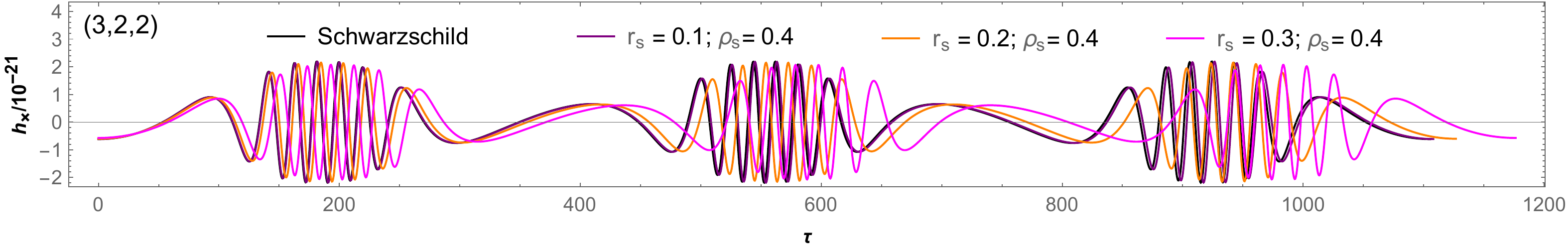}
			\caption*{$\rho_{\mathrm{s}}=0.4$}
		\end{subfigure}
		
		\caption{The gravitational waveforms emitted by a test particle of mass $m = 10\,M_\odot$ moving along $(3,2,2)$ periodic orbits around a Schwarzschild black hole of mass $M = 10^7\,M_\odot$, embedded in a Hernquist dark matter halo. The particle energy is fixed at $E = 0.96$. Top panel: the waveforms for varying central density $\rho_{\mathrm{s}}$ with scale radius fixed at $r_{\mathrm{s}} = 0.2$. Bottom panel: the waveforms for varying scale radius $r_{\mathrm{s}}$ with central density fixed at $\rho_{\mathrm{s}} = 0.4$.}
		\label{GW2}
	\end{figure*}

	The gravitational waveforms presented here correspond to an EMRI system in which a $10\,M_\odot$ compact object orbits a $10^7\,M_\odot$ Schwarzschild BH embedded in a Hernquist DM halo. The source is situated at a luminosity distance of $200~\mathrm{Mpc}$, and we fix the orbital inclination and longitude of pericenter to $\iota = \pi/4$ and $\omega = \pi/4$, respectively. Solving the phase evolution equation in terms of the particle's proper time $\tau$ and inserting the solution into Eqs.~(\ref{hz}) and (\ref{hc}), yields the time-domain waveforms for this configuration. The influence of the DM halo on orbital dynamics is illustrated in Figs.~\ref{GW1} and~\ref{GW2}, which show GW waveforms for the $(3,2,2)$ periodic orbit at different $\rho_{\mathrm{s}}$ and $r_{\mathrm{s}}$. The hallmark zoom-whirl morphology appears within one orbital period. Crucially, as the halo becomes more massive or extended (i.e., as $\rho_{\mathrm{s}}$ or $r_{\mathrm{s}}$ increases), the waveform is progressively delayed in time—shifting rightward relative to the Schwarzschild case. This phase lag directly signals an increase in the proper time per orbit, $\Delta\tau$, demonstrating how the DM environment modifies relativistic orbital motion.
	
	\section{The null Geodesics in a Schwarzschild Black Hole Embedded in a Hernquist Dark Matter Halo}
	\label{section5}
	
\begin{figure*}[htbp]
		\centering
		\subfloat[$\rho_{\mathrm{s}}=0.4, r_{\mathrm{s}}=0.1$]{\includegraphics[width=0.28\textwidth]{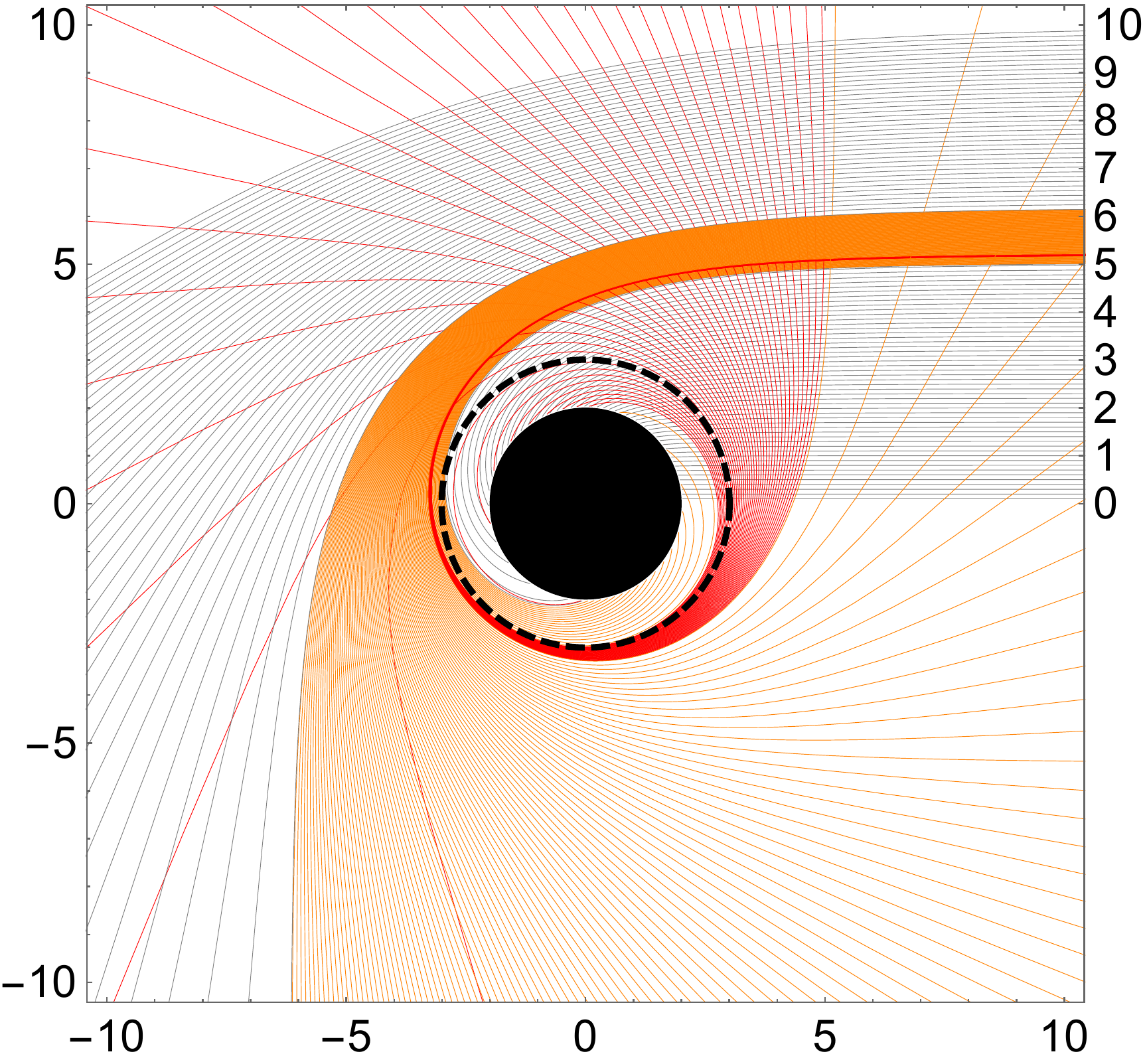}}\hfill
		\subfloat[$\rho_{\mathrm{s}}=0.4, r_{\mathrm{s}}=0.2$]{\includegraphics[width=0.28\textwidth]{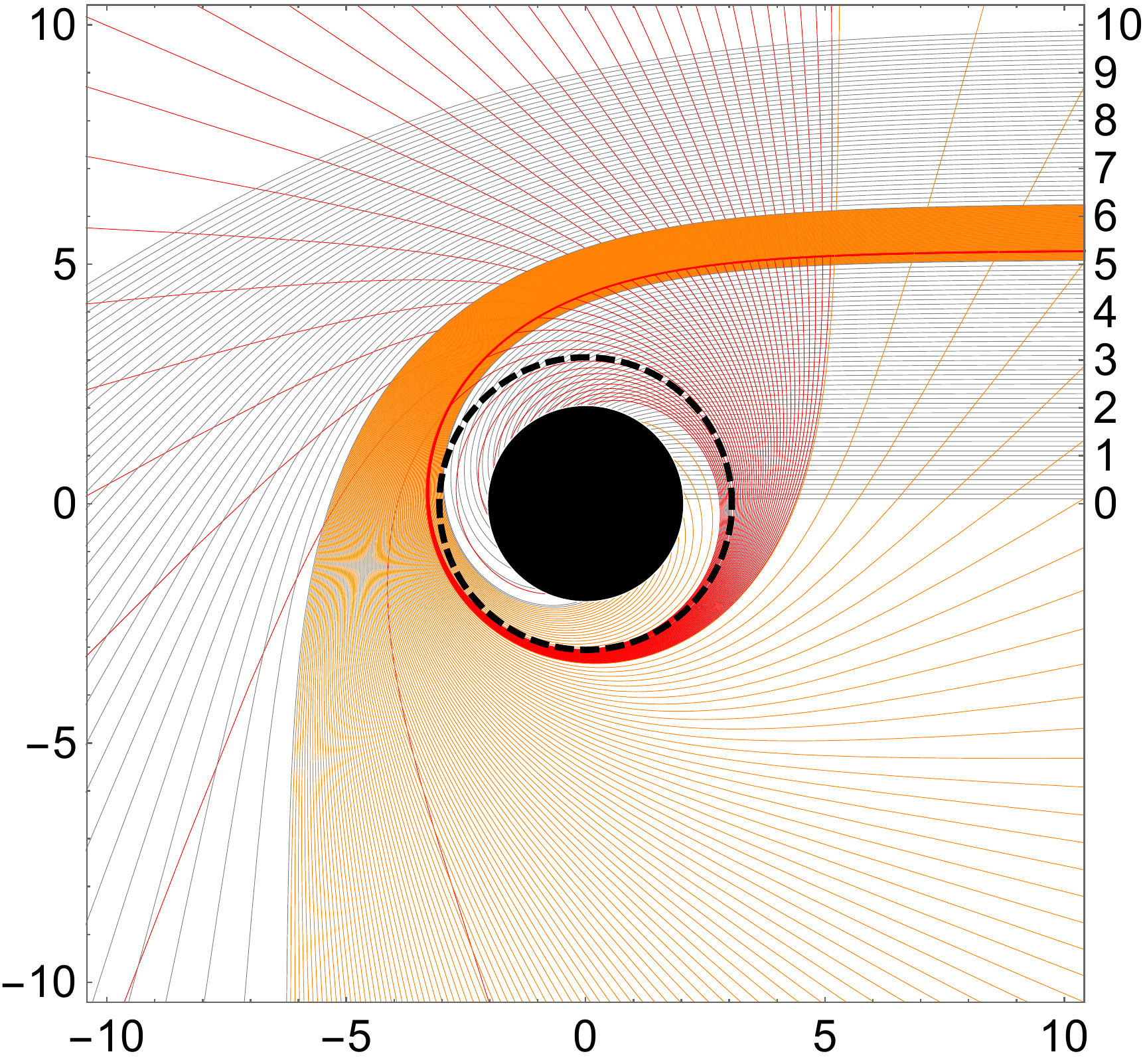}}\hfill
		\subfloat[$\rho_{\mathrm{s}}=0.4, r_{\mathrm{s}}=0.3$]{\includegraphics[width=0.28\textwidth]{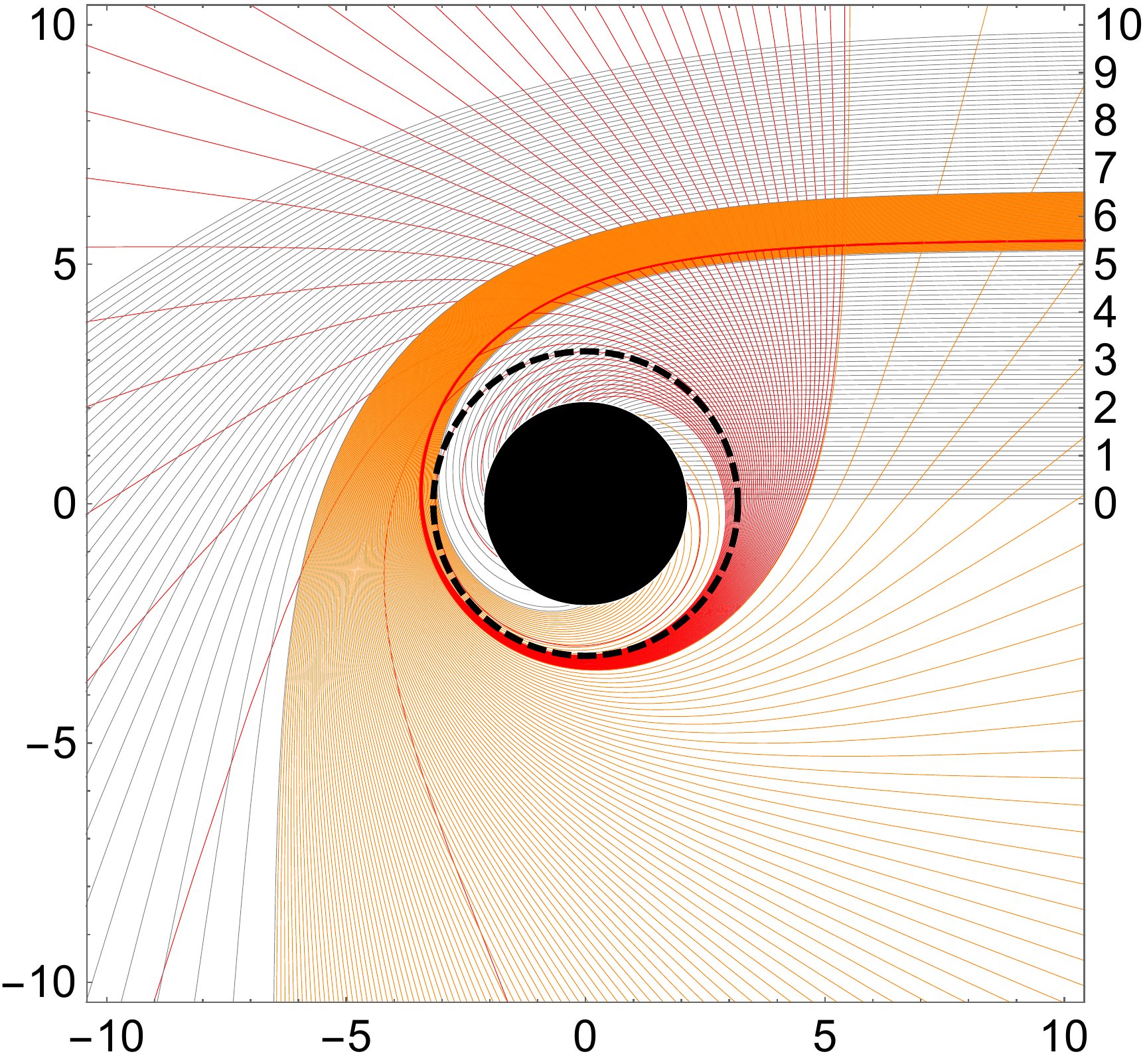}}\\
		\subfloat[$r_{\mathrm{s}}=0.2, \rho_{\mathrm{s}}=0.3$]{\includegraphics[width=0.28\textwidth]{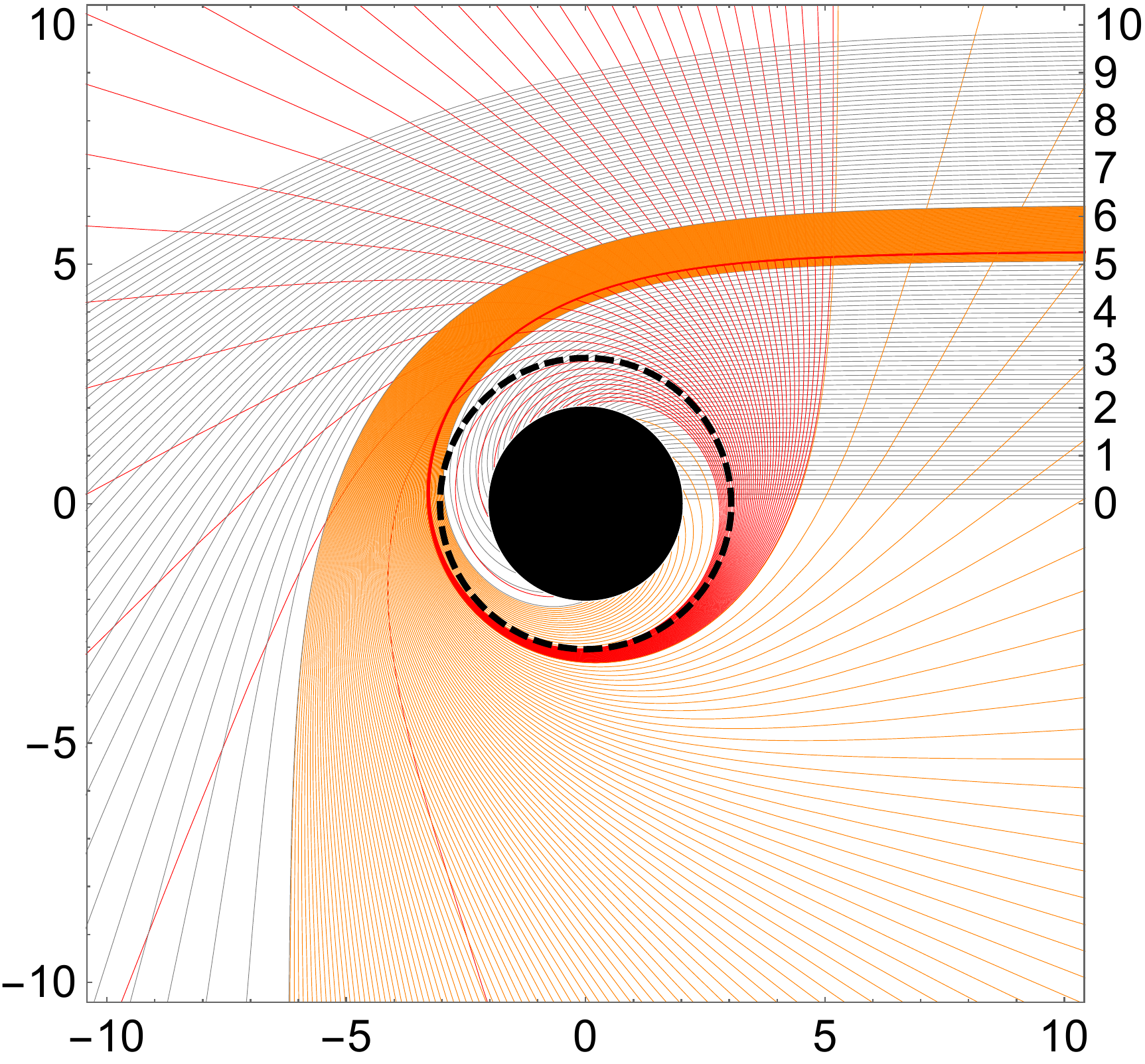}}\hfill
		\subfloat[$r_{\mathrm{s}}=0.2, \rho_{\mathrm{s}}=0.6$]{\includegraphics[width=0.28\textwidth]{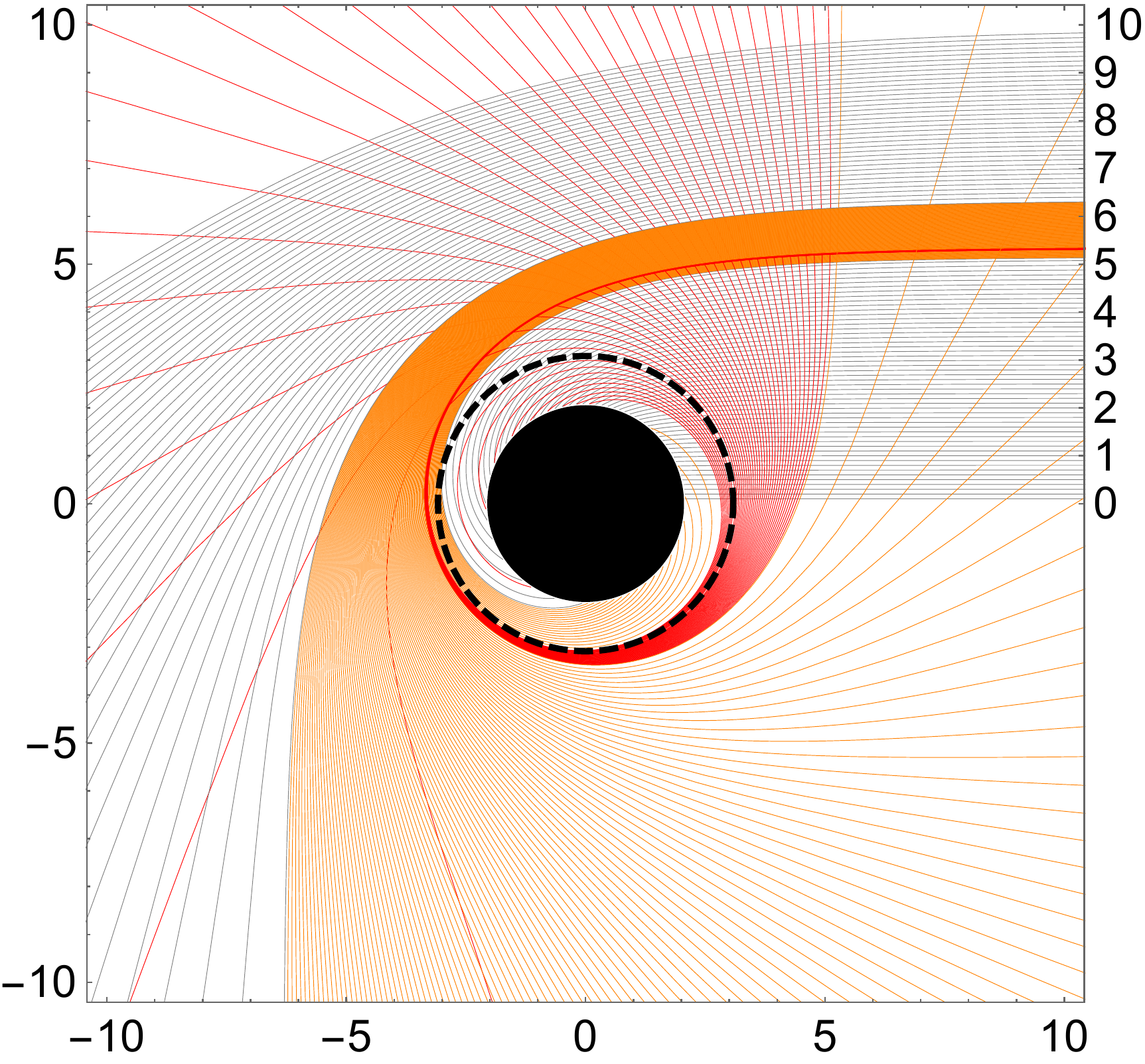}}\hfill
		\subfloat[$r_{\mathrm{s}}=0.2, \rho_{\mathrm{s}}=0.9$]{\includegraphics[width=0.28\textwidth]{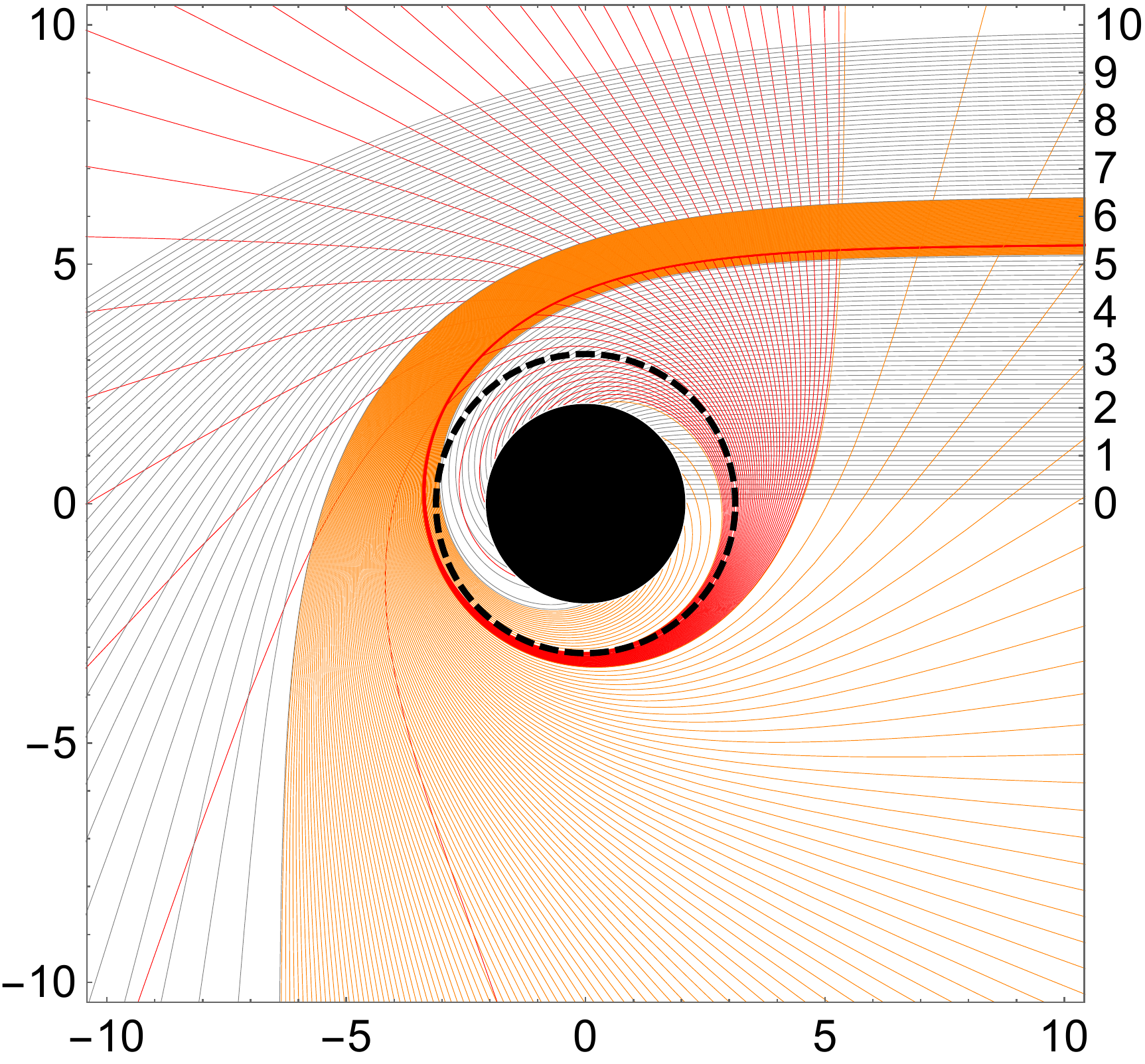}}
		\caption{The photon trajectories around a Schwarzschild black hole immersed in a Hernquist dark matter halo. Top: $\rho_{\mathrm{s}}=0.4$, $r_{\mathrm{s}}=0.1,~ 0.3,~ 0.5$. Bottom: $r_{\mathrm{s}}=0.2$, $\rho_{\mathrm{s}}=0.3,~ 0.6,~ 0.9$.}
		\label{wanqu}
	\end{figure*}

	The photon geodesic equations, originally given in Eq. (\ref{meq}), are reformulated by introducing the rescaled affine parameter $\tau=\tau/L$. In terms of this new parameter, the equations take the following form
	\begin{equation}
		\dot{t}=\frac{1}{bf(r)},\label{tdian}
	\end{equation}
	\begin{equation}
		\dot{\phi}=\frac{1}{r^{2}},\label{fdian}
	\end{equation}
	\begin{equation}
		\dot{r}^{2}=\frac{1}{b^{2}}-\frac{f(r)}{r^{2}},\label{rdian}
	\end{equation}
	where $b\equiv L/E$ represents the impact parameter of the light ray. The orbital equation for photons in the equatorial plane is obtained by combining Eqs. (\ref{fdian}) and (\ref{rdian}), resulting in
	\begin{equation}
		\left(\frac{\mathrm{d}r}{\mathrm{d}\phi}\right)^{2}=r^{4}\left[\frac{1}{b^{2}}-\frac{f(r)}{r^{2}}\right]=V_{\mathrm{eff}},\label{veff}
	\end{equation}
	where $V_{\mathrm{eff}}$ is the effective potential for photons. A light ray's trajectory is fully determined by its impact parameter $b$. The photon sphere—corresponding to an unstable circular null orbit at radius $r_{\mathrm{ph}}$—is located by solving $V_{\mathrm{eff}}(r_{\mathrm{ph}})=0$ and $V_{\mathrm{eff}}^{'}(r_{\mathrm{ph}})=0$. The resulting solution provides the critical impact parameter $b_{\mathrm{c}}$ associated with this orbit
	\begin{equation}
		b_{\mathrm{c}}=\frac{r_{\mathrm{ph}}}{\sqrt{f(r_{\mathrm{ph}})}}.\label{bc}
	\end{equation}
	The study of light deflection near a BH is greatly simplified by the substitution $u=1/r$. Applying this change of variable to the orbital equation (\ref{veff}), we obtain
	\begin{equation}
		\left(\frac{\mathrm{d}u}{\mathrm{d}\phi}\right)^{2}=\frac{1}{b^{2}}-u^{2}f\left(\frac{1}{u}\right)\equiv G(u).\label{guiji2}
	\end{equation}
	For photons with an impact parameter $b<b_{c}$, the trajectory remains entirely outside the event horizon. The accumulated azimuthal angle $\varphi$, representing the net change in the photon's azimuthal coordinate $\phi$, is computed from the external portion of the orbital trajectory
	\begin{equation}
		\varphi = \int_{0}^{u_{\mathrm{h}}} \frac{1}{\sqrt{G(u)}} \, \mathrm{d}u, \quad b < b_{\mathrm{c}}.\label{fai1}
	\end{equation}
	where $u_{\mathrm{h}}=1/r_{\mathrm{h}}$ and $r_{\mathrm{h}}$ is the radius of the outermost event horizon. When $b>b_{\mathrm{c}}$, the photon trajectory reaches a distance of closest approach corresponding to $u_{\mathrm{m}}$, the smallest positive solution of $G(u)=0$. The total azimuthal angle accumulated along the orbit is expressed as
	\begin{equation}
		\label{12}
		\varphi=2\int_{0}^{u_{\mathrm{m}}}\frac{1}{\sqrt{G(u)}} \, \mathrm{d}u, \quad b > b_{c}.
	\end{equation}
	Figure~\ref{wanqu} displays the photon trajectories for various parameter choices, color-coded by the total azimuthal angle $\varphi$: black for $\varphi<1.5\pi$, orange for $1.5\pi<\varphi<2.5\pi$, and red for $\varphi>2.5\pi$, 
	
	\section{The Impact of a Hernquist Dark Matter Halo on the Accretion Disk Images around a Schwarzschild Black Hole}
	\label{section6}
	
	In this Section, we will discuss the observational signatures of the accretion disk images around a Schwarzschild Black Hole and the observed flux. We will firstly introduce the observation coodinate system.
	
	\subsection{Observation coordinate system}
	
\begin{figure}[htbp]
		\centering
		\begin{subfigure}{0.49\textwidth}
			\includegraphics[width=3.5in, height=3.5in, keepaspectratio]{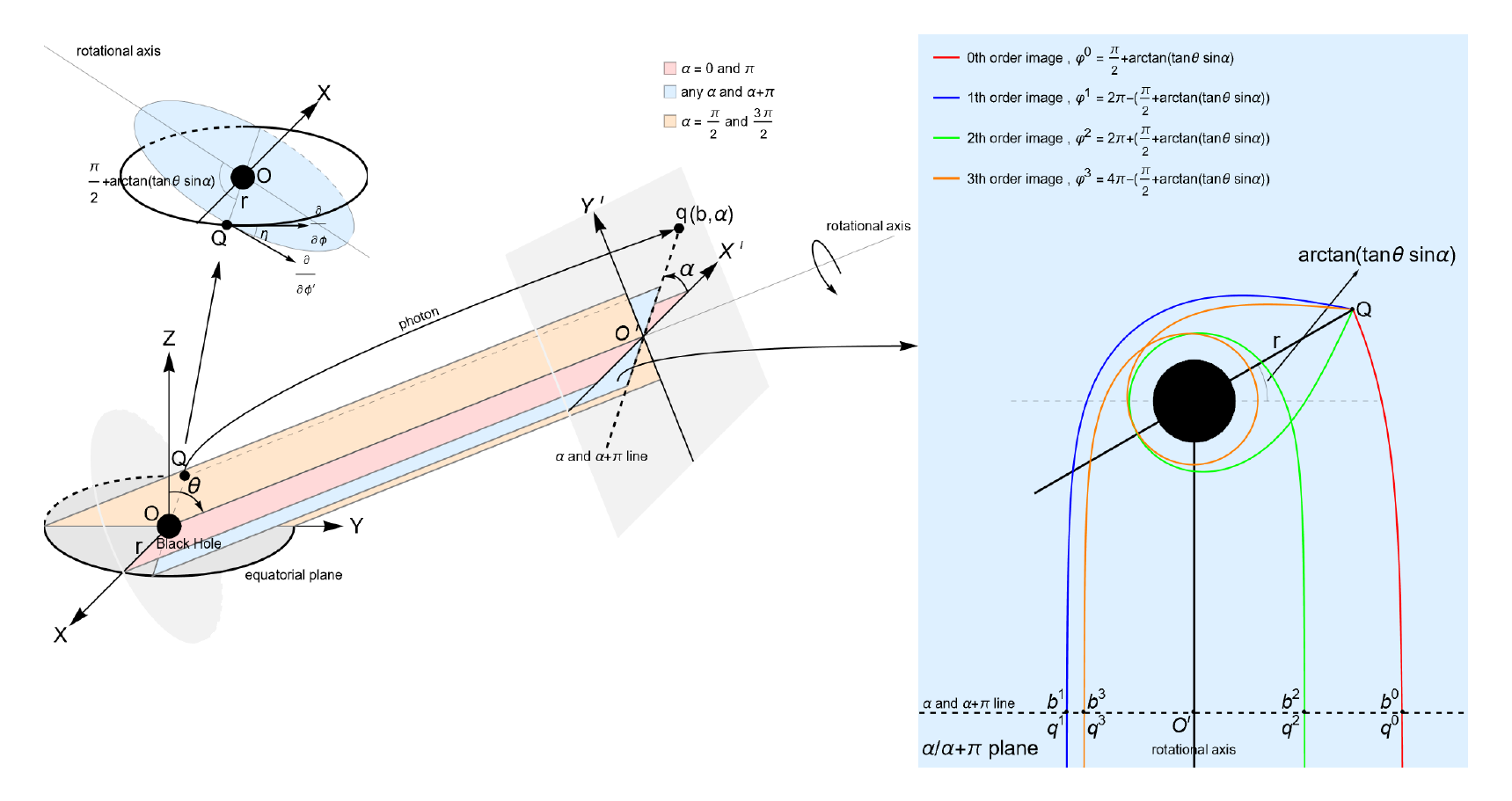}
		\end{subfigure}
		\caption{Coordinate system is indicated in Ref. \cite{You:2024uql}.}
		\label{coordinate}
	\end{figure}

	Figure~\ref{coordinate} defines the observational coordinate system employed throughout this work. The observer is assumed to be static at spatial infinity, viewing the spacetime from a direction specified by the polar angle $\theta$ (with $\phi = 0$) in the BH's spherical coordinates $(r, \theta, \phi)$, where $r = 0$ marks the central singularity. Synthetic images are constructed via backward ray tracing: rays are traced from the observer’s image plane—labeled by $(b, \alpha)$—into the spacetime. Starting at $q(b, \alpha)$, each null geodesic is integrated backward until it strikes the equatorial disk at $Q(r, \pi/2, \phi)$. By time-reversal invariance, the inverse path corresponds to a photon emitted from $Q$ that arrives at the observer, thereby determining the specific intensity at the image pixel $(b, \alpha)$. A trajectory at fixed radius $r$ represents a circular orbit in the equatorial plane. The left panel of Fig.~\ref{coordinate} illustrates that each meridional plane—specified by a given $\alpha$ (and its counterpart $\alpha + \pi$)—intersects this orbit at two diametrically opposed points, separated by $\Delta\phi = \pi$. By construction, the $X'$-axis is defined by $\alpha = 0$, whereas the $X$-axis coincides with $\phi = 0$. Within this framework, the angle $\varphi$ between the symmetry axis and the vector $\overrightarrow{OQ}$ is uniquely determined as
	\begin{equation}
		\varphi=\frac{\pi}{2}+\arctan(\tan\theta\sin\alpha).\label{fai}
	\end{equation}
	
	When the impact parameter $b$ tends toward the critical value $b_c$ associated with the photon sphere, the gravitational deflection angle becomes unbounded. As a result, a unique emission point $Q$ on the accretion disk is mapped to infinitely many images in the observer’s sky. These are labeled $q^{(n)}$ ($n = 0, 1, 2, \dots$), where $n$ counts the number of orbital windings completed by the photon en route to the observer. The image with $n = 0$ is the primary (direct) image, while those with $n \geq 1$ constitute higher-order relativistic images formed by increasingly bent light rays. Due to the near-critical bending of light near the photon sphere, successive relativistic images of a source point $Q$ are distributed azimuthally in an alternating pattern. As shown in the right panel of Fig.~\ref{coordinate}, the even-order images ($n = 0, 2, 4, \ldots$) remain on the same side of the BH as $Q$, while the odd-order images ($n = 1, 3, 5, \ldots$) emerge on the opposite side, roughly $\pi$ apart in $\phi$. This alternation arises because each additional half-orbit flips the apparent direction of arrival. The observed azimuth $\varphi^{(n)}$ thus increases with $n$, reflecting the photon’s cumulative deflection
	\begin{equation}
		\varphi^{n}= 
		\begin{cases} 
			\frac{n}{2}2\pi+(-1)^{n}[\frac{\pi}{2}+\arctan(\tan\theta\sin\alpha)], & \text{when $n$ is even,} \\
			\frac{n+1}{2}2\pi+(-1)^{n}[\frac{\pi}{2}+\arctan(\tan\theta\sin\alpha)], & \text{when $n$ is odd.}\label{fain}
		\end{cases}
	\end{equation}

	\subsection{Observational signatures of direct and secondary images}
	
\begin{figure*}[htbp]
		\centering
		\begin{subfigure}{0.38\textwidth}
			\includegraphics[width=\linewidth]{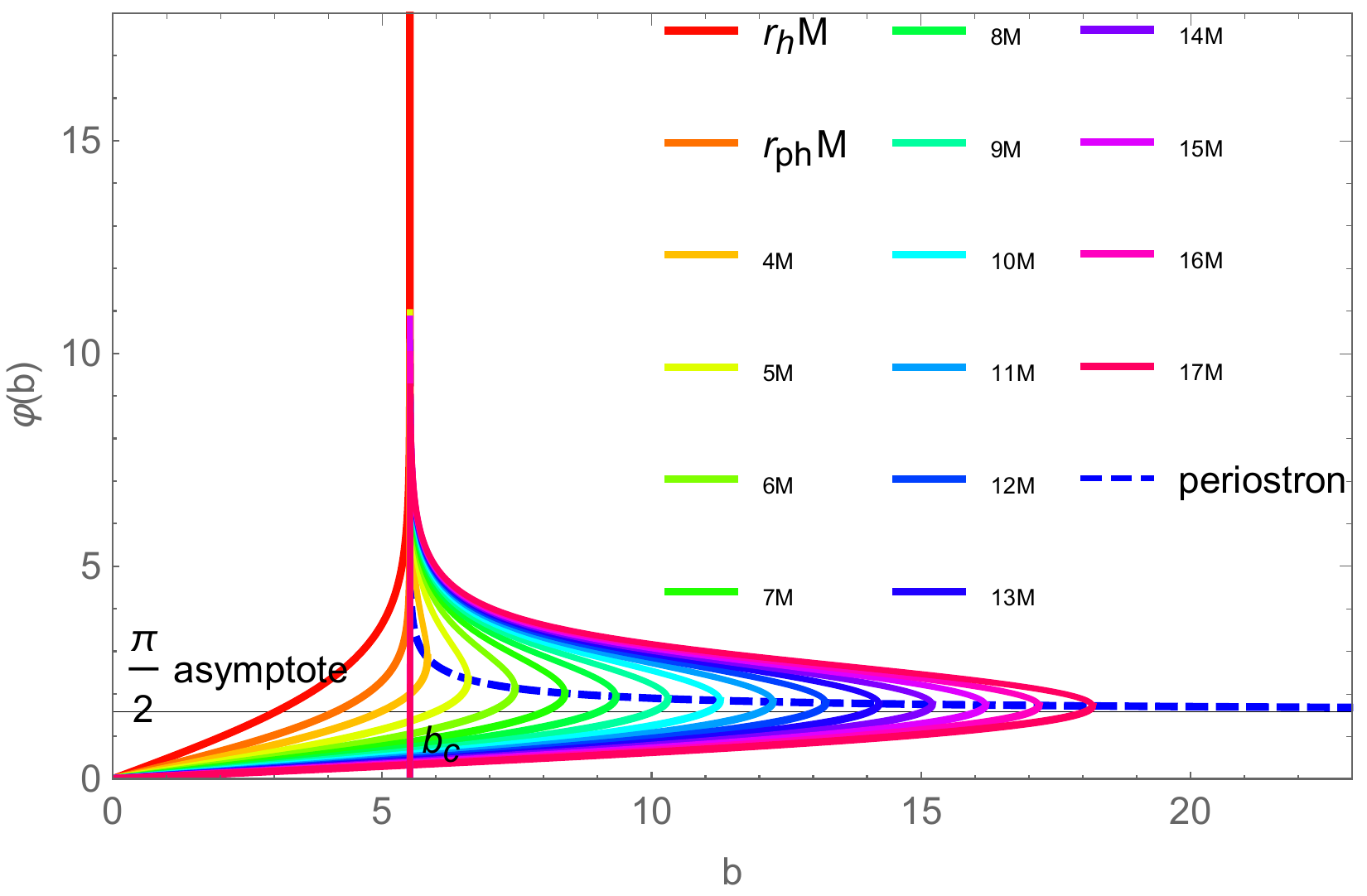}
		\end{subfigure}%
		\hspace{0.13\textwidth}% 
		\begin{subfigure}{0.38\textwidth}
			\includegraphics[width=\linewidth]{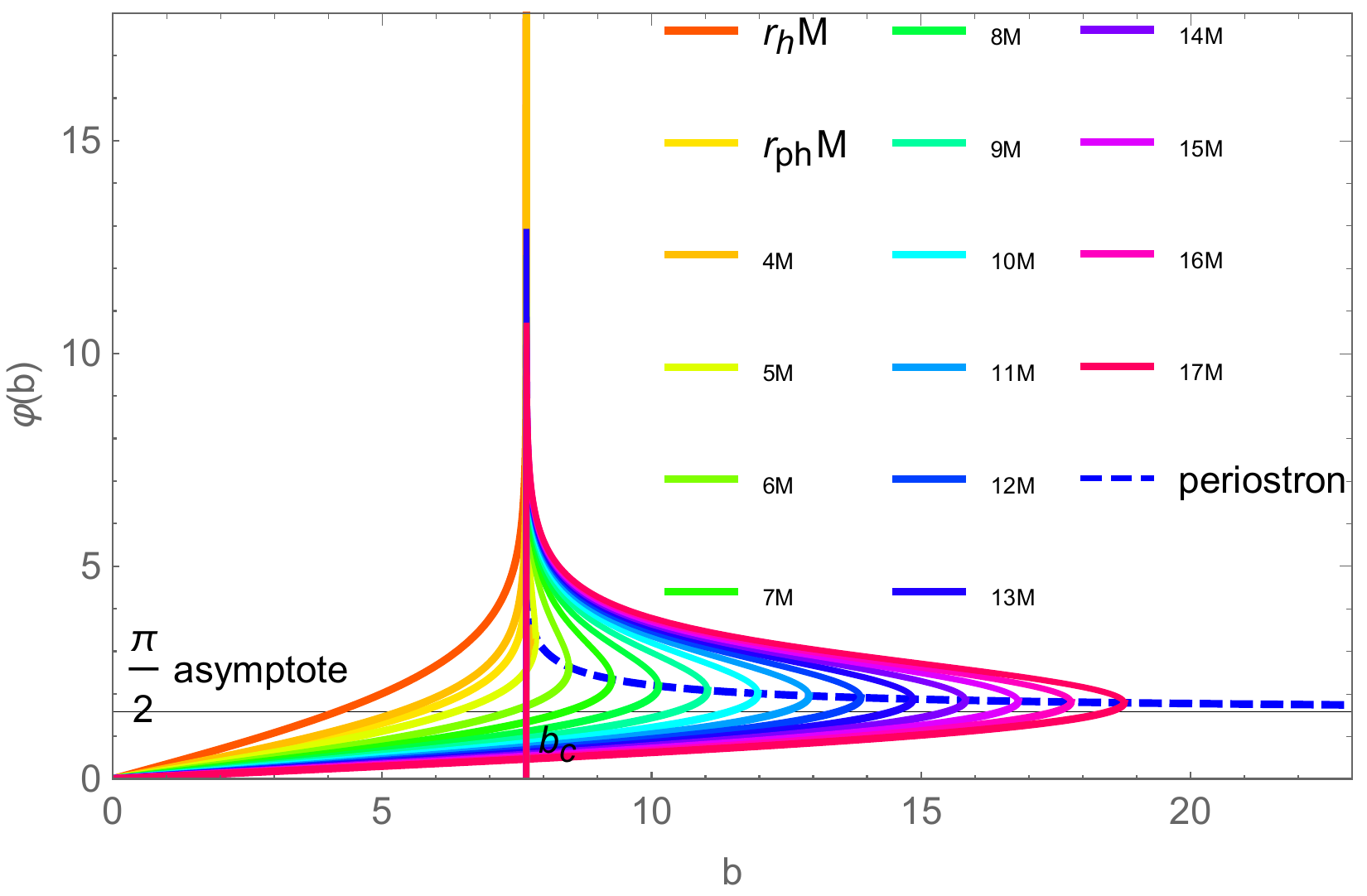}
		\end{subfigure}
		
		\caption{Deflection angle $\varphi(b)$ corresponding to intersections as a function of $b$
			for different $r$. We set $\rho_{\mathrm{s}}=0.4$, $r_{\mathrm{s}}=0.3$ for the left side and $\rho_{\mathrm{s}}=0.4$, $r_{\mathrm{s}}=0.6$ for the right side.}
		\label{faibb}
	\end{figure*}

	Each photon trajectory from infinity, labeled by its impact parameter $b$, pierces a constant-$r$ orbit at a specific azimuth $\varphi$. The functional form $\varphi(b)$, plotted in Fig.~\ref{faibb}, depends on the emission radius $r_{\mathrm{s}}$; for fixed $\rho_{\mathrm{s}} = 0.4$, larger $r_{\mathrm{s}}$ results in a rightward displacement of the curve. To distinguish qualitatively different imaging regimes, we introduce a fiducial boundary $\varphi_1(b)$ (blue dashed line). Solutions below this threshold constitute the $\varphi_2(b)$ branch, while those above form the $\varphi_3(b)$ branch. The following definitions follow naturally
	\begin{equation}
		\varphi_{1}(b)=\int_{0}^{u_{min}}\frac{1}{\sqrt{G(u)}}\mathrm{d}u,\label{faib1}
	\end{equation}
	\begin{equation}
		\varphi_{2}(b)=\int_{0}^{u_{r}}\frac{1}{\sqrt{G(u)}}\mathrm{d}u,\label{faib2}
	\end{equation}
	\begin{equation}
		\varphi_{3}(b)=2\int_{0}^{u_{min}}\frac{1}{\sqrt{G(u)}}\mathrm{d}u-\int_{0}^{u_{r}}\frac{1}{\sqrt{G(u)}}\mathrm{d}u.\label{faib3}
	\end{equation}
	  
	Each colored curve in the figures corresponds to a constant-radius orbit $r$, with coordinates $(b, \varphi)$ indicating the azimuthal angle $\varphi$ where a photon of impact parameter $b$ crosses the orbit. The blue dashed line intersects every curve at its peak, marking the perihelion passage at radius $r_{\rm pe}$ (the point of closest approach). In the limit $b \to \infty$, the photon trajectory becomes rectilinear, grazing circles at ever-increasing radii; consequently, the perihelion azimuth tends to $\varphi \to \pi/2$, as shown by the asymptotic behavior of the dashed line.
	
\begin{figure*}[htbp]
		\centering
		\begin{subfigure}{0.31\textwidth}
			\includegraphics[width=\linewidth]{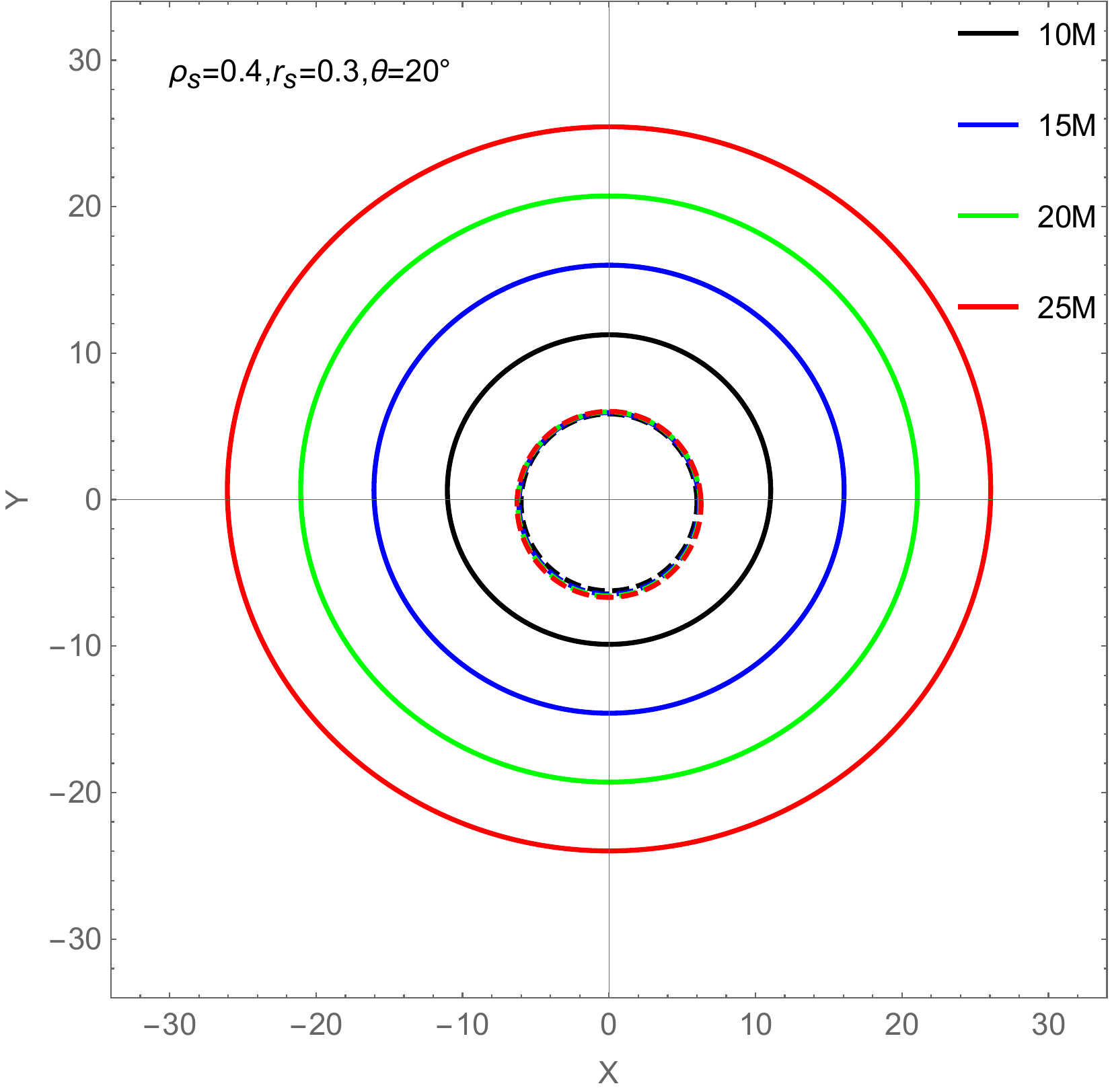}
		\end{subfigure}
		\hspace{0.02\textwidth}
		\begin{subfigure}{0.31\textwidth}
			\includegraphics[width=\linewidth]{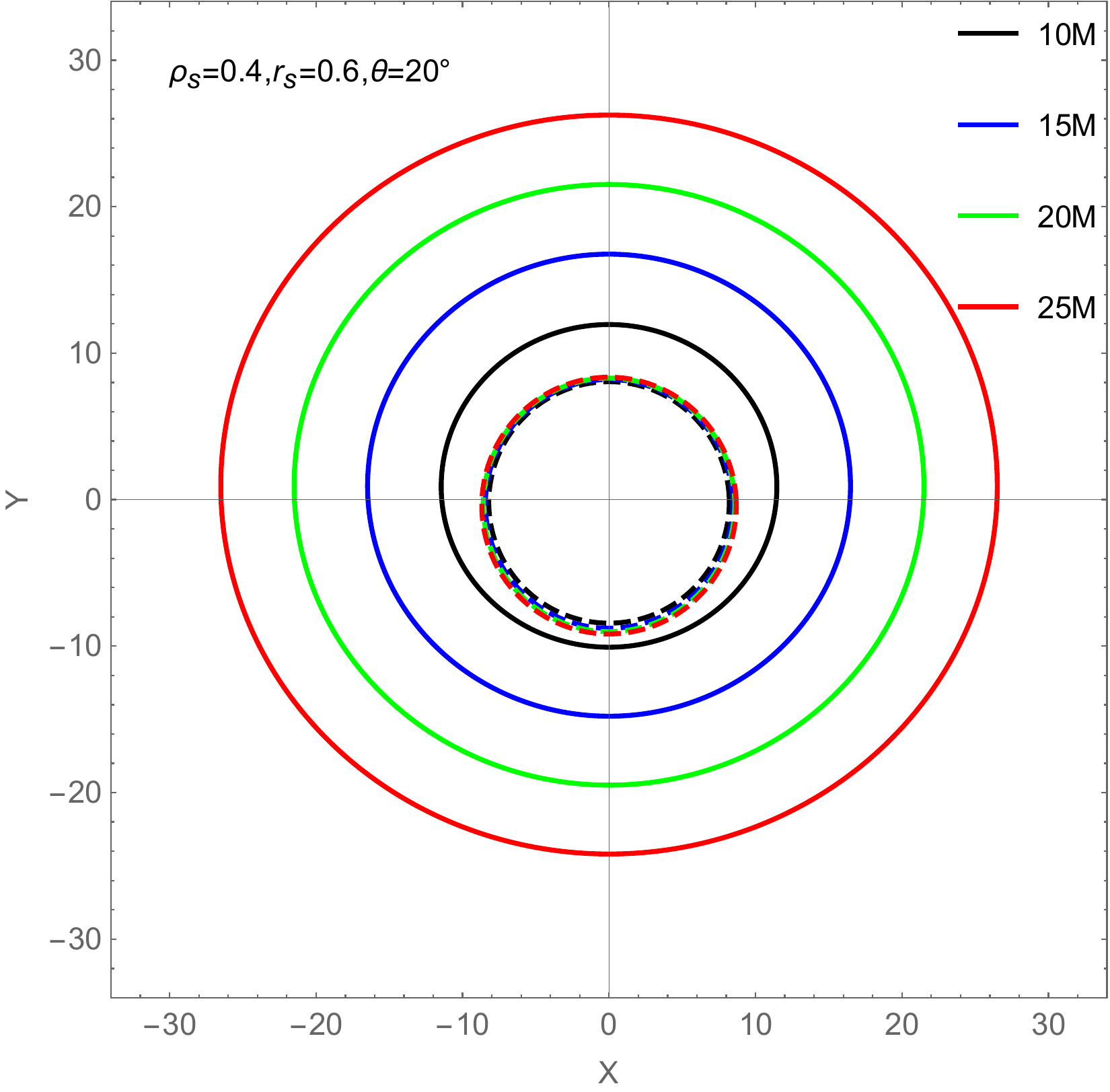}
		\end{subfigure}
		\hspace{0.02\textwidth}
		\begin{subfigure}{0.31\textwidth}
			\includegraphics[width=\linewidth]{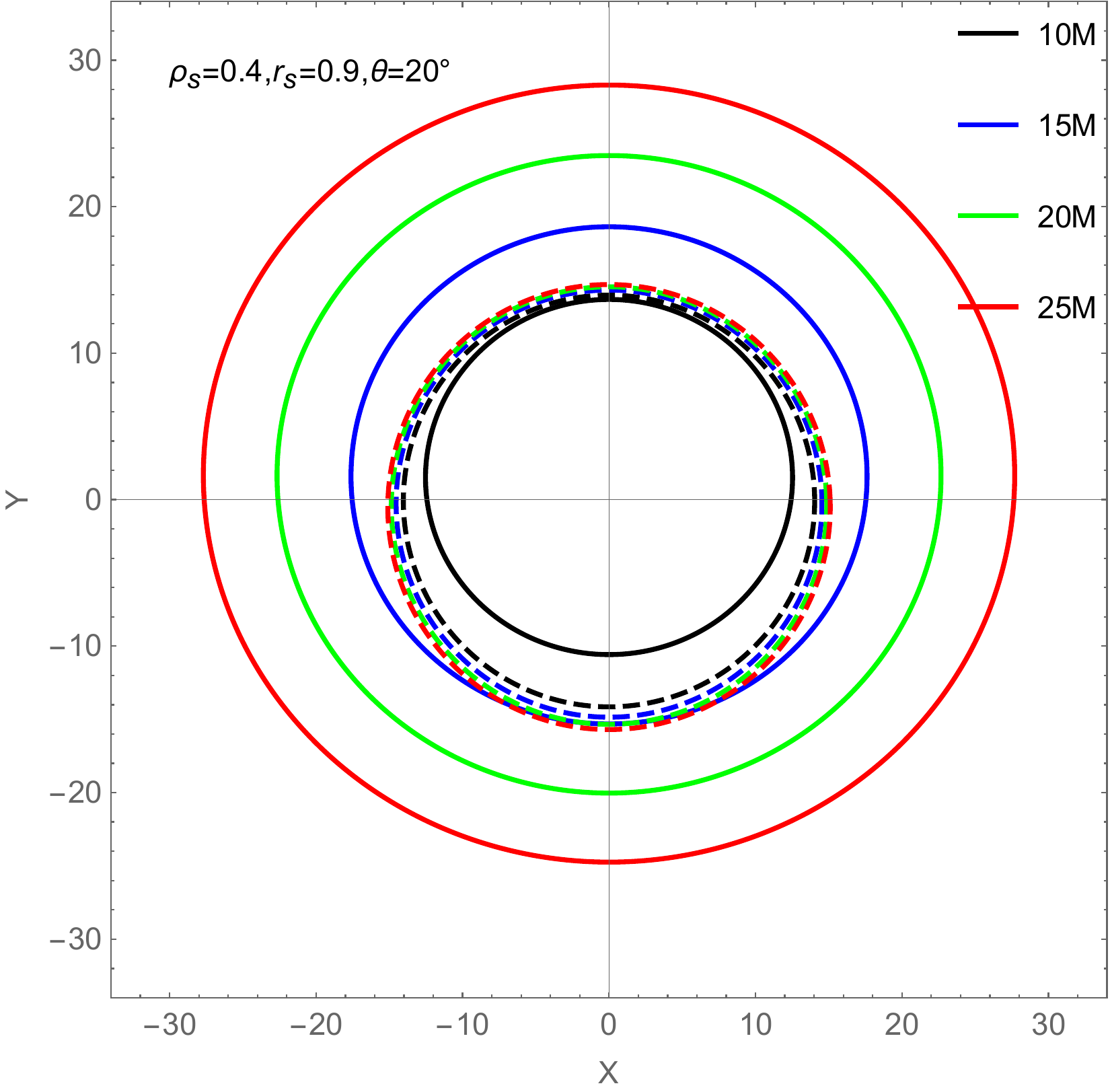}
		\end{subfigure}
		
		\vspace{0.3cm}
		
		\begin{subfigure}{0.31\textwidth}
			\includegraphics[width=\linewidth]{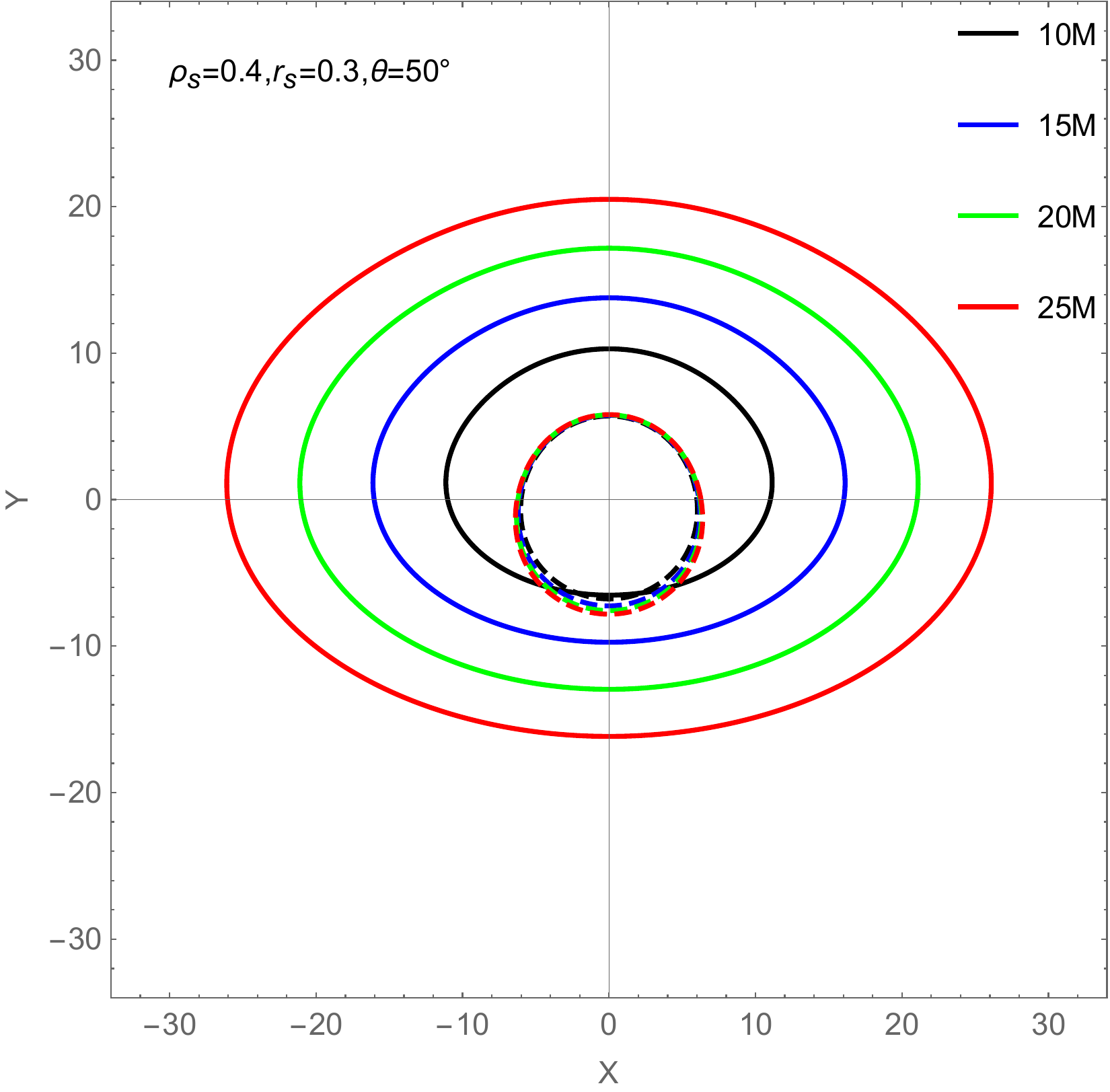}
		\end{subfigure}
		\hspace{0.02\textwidth}
		\begin{subfigure}{0.31\textwidth}
			\includegraphics[width=\linewidth]{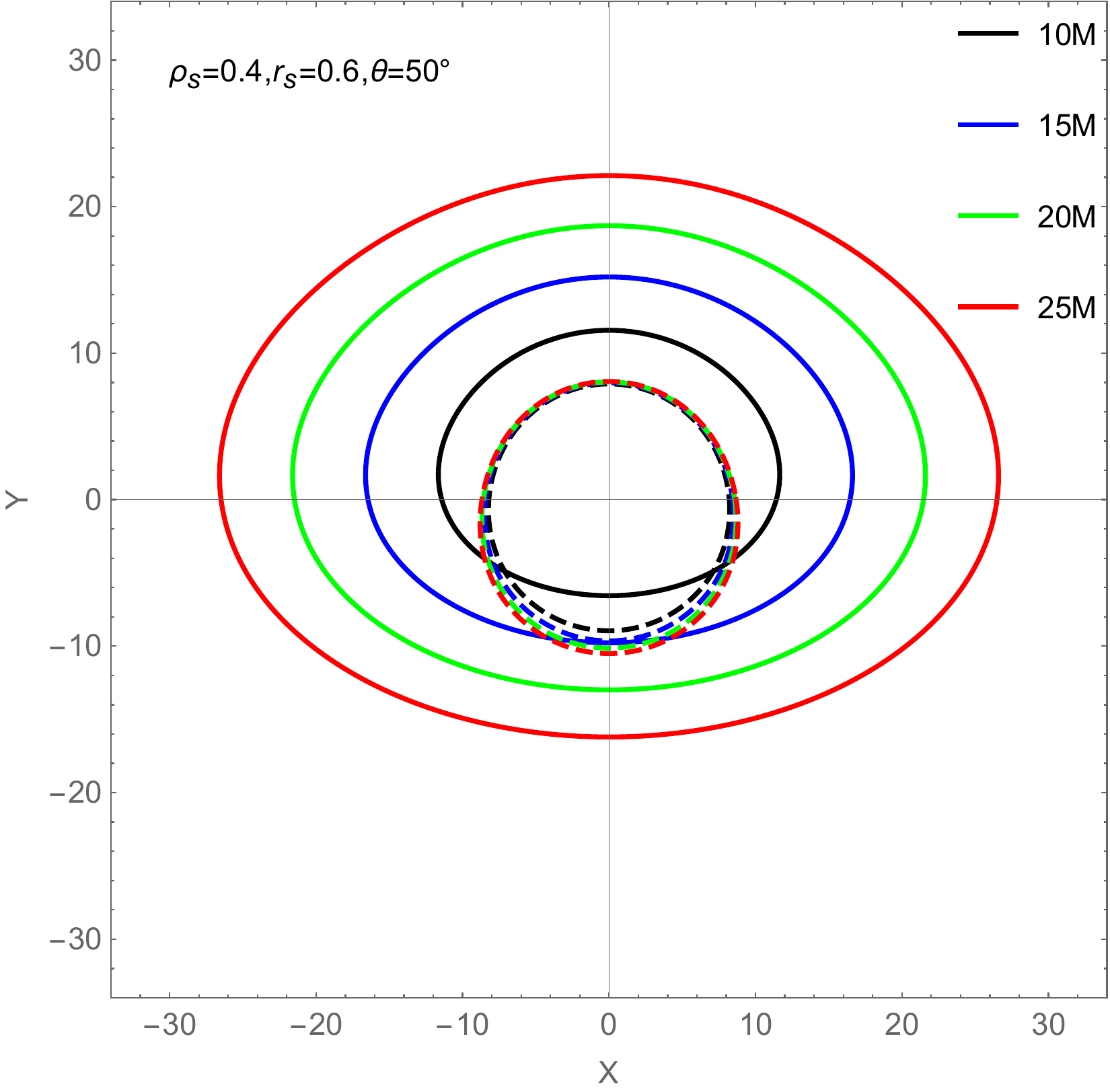}
		\end{subfigure}
		\hspace{0.02\textwidth}
		\begin{subfigure}{0.31\textwidth}
			\includegraphics[width=\linewidth]{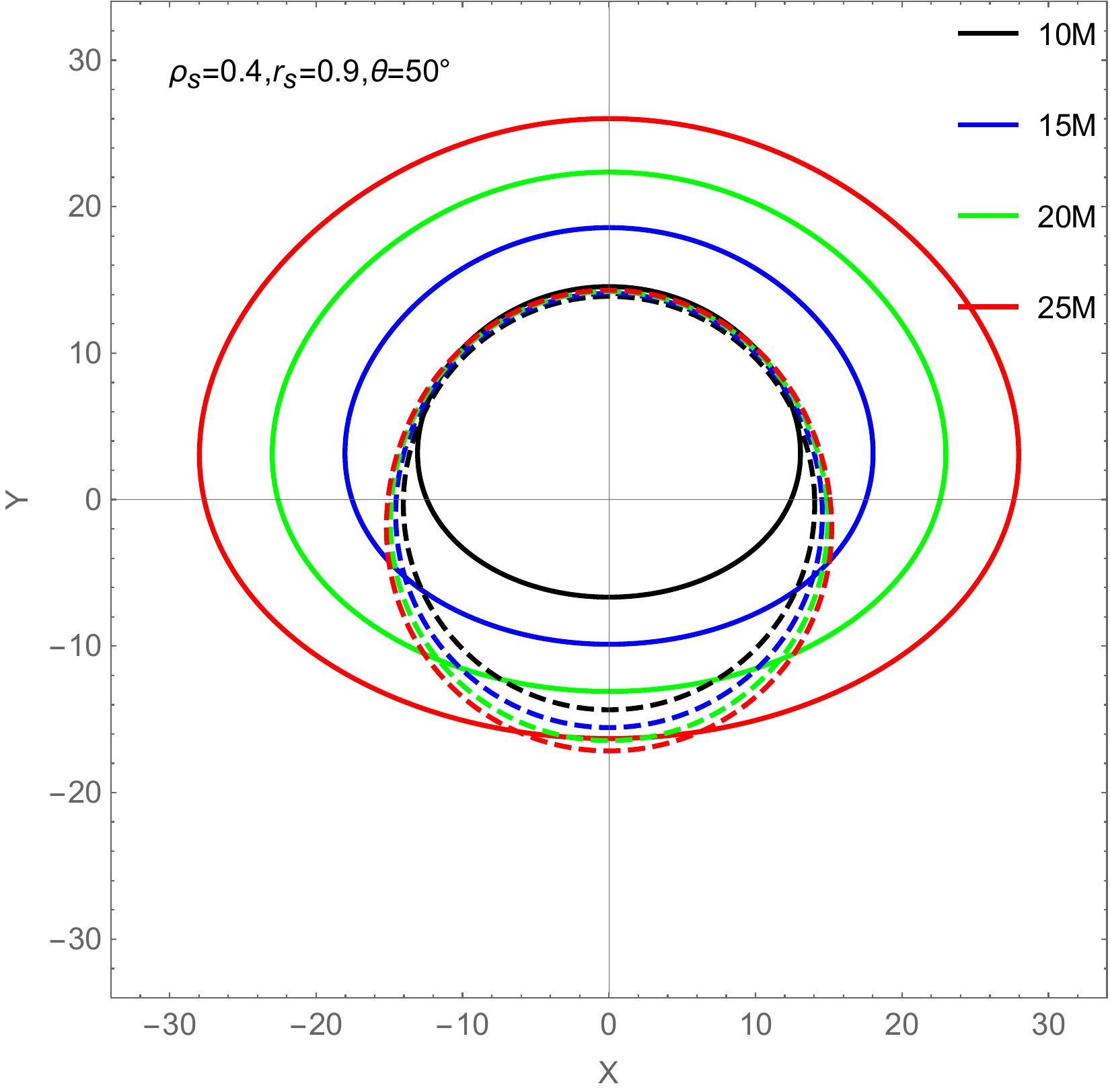}
		\end{subfigure}
		
		\vspace{0.3cm}
		
		\begin{subfigure}{0.31\textwidth}
			\includegraphics[width=\linewidth]{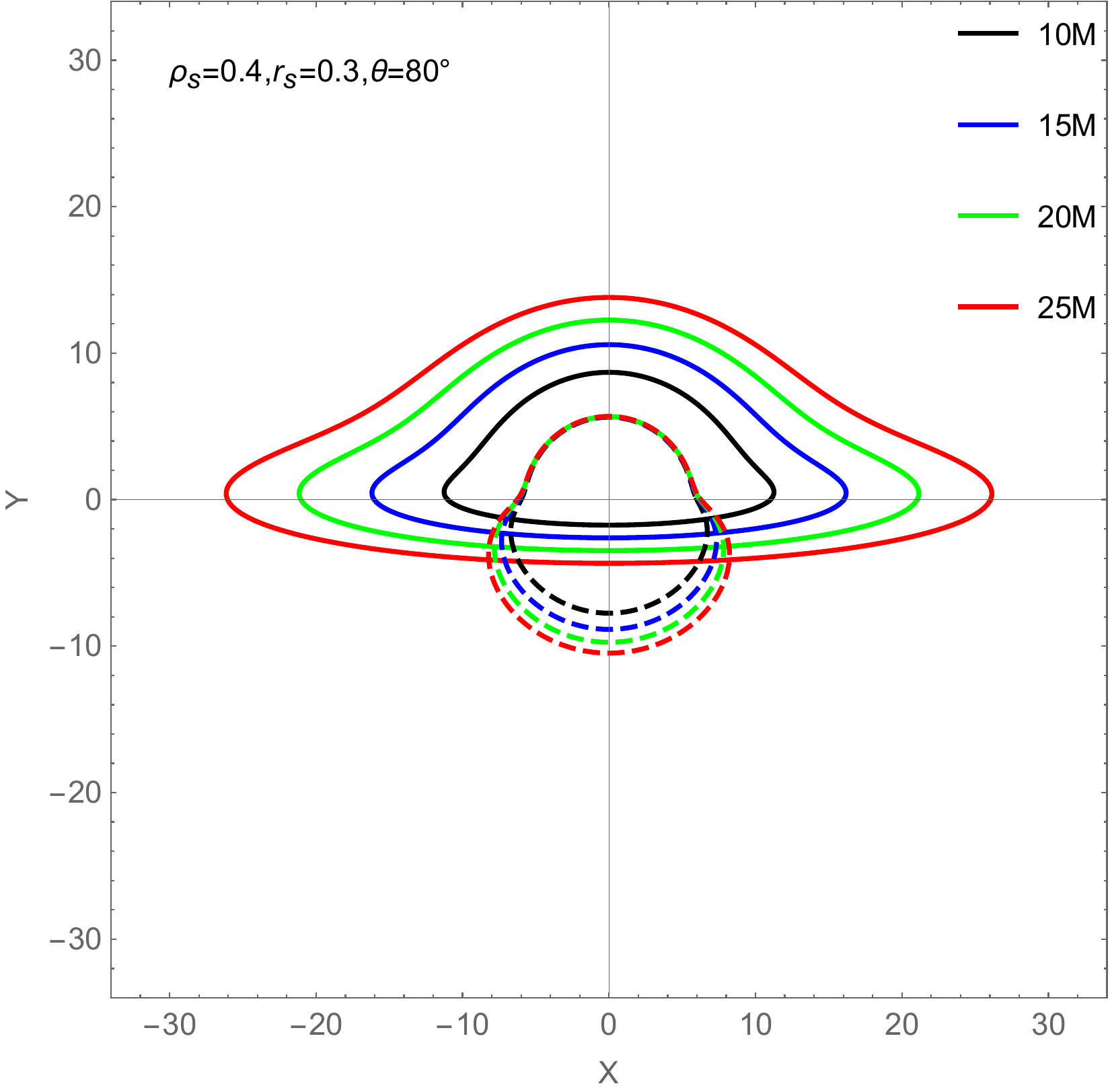}
		\end{subfigure}
		\hspace{0.02\textwidth}
		\begin{subfigure}{0.31\textwidth}
			\includegraphics[width=\linewidth]{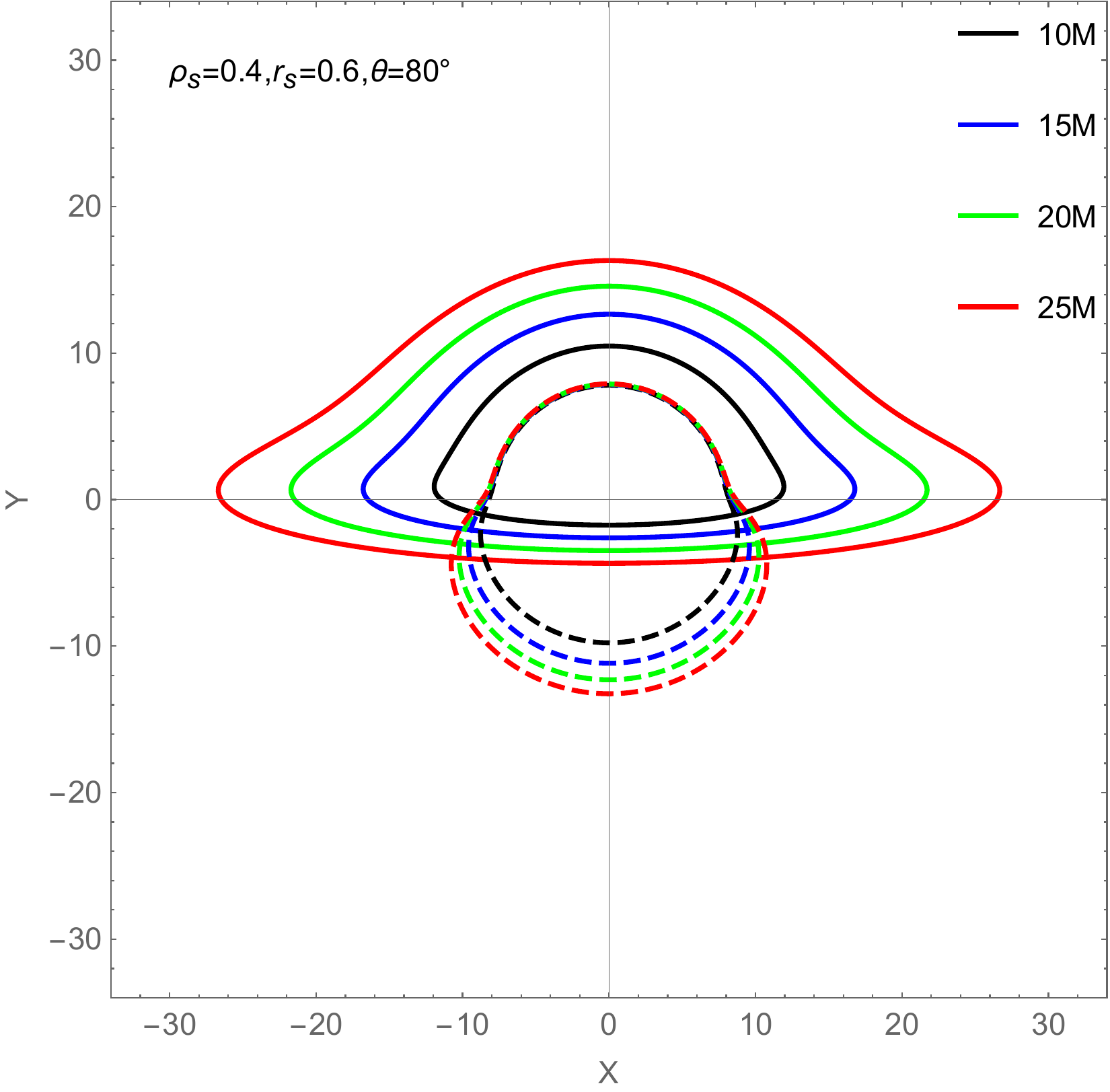}
		\end{subfigure}
		\hspace{0.02\textwidth}
		\begin{subfigure}{0.31\textwidth}
			\includegraphics[width=\linewidth]{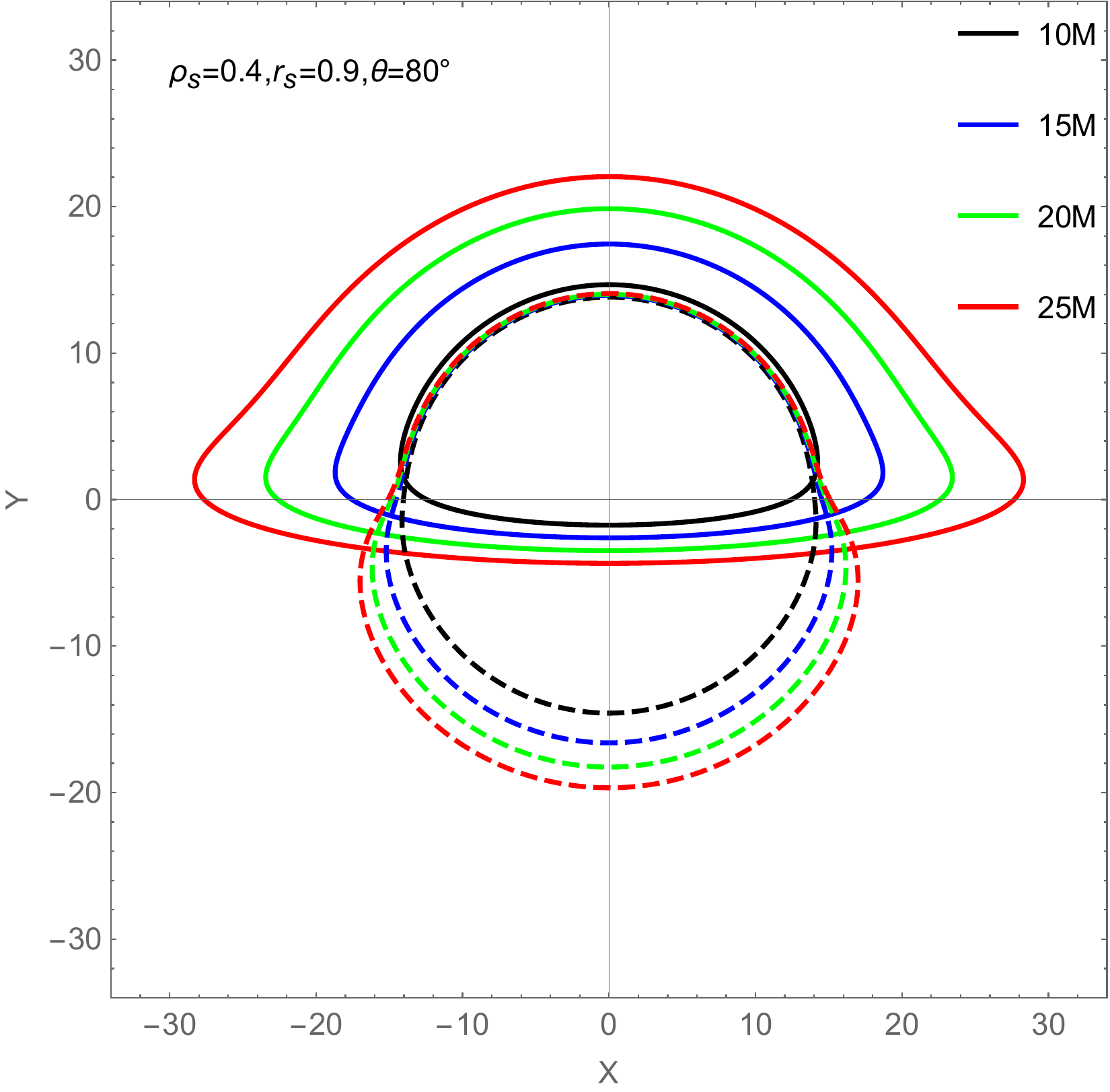}
		\end{subfigure}
		
		\caption{The direct (solid line) and secondary (dashed line) image of the thin accretion disk.}
		\label{dengr1}
	\end{figure*}

\begin{figure*}[htbp]
		\centering
		\begin{subfigure}{0.31\textwidth}
			\includegraphics[width=\linewidth]{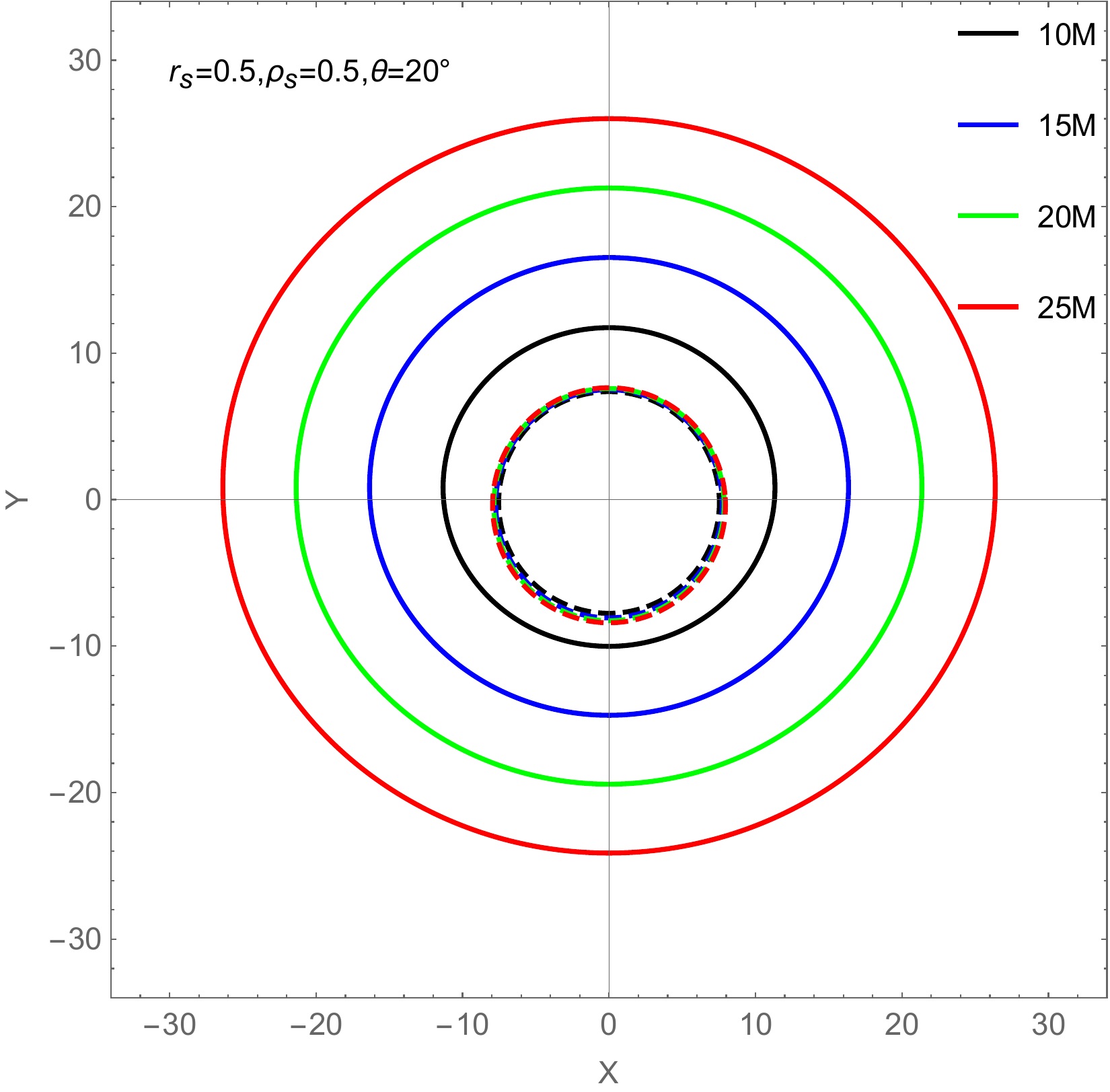}
		\end{subfigure}
		\hspace{0.02\textwidth}
		\begin{subfigure}{0.31\textwidth}
			\includegraphics[width=\linewidth]{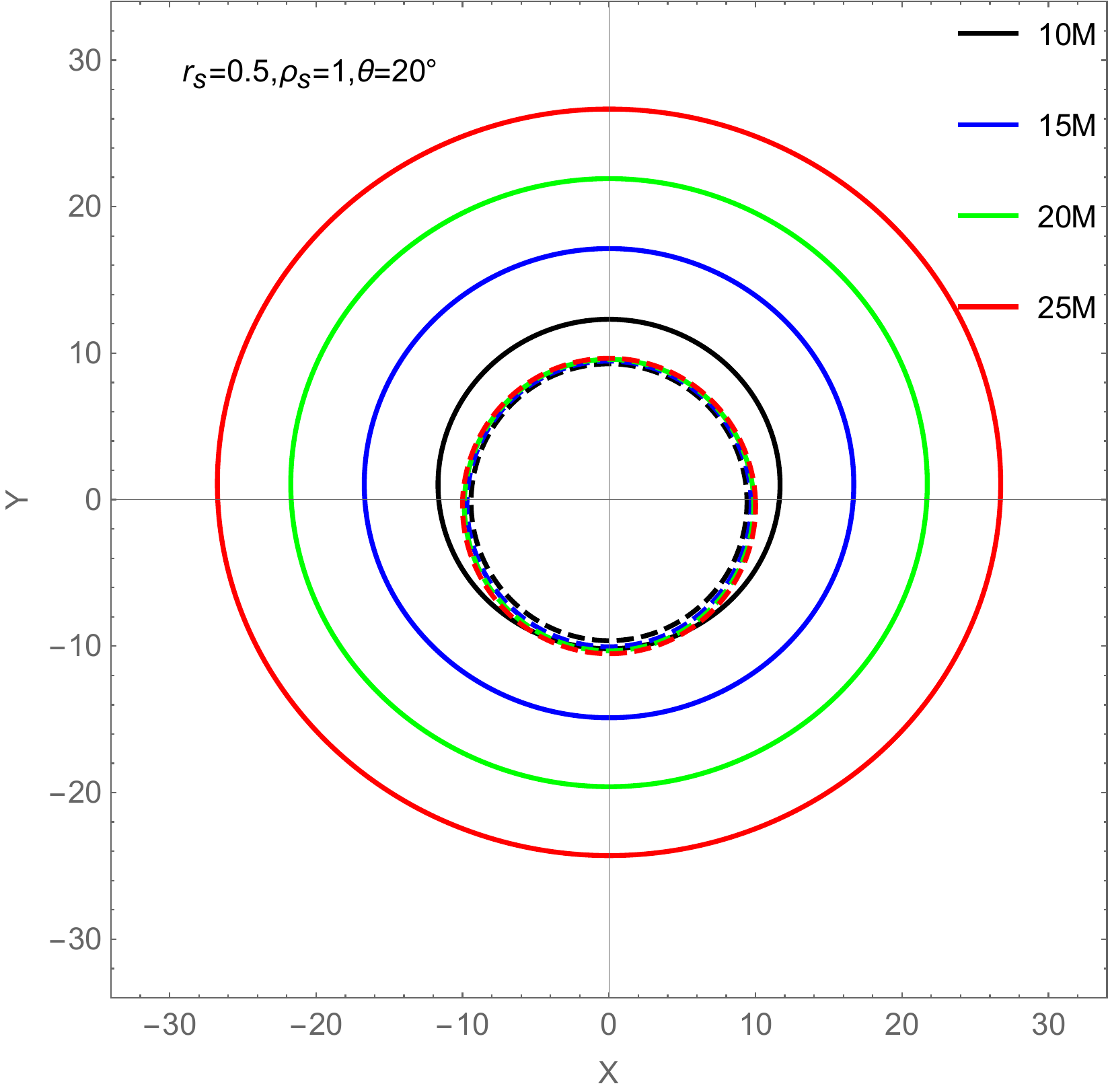}
		\end{subfigure}
		\hspace{0.02\textwidth}
		\begin{subfigure}{0.31\textwidth}
			\includegraphics[width=\linewidth]{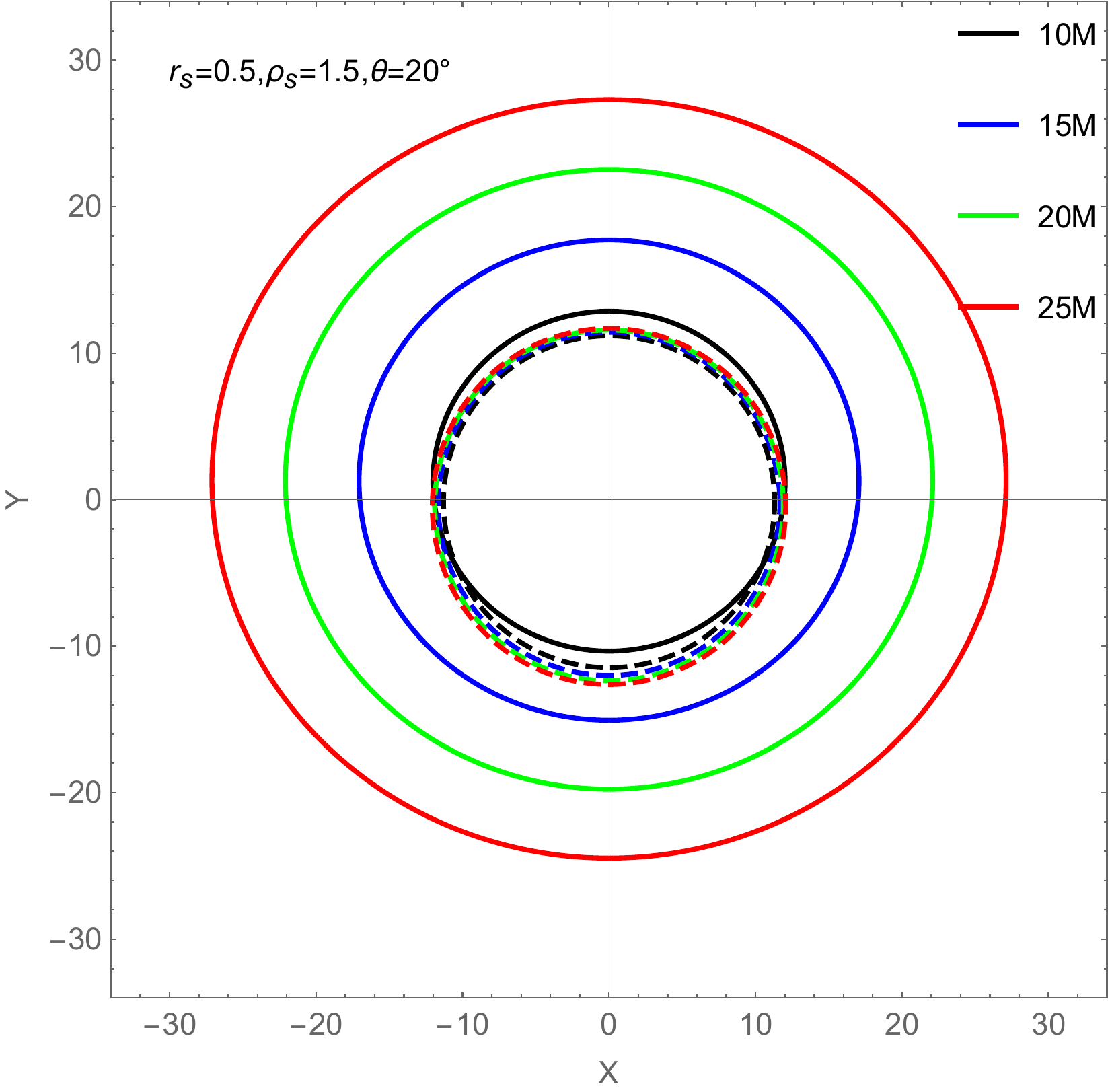}
		\end{subfigure}
		
		\vspace{0.3cm}
		
		\begin{subfigure}{0.31\textwidth}
			\includegraphics[width=\linewidth]{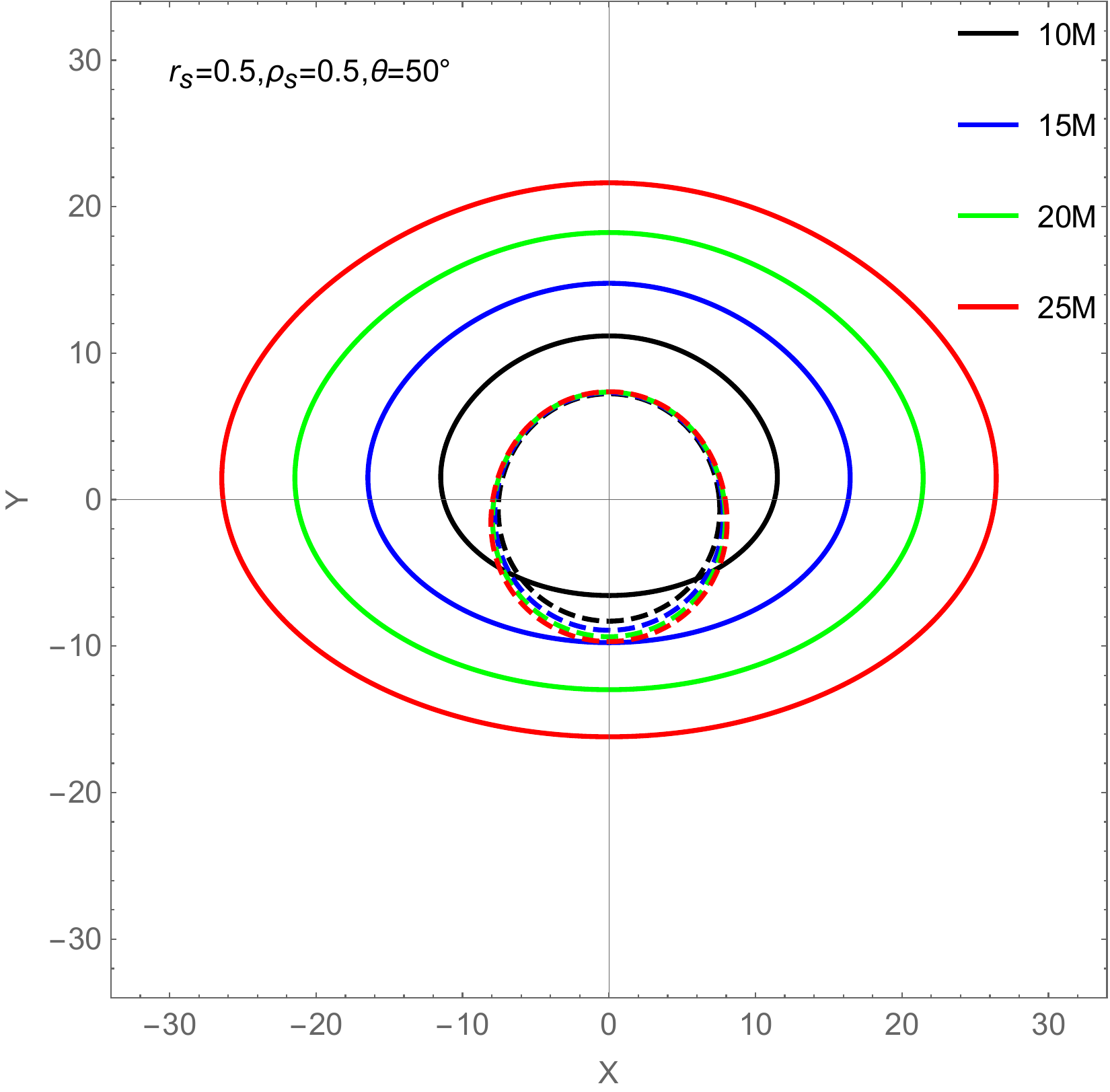}
		\end{subfigure}
		\hspace{0.02\textwidth}
		\begin{subfigure}{0.31\textwidth}
			\includegraphics[width=\linewidth]{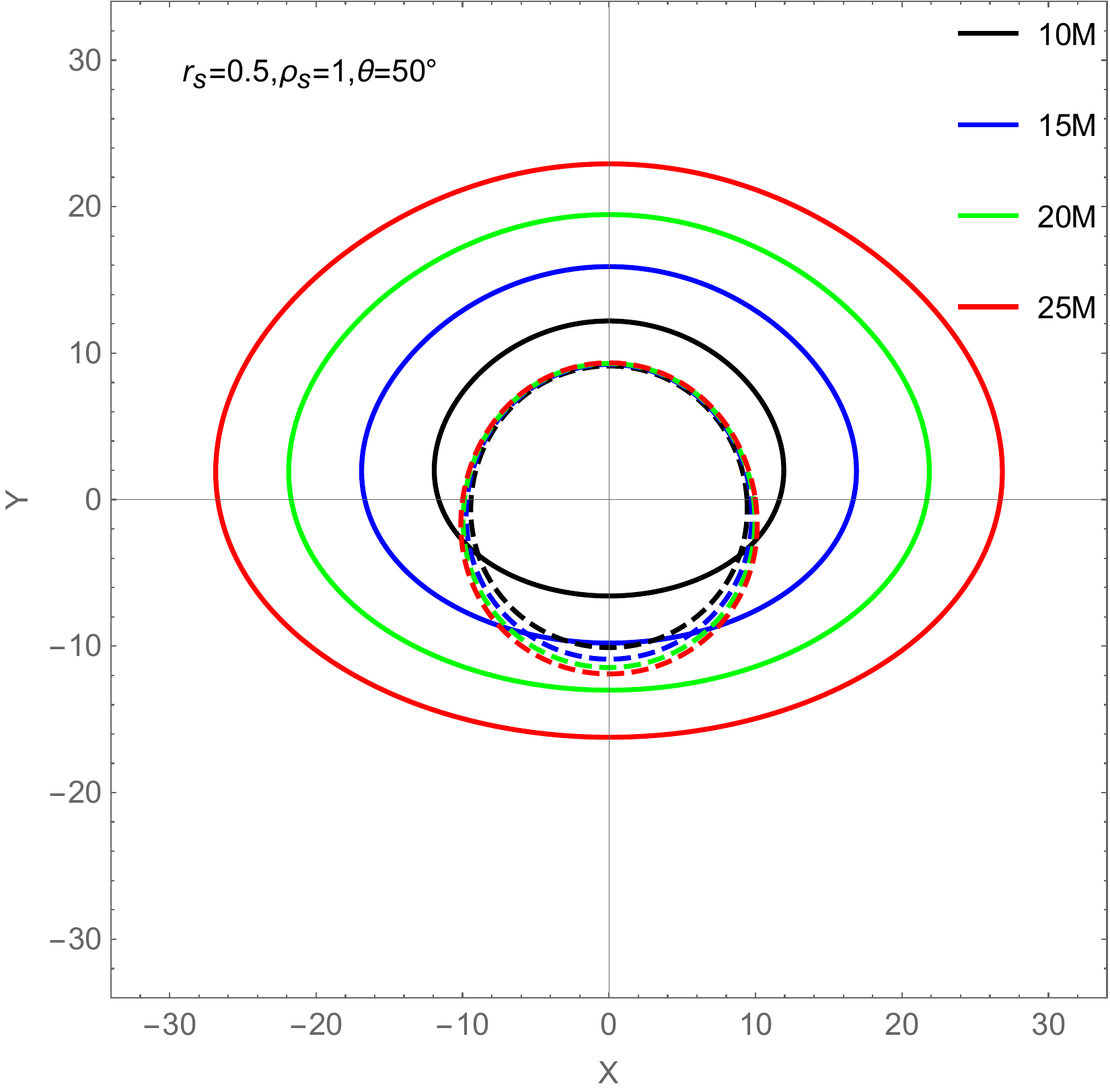}
		\end{subfigure}
		\hspace{0.02\textwidth}
		\begin{subfigure}{0.31\textwidth}
			\includegraphics[width=\linewidth]{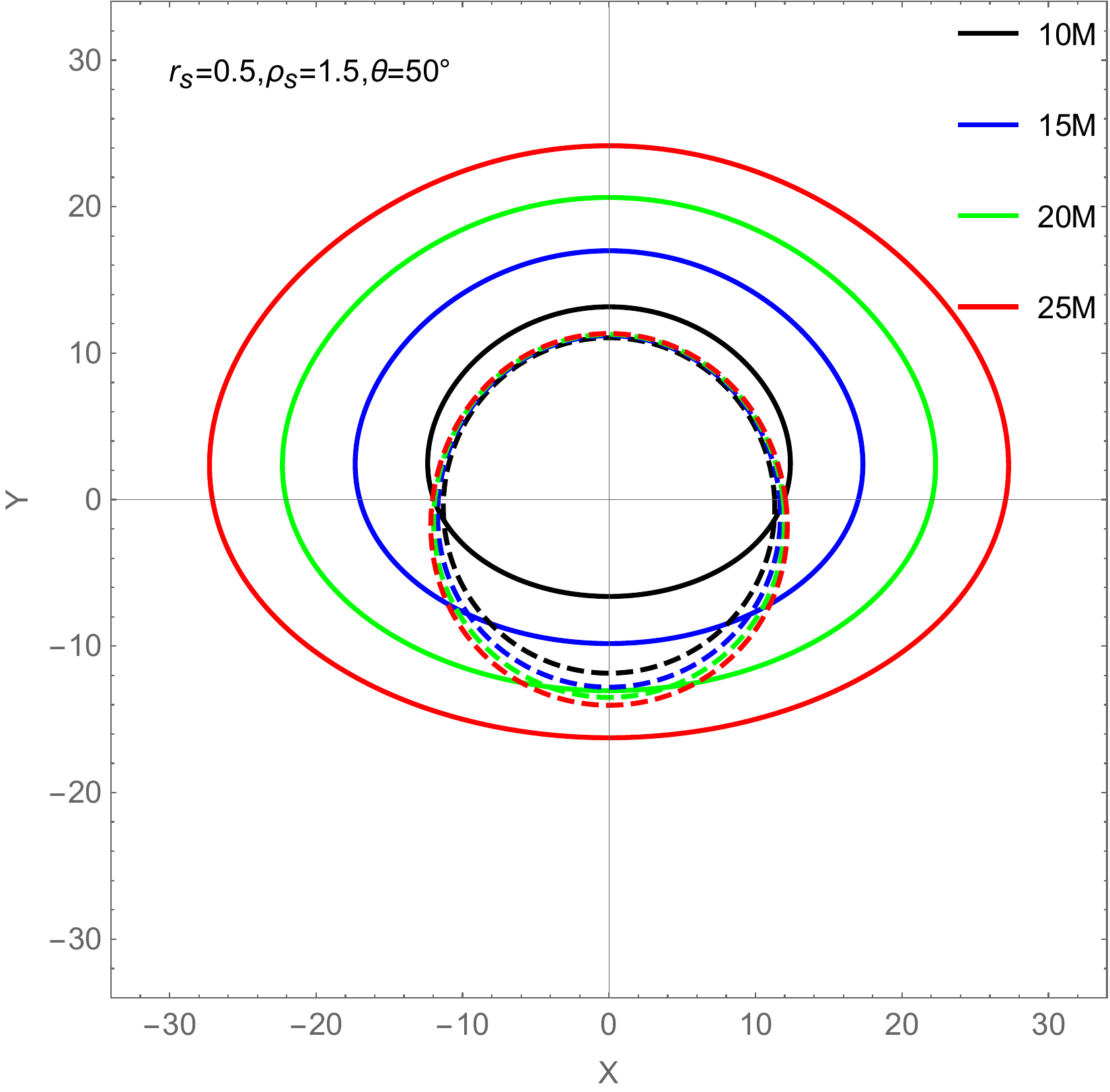}
		\end{subfigure}
		
		\vspace{0.3cm}
		
		\begin{subfigure}{0.31\textwidth}
			\includegraphics[width=\linewidth]{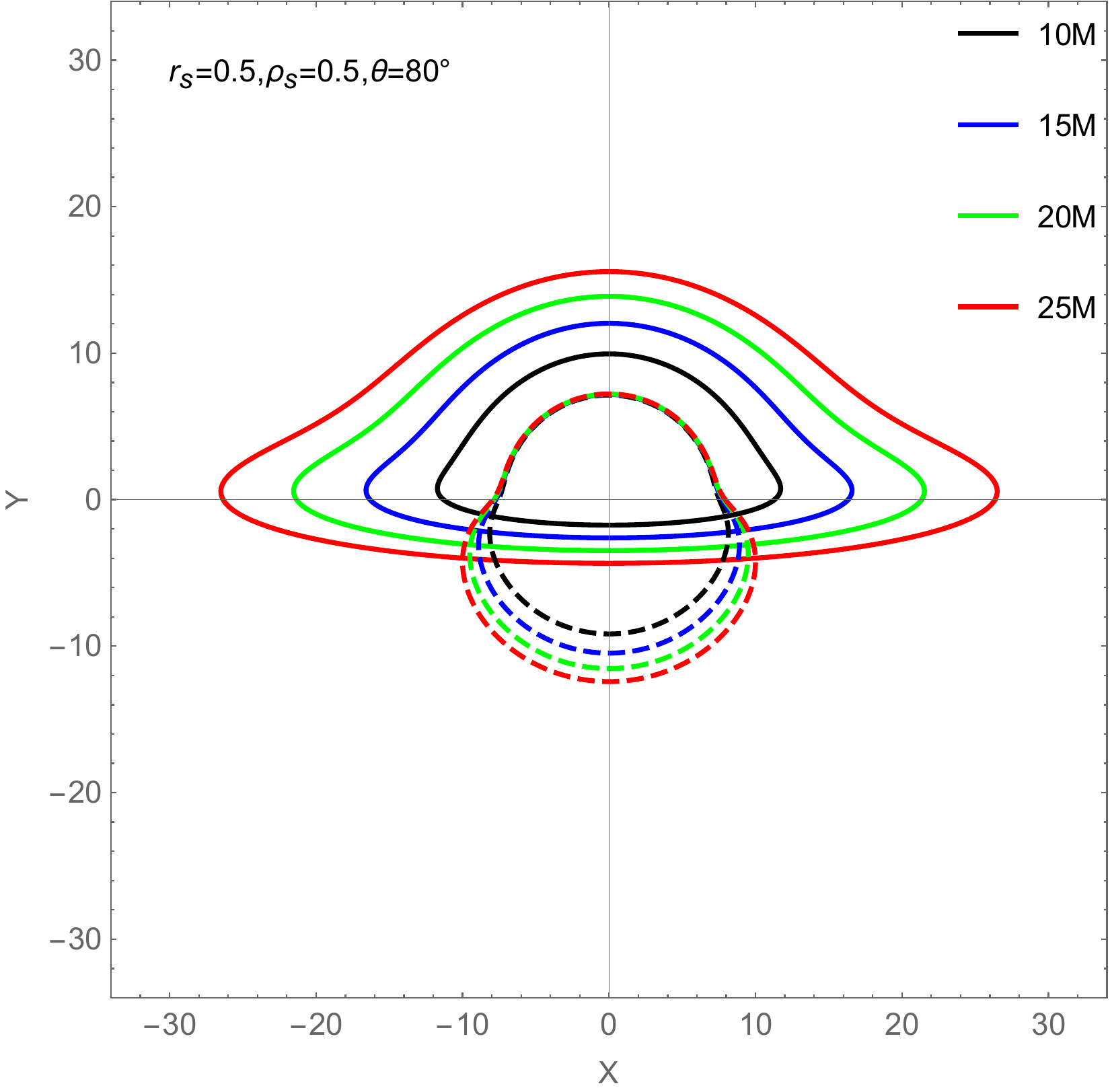}
		\end{subfigure}
		\hspace{0.02\textwidth}
		\begin{subfigure}{0.31\textwidth}
			\includegraphics[width=\linewidth]{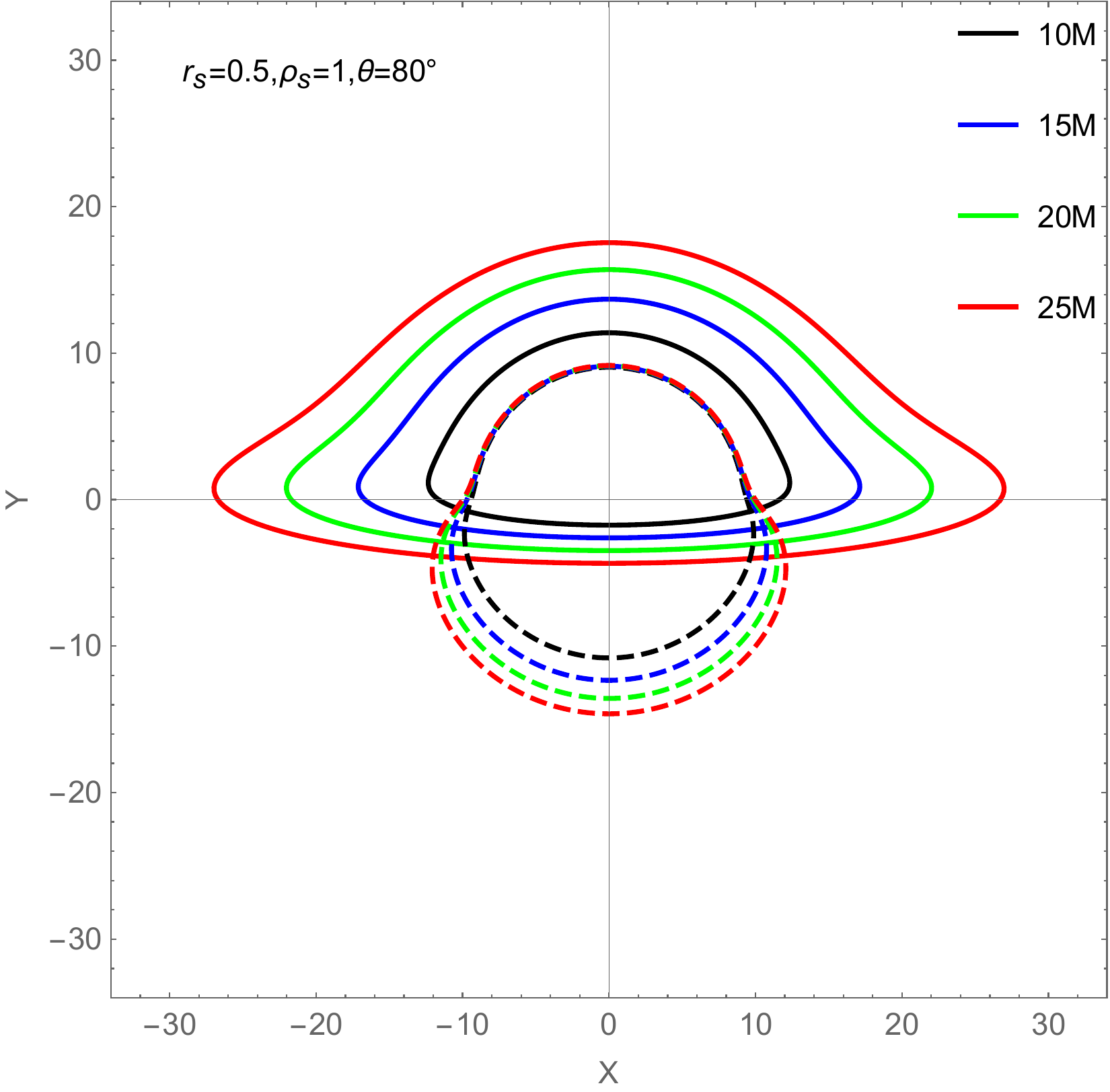}
		\end{subfigure}
		\hspace{0.02\textwidth}
		\begin{subfigure}{0.31\textwidth}
			\includegraphics[width=\linewidth]{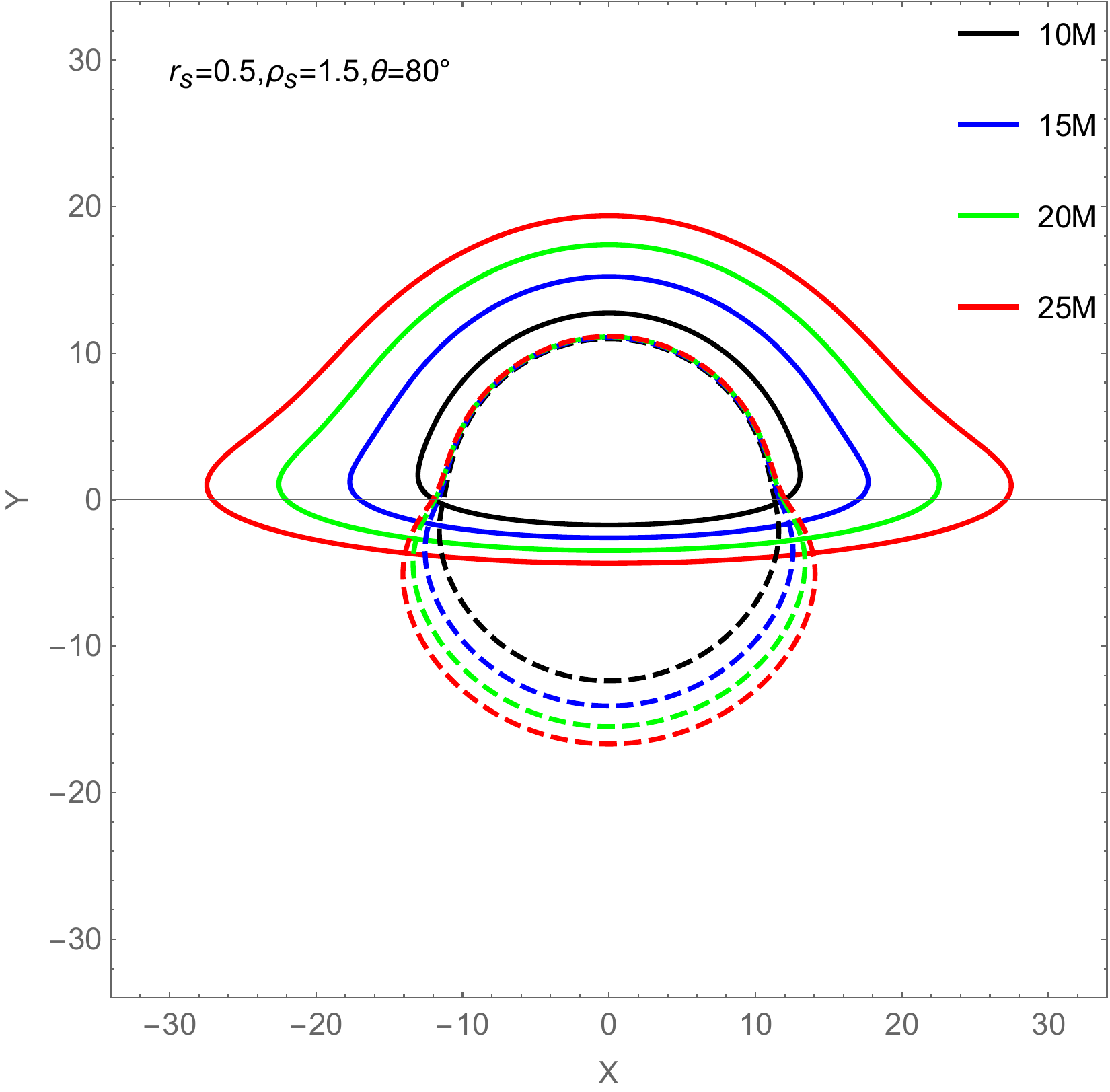}
		\end{subfigure}
		
		\caption{The direct (solid line) and secondary (dashed line) images of the thin accretion disk.}
		\label{dengr2}
	\end{figure*}

	To construct the observed image of the accretion disk, we solve Eqs.~(\ref{fai}), (\ref{faib1}), and (\ref{faib2}) numerically, obtaining the complete mapping from emission coordinates to impact parameters $(b, \alpha)$. The resulting direct and secondary images of stable circular orbits around a Schwarzschild BH—surrounded by a Hernquist DM halo—are shown in Fig.~\ref{dengr1} for three observer inclinations: $20^\circ$, $50^\circ$, and $80^\circ$ (columns, top to bottom). Each row (left to right) corresponds to a fixed $\rho_{\mathrm{s}} = 0.4$ and increasing halo scale radius $r_{\mathrm{s}} = 0.3$, $0.6$, $0.9$. The four concentric orbits, with radii $r = 10$–$25$, are plotted from innermost to outermost. Two clear trends emerge: (i) higher inclination angles enhance the separation between the primary and secondary images; (ii) increasing $r_{\mathrm{s}}$ causes both image families to expand outward in all directions on the observer plane. The influence of the DM density on strong-field lensing is illustrated in Fig.~\ref{dengr2}, which mirrors the setup of Fig.~\ref{dengr2} but fixes $r_{\mathrm{s}} = 0.5$ and varies $\rho_{\mathrm{s}}$ across columns ($0.5$, $1$, $1.5$). As the central density of the Hernquist halo increases, photon trajectories are deflected more strongly, causing both the direct and secondary images to expand. This expansion is anisotropic: at high observer inclinations, the vertical (polar) dimension stretches more significantly than the horizontal (azimuthal) one, reflecting the non-uniform enhancement of spacetime curvature near the equatorial plane. 
	
	\subsection{Observed flux}
	Following the relativistic accretion disk model of Page and Thorne~\cite{Page:1974he}, the emitted local energy flux $F(r)$ is derived self-consistently within the Novikov–Thorne paradigm, which incorporates general relativistic effects on radiative transfer in a geometrically thin, optically thick disk and takes the form \cite{Page:1974he,Collodel:2021gxu}
	\begin{equation}
		F(r)=-\frac{\dot{M}_{0}\Omega_{,r}}{4\pi\sqrt{-g/g_{\theta \theta}}(E-\Omega L)^{2}}\int^{r}_{r_{\mathrm{ISCO}}}(E-\Omega L)L_{,r}\mathrm{d}r,\label{flux}
	\end{equation}
	where $g$ is the metric determinant; $E$, $L$, and $\Omega$ are the specific energy, specific angular momentum, and angular velocity of a circularly orbiting particle, respectively; and $\dot{M}_0$ denotes the mass accretion rate. The specific energy $E$, the specific angular momentum $L$, and the angular velocity $\Omega$ for circular orbits in an arbitrary static, spherically symmetric spacetime are expressed as
	\begin{equation}
		E=-\frac{g_{tt}}{\sqrt{-g_{tt}-g_{\phi\phi}\Omega^{2}}},\label{En}
	\end{equation}
	\begin{equation}
		L=\frac{g_{\phi\phi}\Omega}{\sqrt{-g_{tt}-g_{\phi\phi}\Omega^{2}}},\label{Ln}
	\end{equation}
	\begin{equation}
		\Omega=\frac{\mathrm{d}\phi}{\mathrm{d}t}=\sqrt{-\frac{g_{tt,r}}{g_{\phi\phi,r}}}.\label{An}
	\end{equation}
	
\begin{figure*}[htbp]
		\centering
		\begin{tabular}{ccc}
			\begin{minipage}[t]{0.3\textwidth}
				\centering
				\begin{overpic}[width=0.75\textwidth]{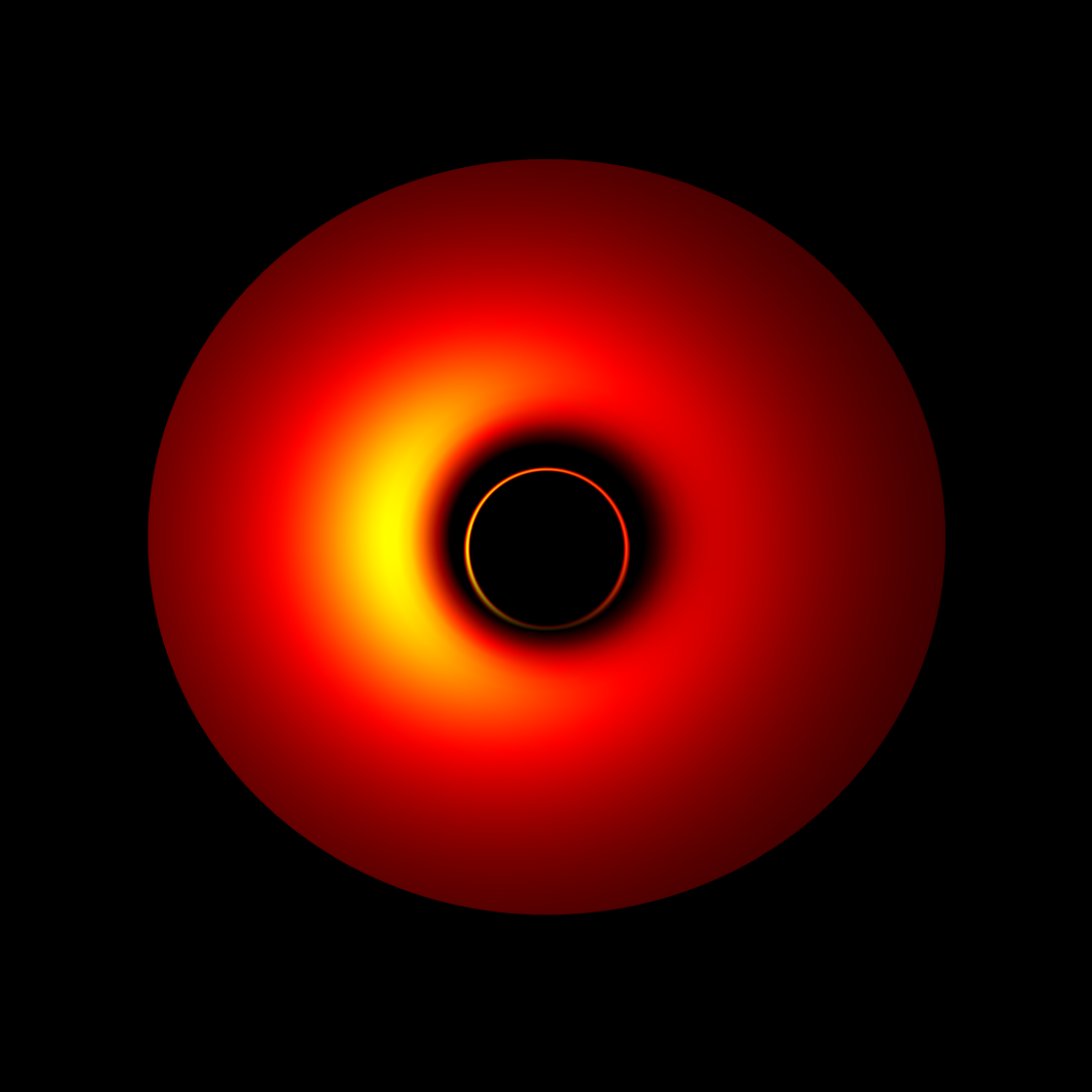} 
					\put(0,103){\color{black}\large $\rho_{\mathrm{s}}=0.4,r_{\mathrm{s}}=0.3,\theta=20^{\circ}$} 
					\put(-10,48){\color{black} Y}
					\put(48,-10){\color{black} X}
				\end{overpic}
			\end{minipage}
			&
			\begin{minipage}[t]{0.3\textwidth}
				\centering
				\begin{overpic}[width=0.75\textwidth]{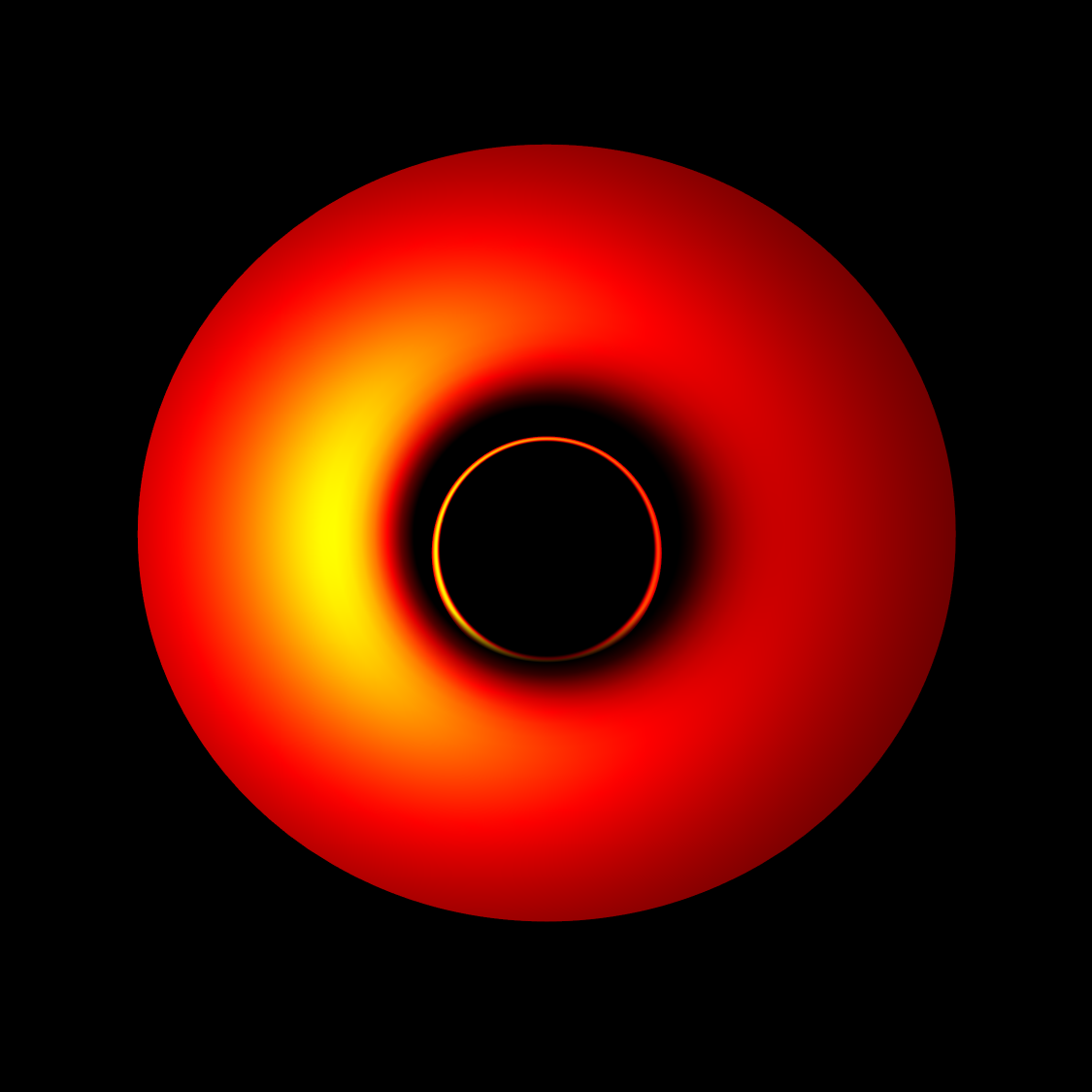} 
					\put(0,103){\color{black}\large $\rho_{\mathrm{s}}=0.4,r_{\mathrm{s}}=0.6,\theta=20^{\circ}$} 
					\put(-10,48){\color{black} Y}
					\put(48,-10){\color{black} X}
				\end{overpic}
			\end{minipage}
			&
			\begin{minipage}[t]{0.3\textwidth}
				\centering
				\begin{overpic}[width=0.75\textwidth]{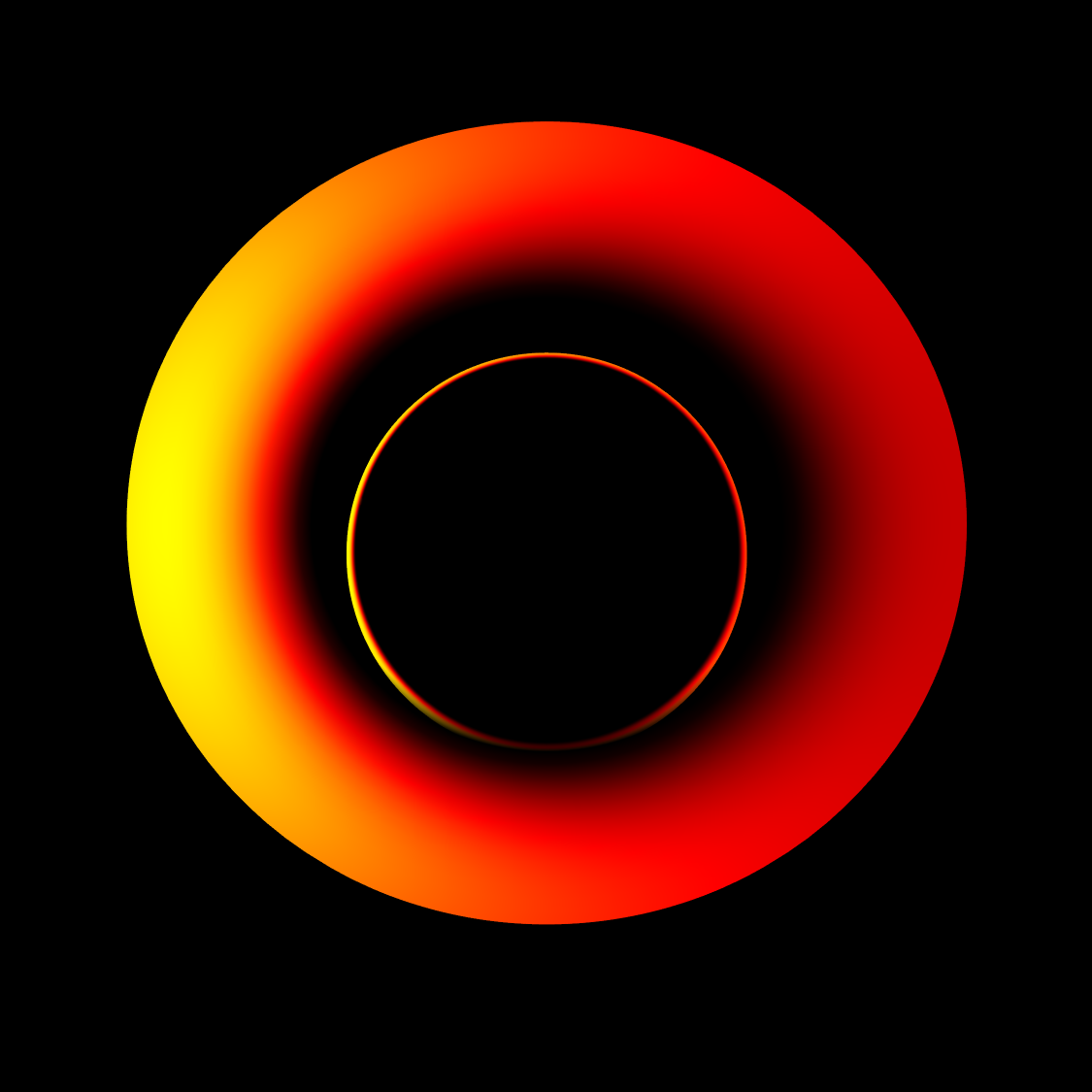}
					\put(0,103){\color{black}\large $\rho_{\mathrm{s}}=0.4,r_{\mathrm{s}}=0.9,\theta=20^{\circ}$} 
					\put(-10,48){\color{black} Y}
					\put(48,-10){\color{black} X}
				\end{overpic}
			\end{minipage}
			\vspace{40pt} 
			\\ 
			\begin{minipage}[t]{0.3\textwidth}
				\centering
				\begin{overpic}[width=0.75\textwidth]{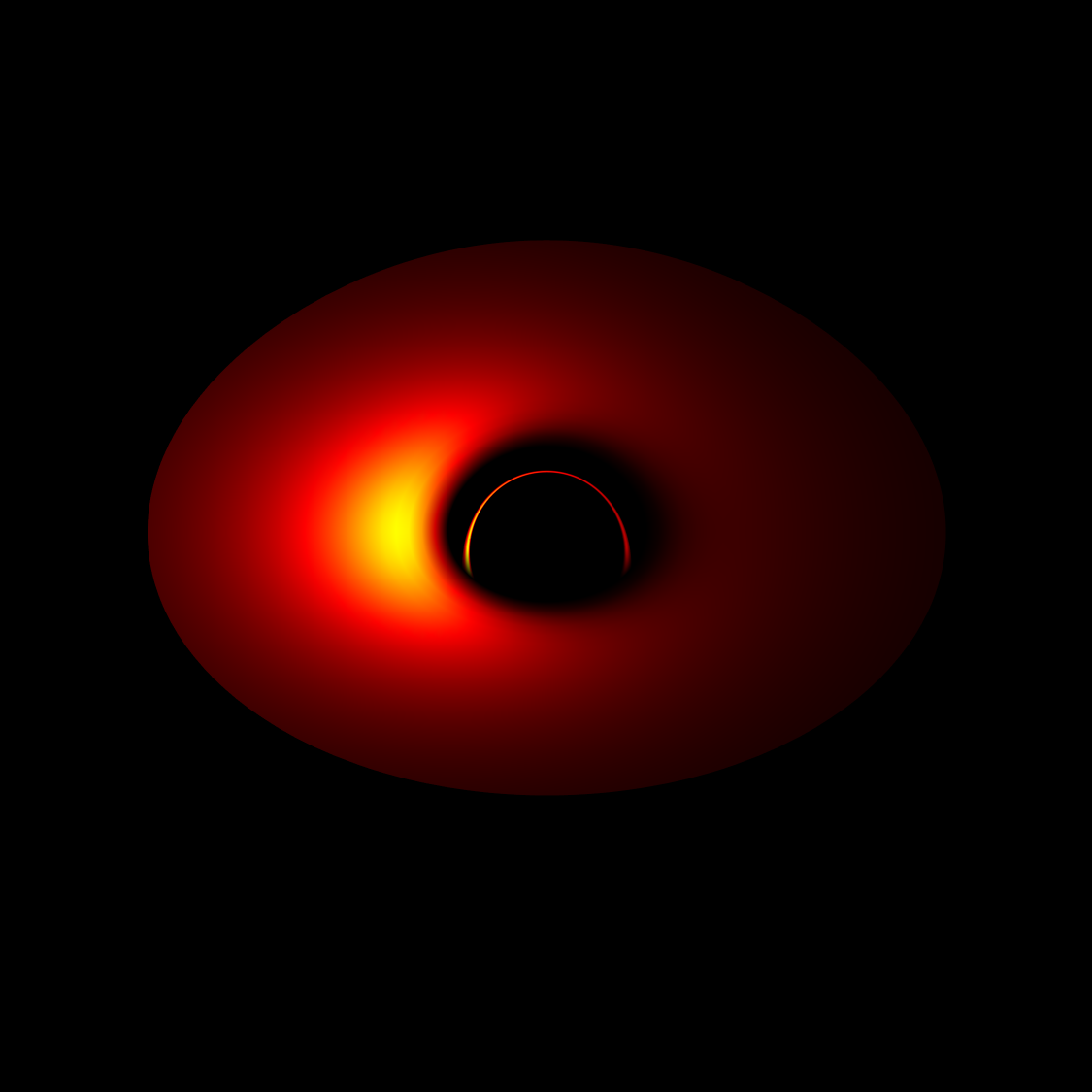} 
					\put(0,103){\color{black}\large $\rho_{\mathrm{s}}=0.4,r_{\mathrm{s}}=0.3,\theta=50^{\circ}$} 
					\put(-10,48){\color{black} Y}
					\put(48,-10){\color{black} X}
				\end{overpic}%\hfill
			\end{minipage}
			&
			\begin{minipage}[t]{0.3\textwidth}
				\centering
				\begin{overpic}[width=0.75\textwidth]{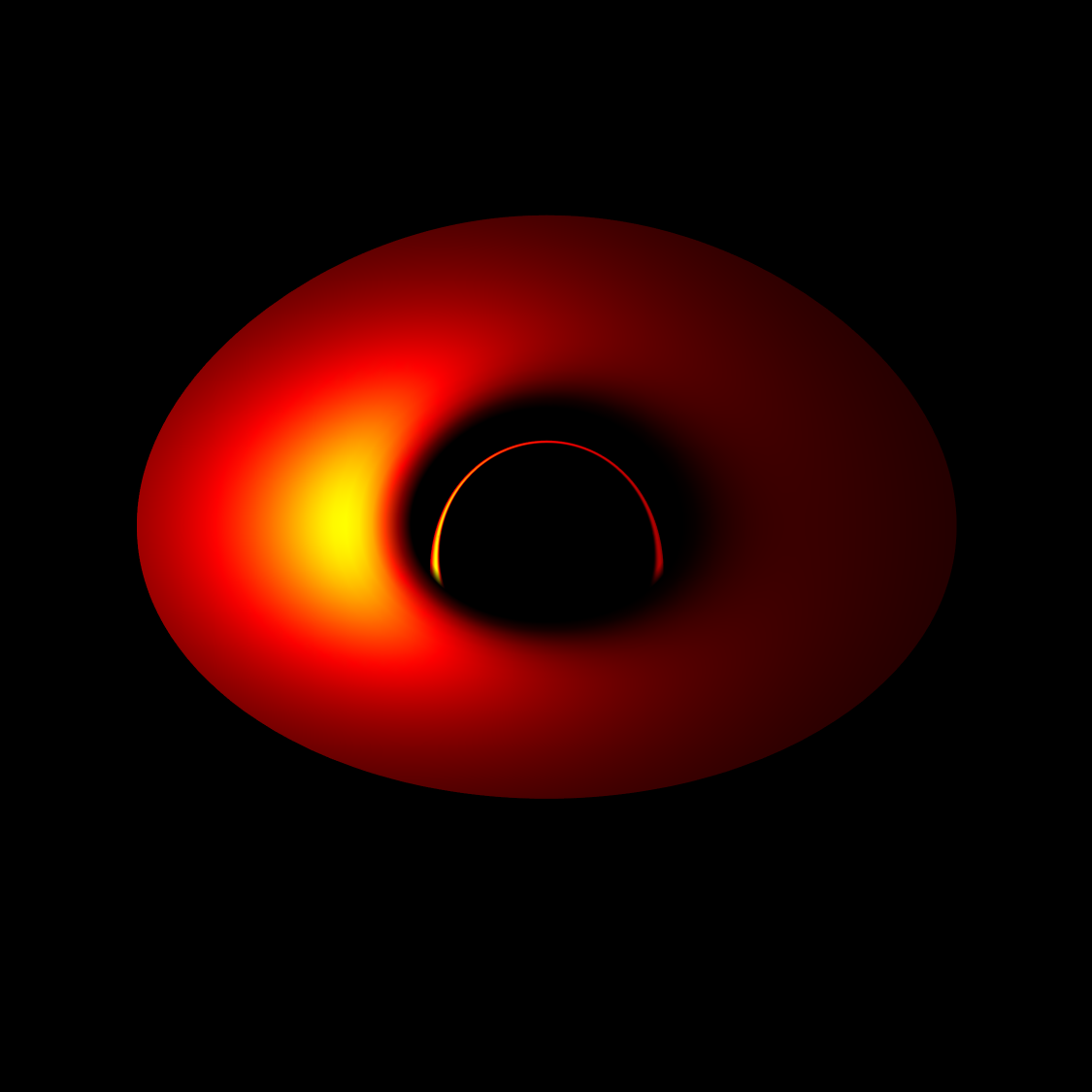} 
					\put(0,103){\color{black}\large $\rho_{\mathrm{s}}=0.4,r_{\mathrm{s}}=0.6,\theta=50^{\circ}$} 
					\put(-10,48){\color{black} Y}
					\put(48,-10){\color{black} X}
				\end{overpic}
			\end{minipage}
			&
			\begin{minipage}[t]{0.3\textwidth}
				\centering
				\begin{overpic}[width=0.75\textwidth]{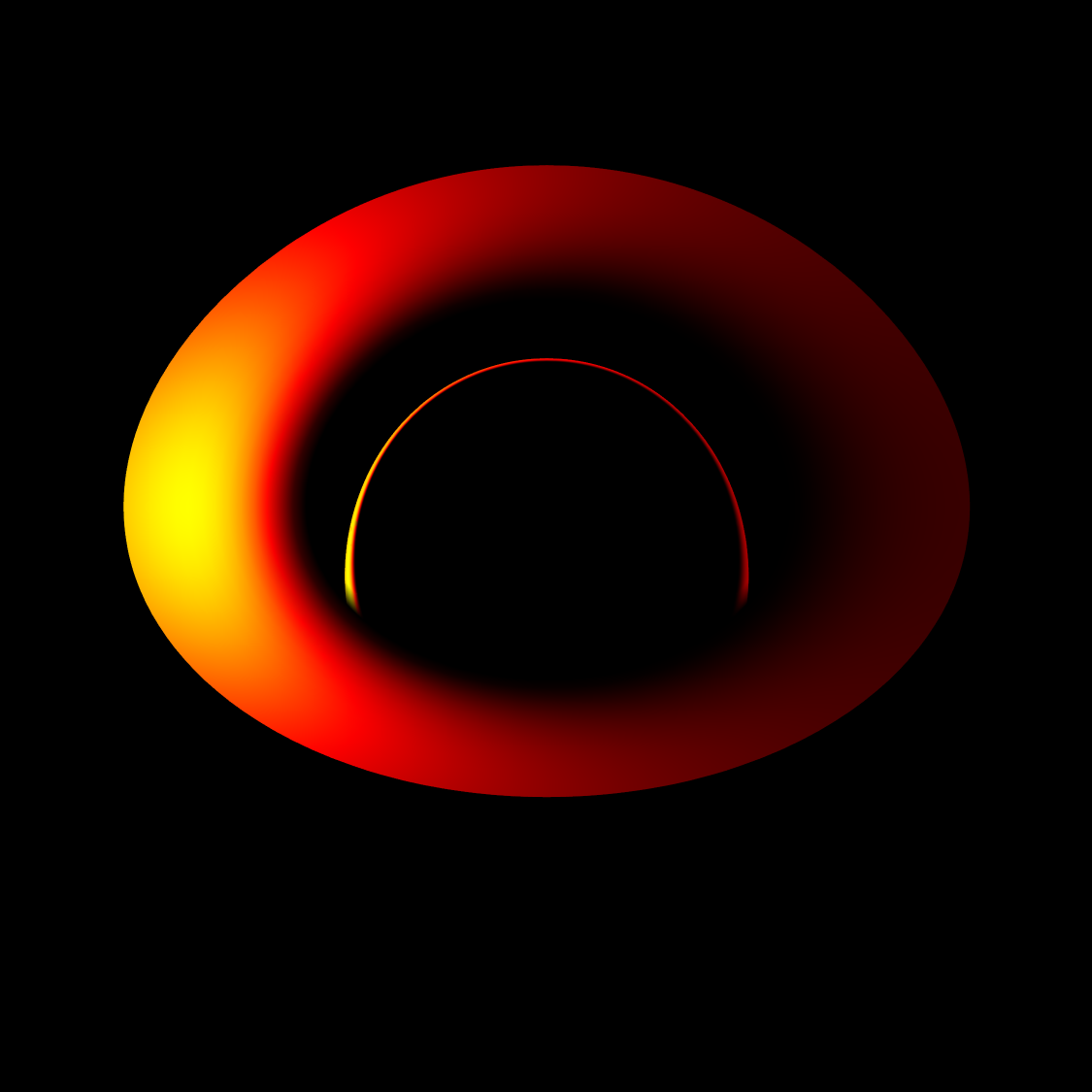} 
					\put(0,103){\color{black}\large $\rho_{\mathrm{s}}=0.4,r_{\mathrm{s}}=0.9,\theta=50^{\circ}$} 
					\put(-10,48){\color{black} Y}
					\put(48,-10){\color{black} X}
				\end{overpic}
			\end{minipage}
			\vspace{40pt} 
			\\
			\begin{minipage}[t]{0.3\textwidth}
				\centering
				\begin{overpic}[width=0.75\textwidth]{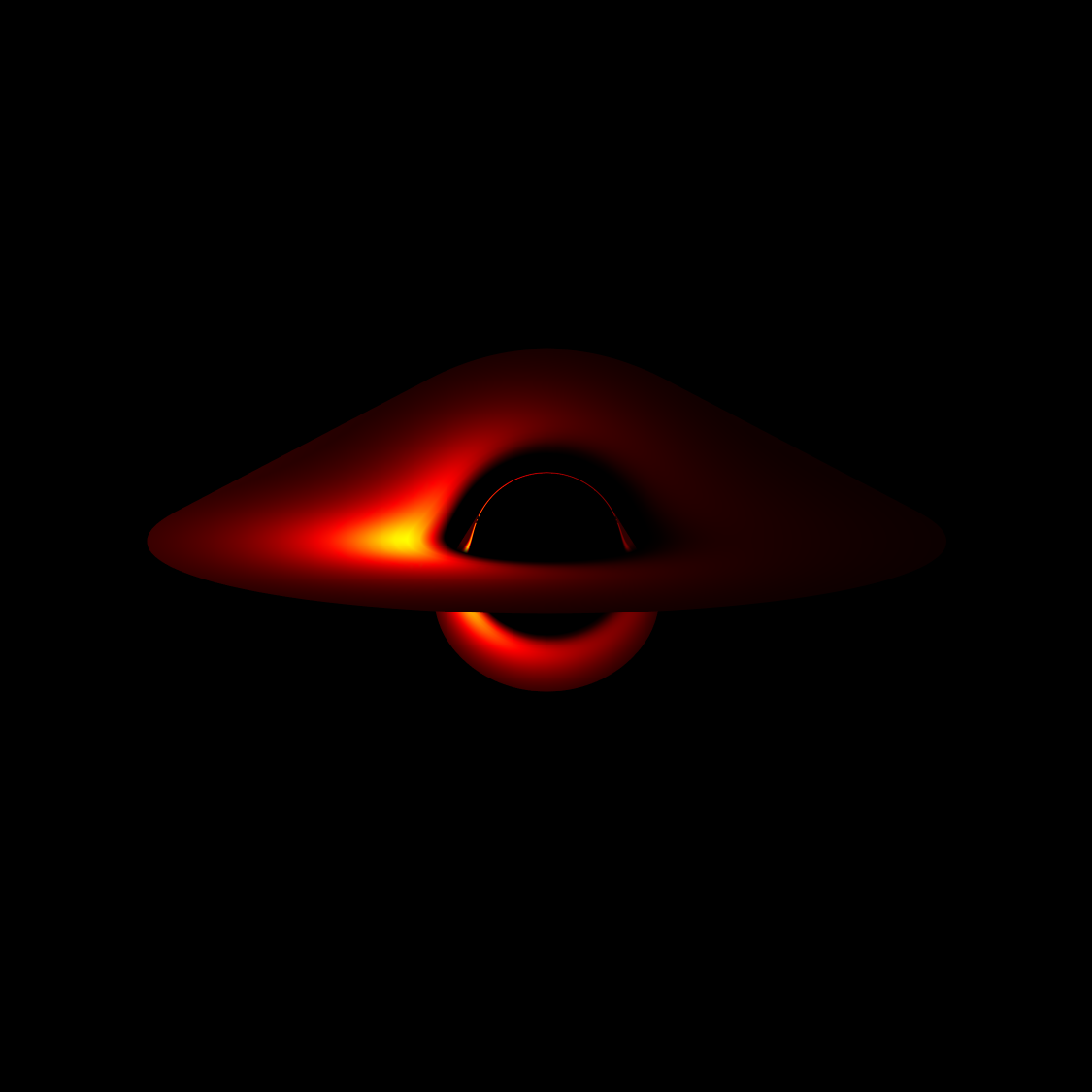}
					\put(0,103){\color{black}\large $\rho_{\mathrm{s}}=0.4,r_{\mathrm{s}}=0.3,\theta=80^{\circ}$}
					\put(-10,48){\color{black} Y}
					\put(48,-10){\color{black} X}
				\end{overpic}
			\end{minipage}
			&
			\begin{minipage}[t]{0.3\textwidth}
				\centering
				\begin{overpic}[width=0.75\textwidth]{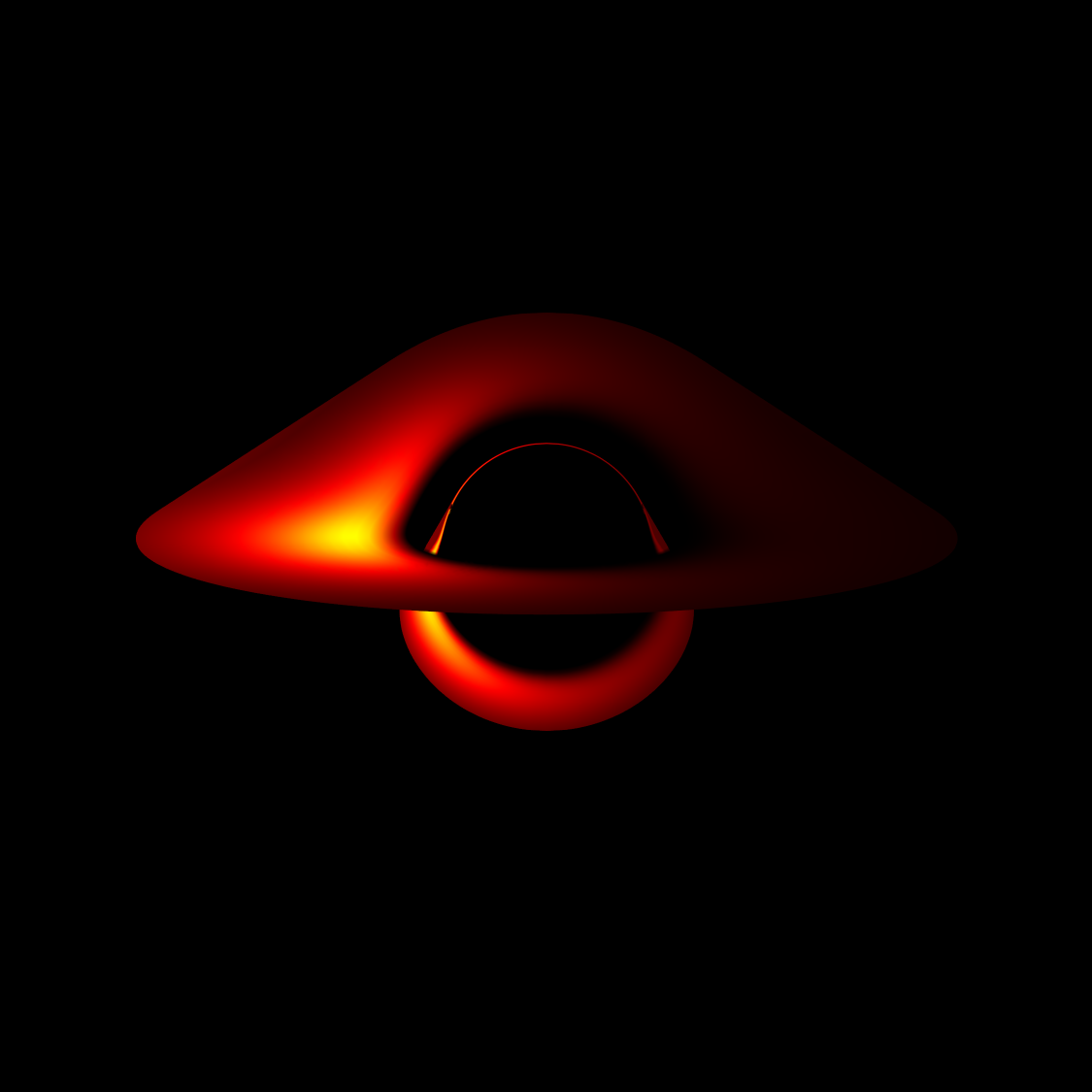}
					\put(0,103){\color{black}\large $\rho_{\mathrm{s}}=0.4,r_{\mathrm{s}}=0.6,\theta=80^{\circ}$} 
					\put(-10,48){\color{black} Y}
					\put(48,-10){\color{black} X}
				\end{overpic}
			\end{minipage}
			&
			\begin{minipage}[t]{0.3\textwidth}
				\centering
				\begin{overpic}[width=0.75\textwidth]{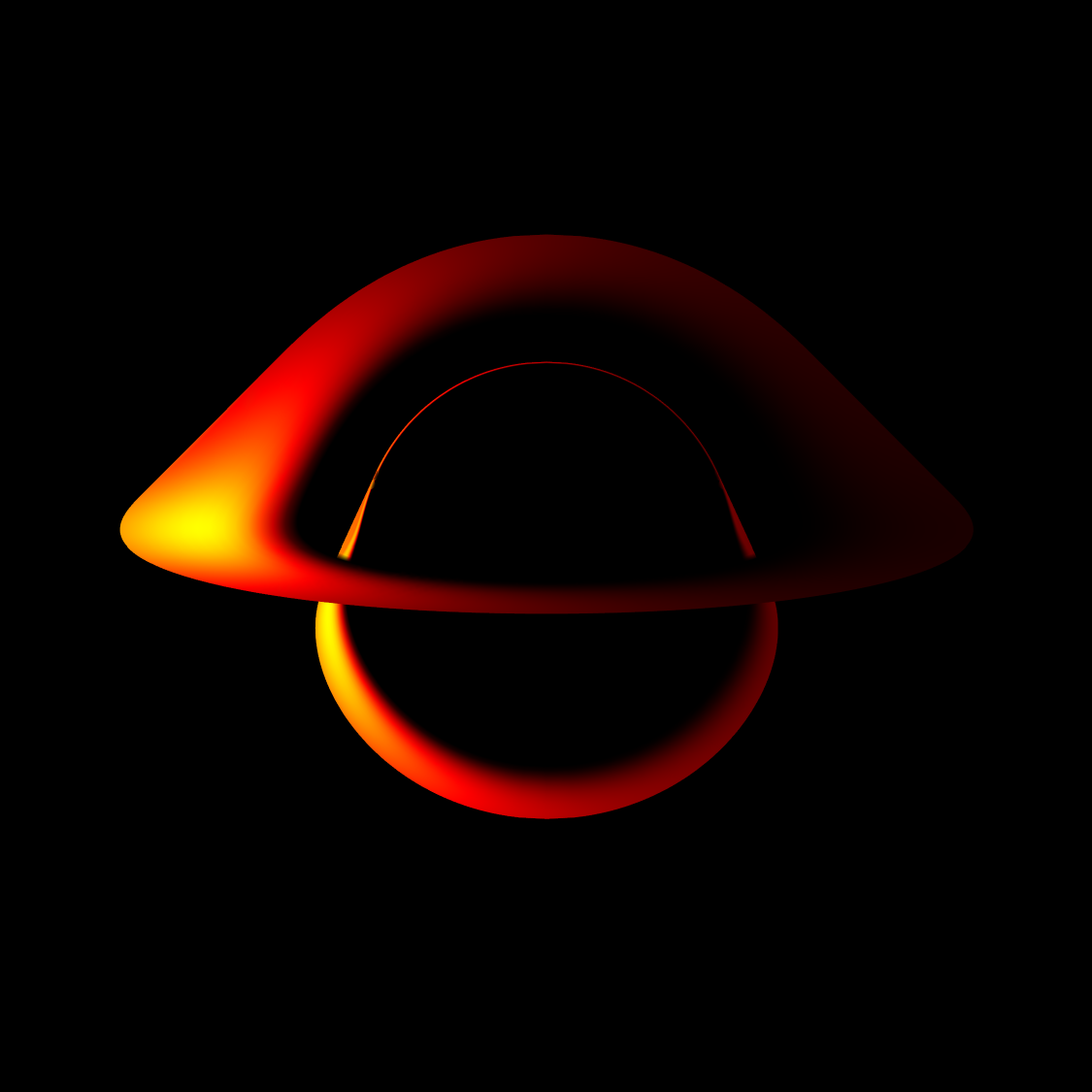}
					\put(0,103){\color{black}\large $\rho_{\mathrm{s}}=0.4,r_{\mathrm{s}}=0.9,\theta=80^{\circ}$}
					\put(-10,48){\color{black} Y}
					\put(48,-10){\color{black} X}
				\end{overpic}
			\end{minipage}
		\end{tabular}
		\caption{The complete apparent images of a thin accretion disk for different values of the parameter $r_{\mathrm{s}}$ and the inclination angle, with $\rho_{\mathrm{s}}=0.4$.}
		\label{xuanran1}
	\end{figure*}

\begin{figure*}[htbp]
		\centering
		\begin{tabular}{ccc}
			\begin{minipage}[t]{0.3\textwidth}
				\centering
				\begin{overpic}[width=0.75\textwidth]{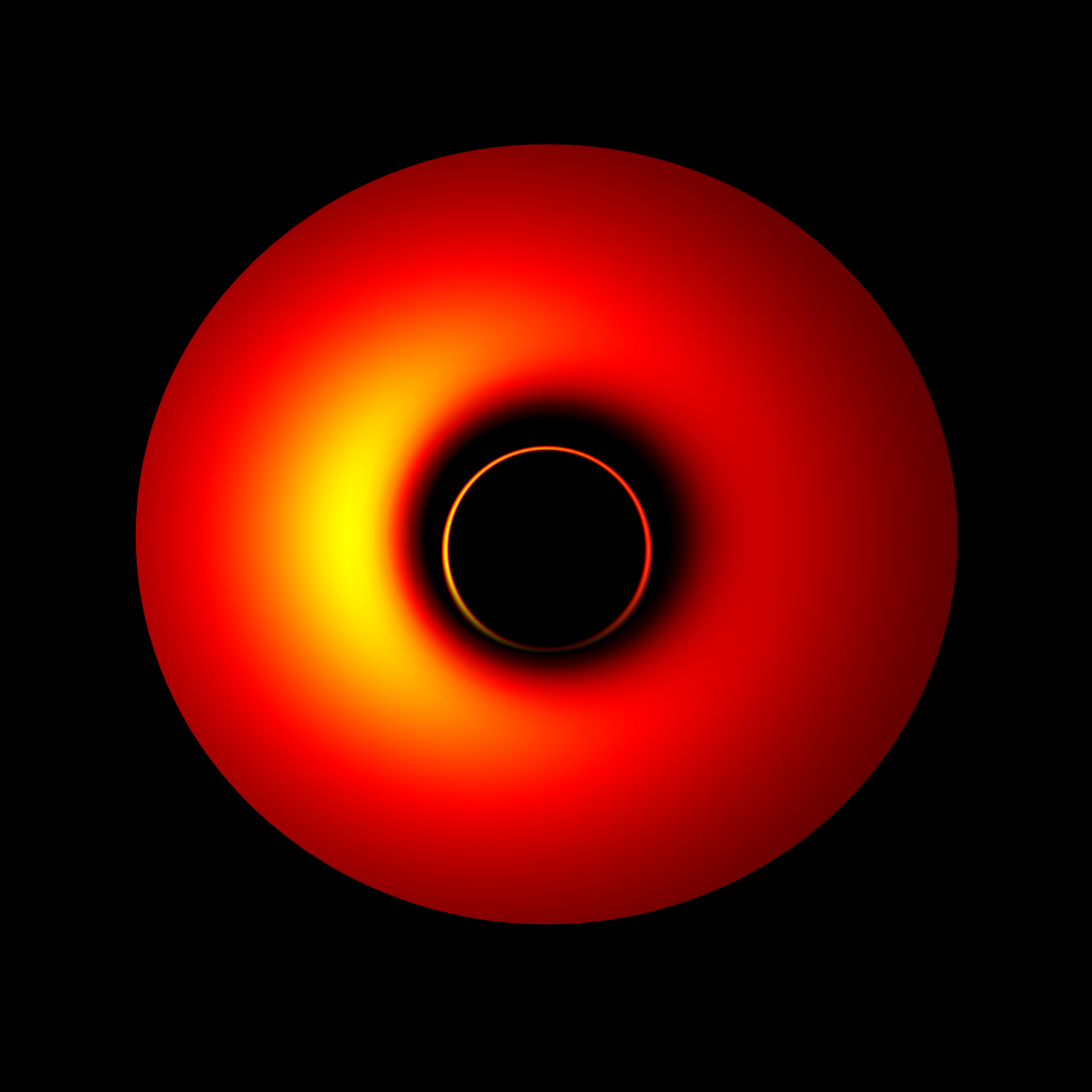} 
					\put(0,103){\color{black}\large $r_{\mathrm{s}}=0.5,\rho_{\mathrm{s}}=0.5,\theta=20^{\circ}$} 
					\put(-10,48){\color{black} Y}
					\put(48,-10){\color{black} X}
				\end{overpic}
			\end{minipage}
			&
			\begin{minipage}[t]{0.3\textwidth}
				\centering
				\begin{overpic}[width=0.75\textwidth]{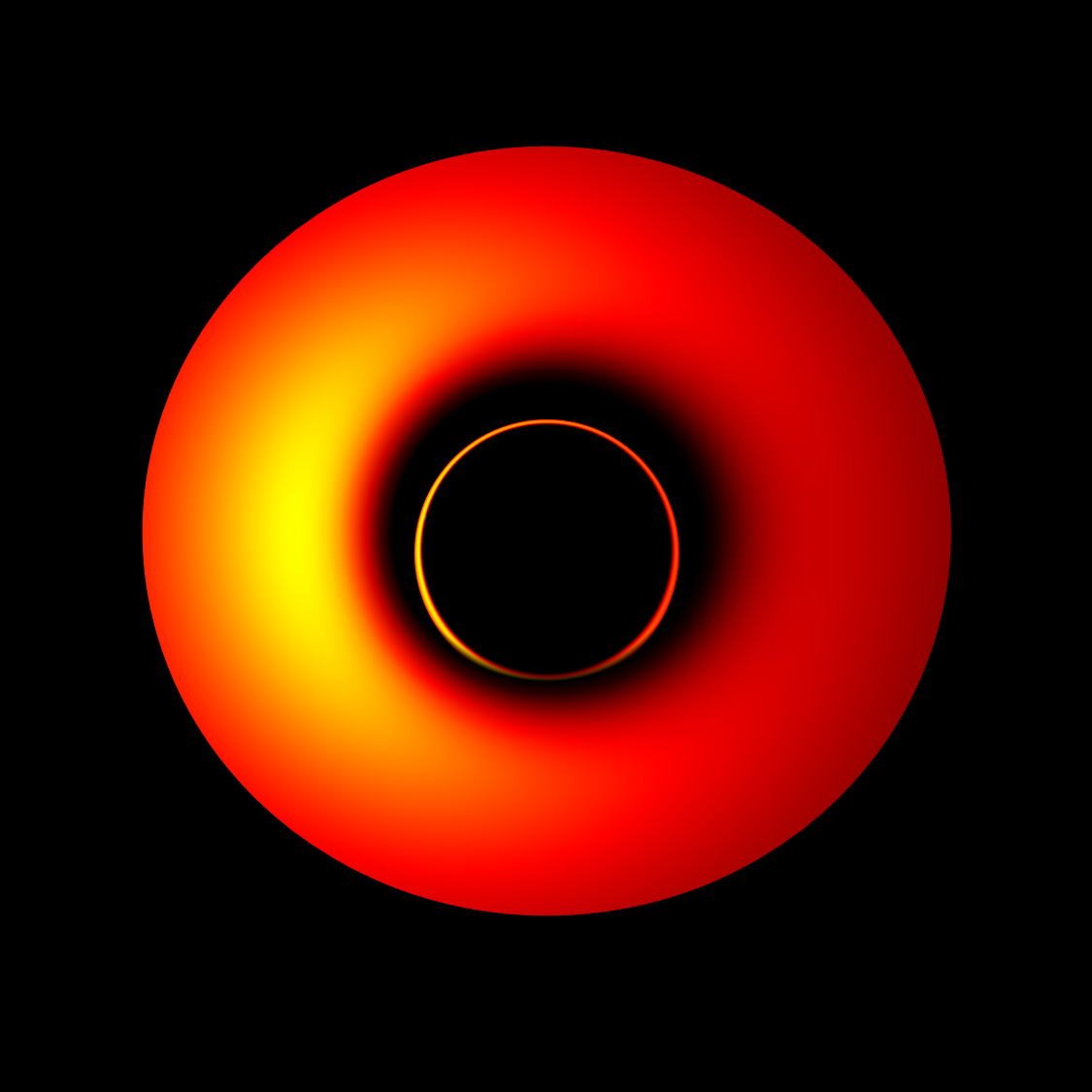} 
					\put(0,103){\color{black}\large $r_{\mathrm{s}}=0.5,\rho_{\mathrm{s}}=1,\theta=20^{\circ}$} 
					\put(-10,48){\color{black} Y}
					\put(48,-10){\color{black} X}
				\end{overpic}
			\end{minipage}
			&
			\begin{minipage}[t]{0.3\textwidth}
				\centering
				\begin{overpic}[width=0.75\textwidth]{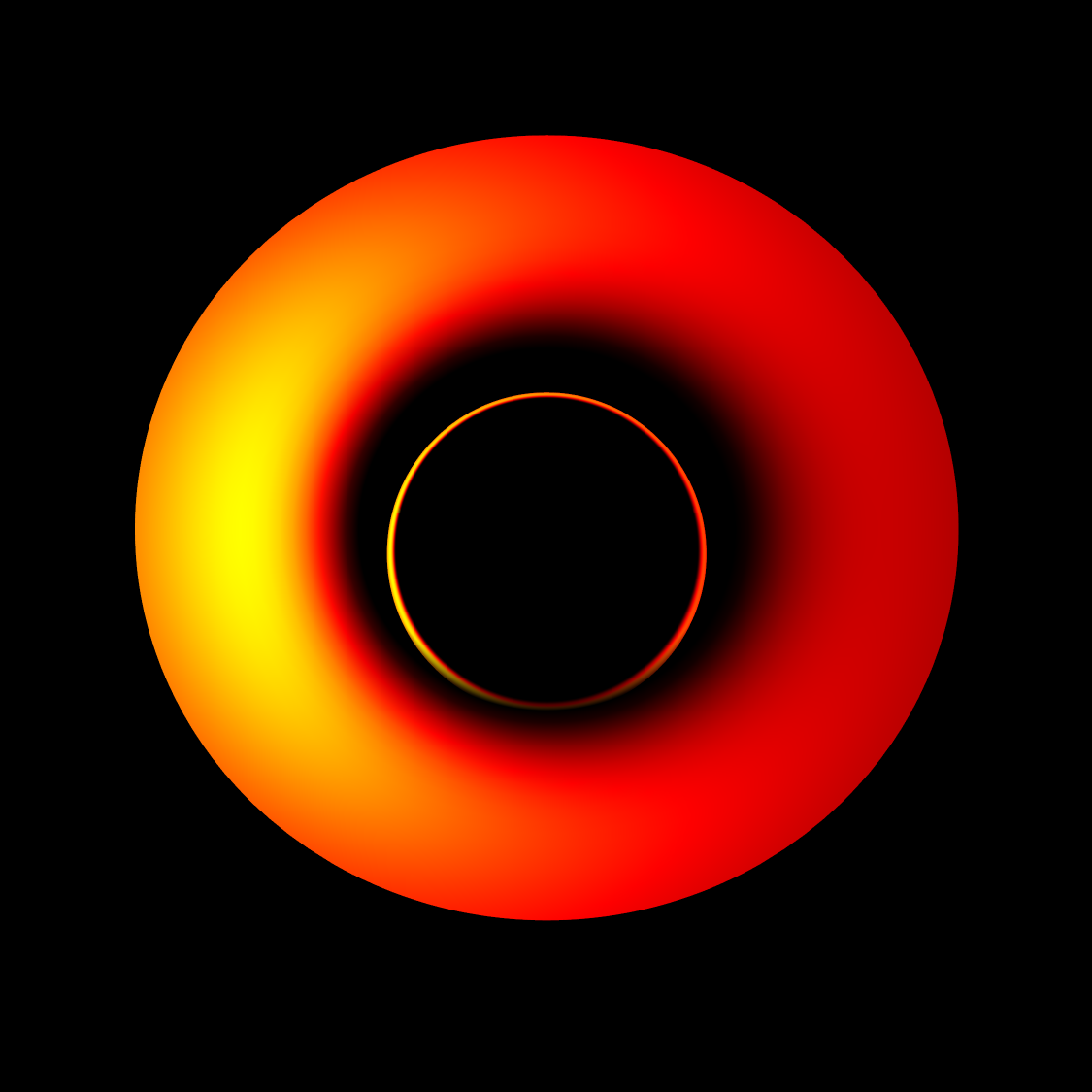}
					\put(0,103){\color{black}\large $r_{\mathrm{s}}=0.5,\rho_{\mathrm{s}}=1.5,\theta=20^{\circ}$} 
					\put(-10,48){\color{black} Y}
					\put(48,-10){\color{black} X}
				\end{overpic}
			\end{minipage}
			\vspace{40pt} 
			\\ 
			\begin{minipage}[t]{0.3\textwidth}
				\centering
				\begin{overpic}[width=0.75\textwidth]{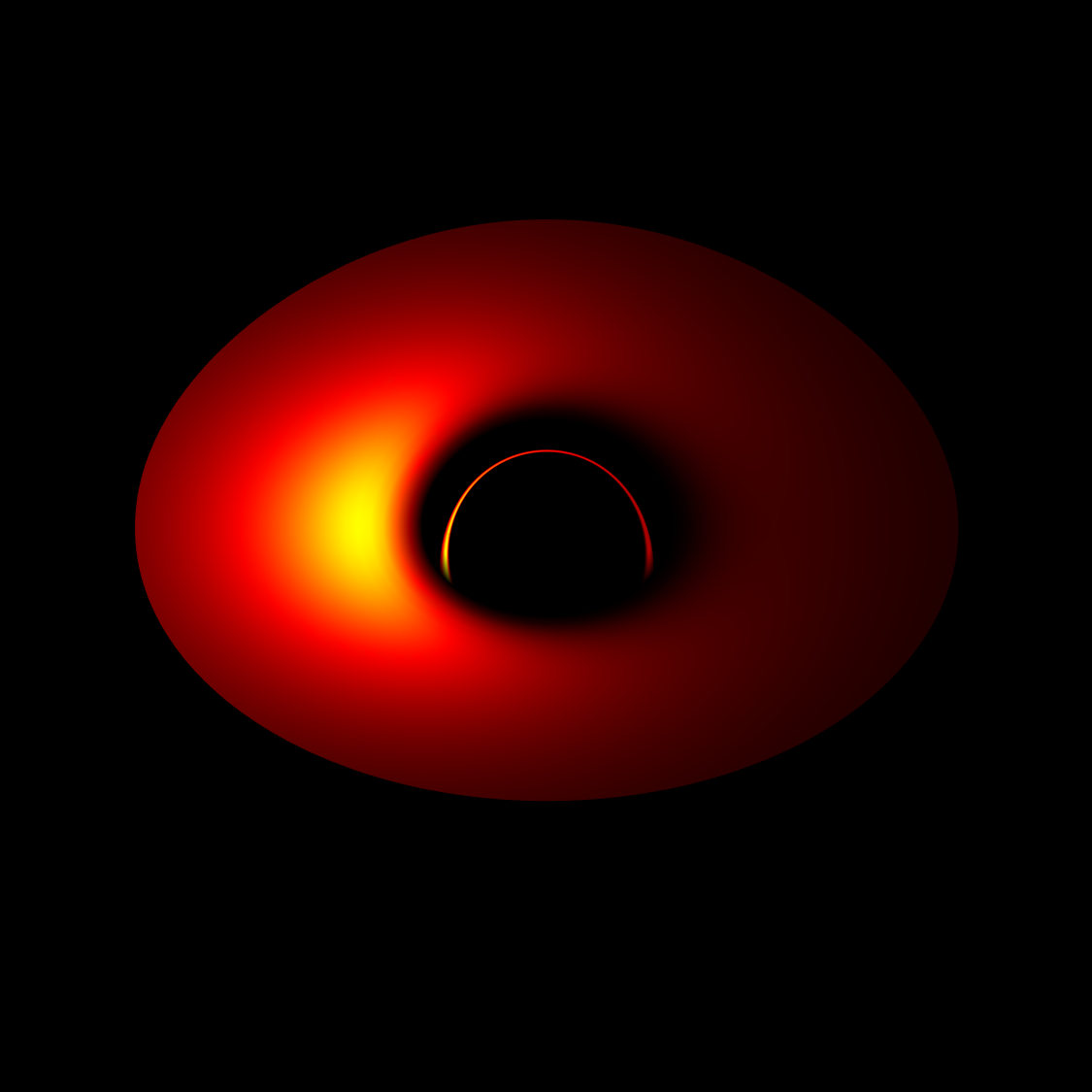} 
					\put(0,103){\color{black}\large $r_{\mathrm{s}}=0.5,\rho_{\mathrm{s}}=0.5,\theta=50^{\circ}$} 
					\put(-10,48){\color{black} Y}
					\put(48,-10){\color{black} X}
				\end{overpic}%\hfill
			\end{minipage}
			&
			\begin{minipage}[t]{0.3\textwidth}
				\centering
				\begin{overpic}[width=0.75\textwidth]{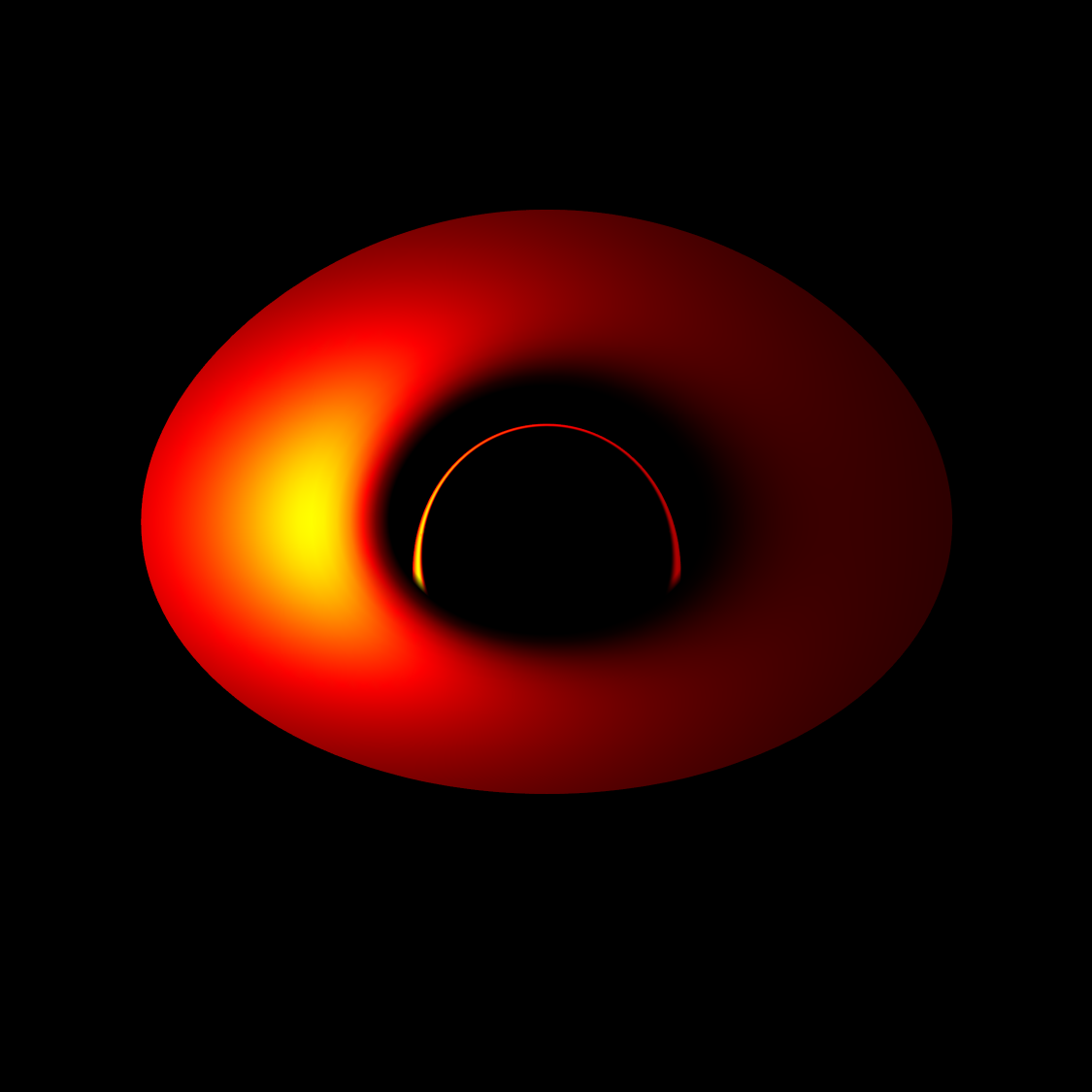} 
					\put(0,103){\color{black}\large $r_{\mathrm{s}}=0.5,\rho_{\mathrm{s}}=1,\theta=50^{\circ}$} 
					\put(-10,48){\color{black} Y}
					\put(48,-10){\color{black} X}
				\end{overpic}
			\end{minipage}
			&
			\begin{minipage}[t]{0.3\textwidth}
				\centering
				\begin{overpic}[width=0.75\textwidth]{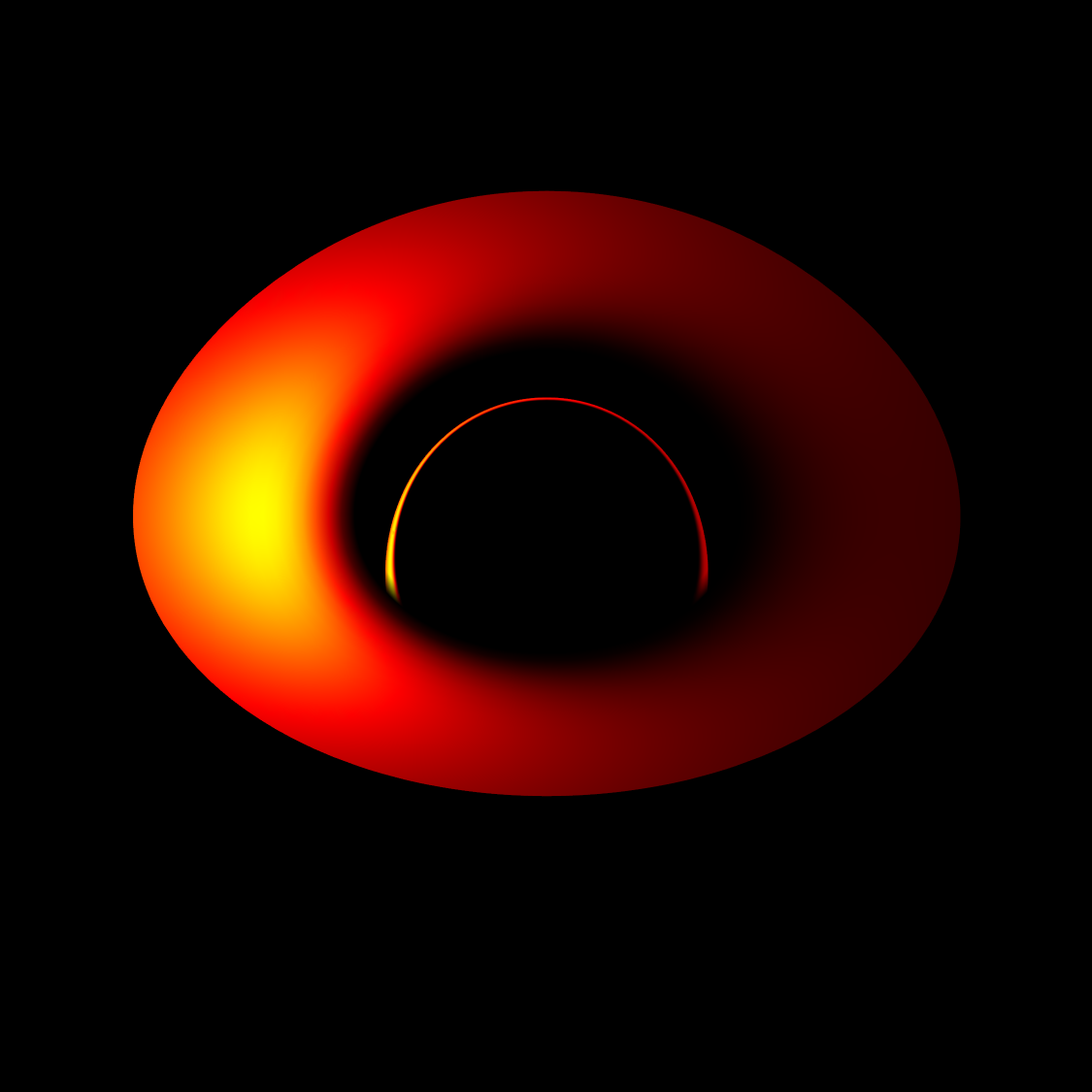} 
					\put(0,103){\color{black}\large $r_{\mathrm{s}}=0.5,\rho_{\mathrm{s}}=1.5,\theta=50^{\circ}$} 
					\put(-10,48){\color{black} Y}
					\put(48,-10){\color{black} X}
				\end{overpic}
			\end{minipage}
			\vspace{40pt} 
			\\
			\begin{minipage}[t]{0.3\textwidth}
				\centering
				\begin{overpic}[width=0.75\textwidth]{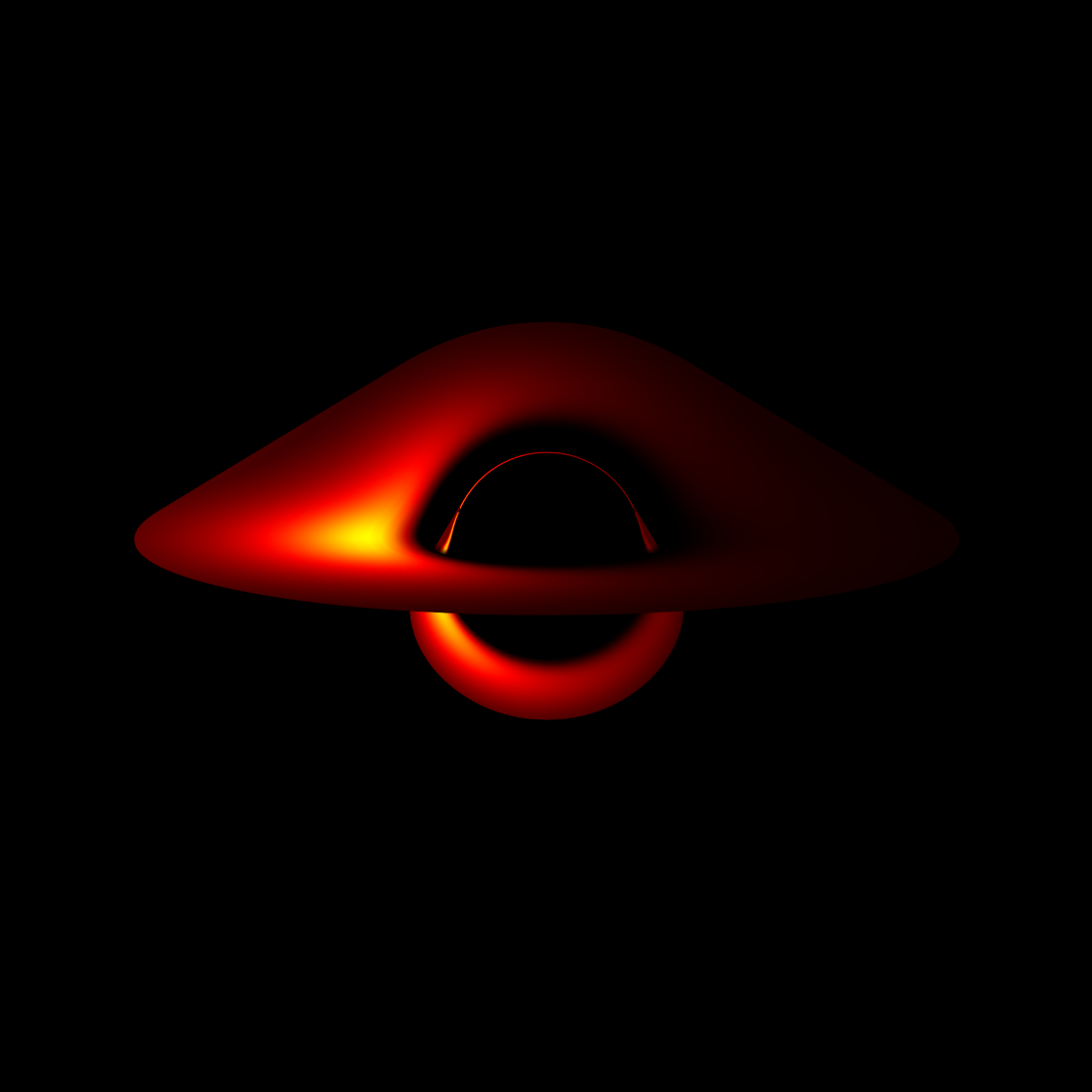}
					\put(0,103){\color{black}\large $r_{\mathrm{s}}=0.5,\rho_{\mathrm{s}}=0.5,\theta=80^{\circ}$}
					\put(-10,48){\color{black} Y}
					\put(48,-10){\color{black} X}
				\end{overpic}
			\end{minipage}
			&
			\begin{minipage}[t]{0.3\textwidth}
				\centering
				\begin{overpic}[width=0.75\textwidth]{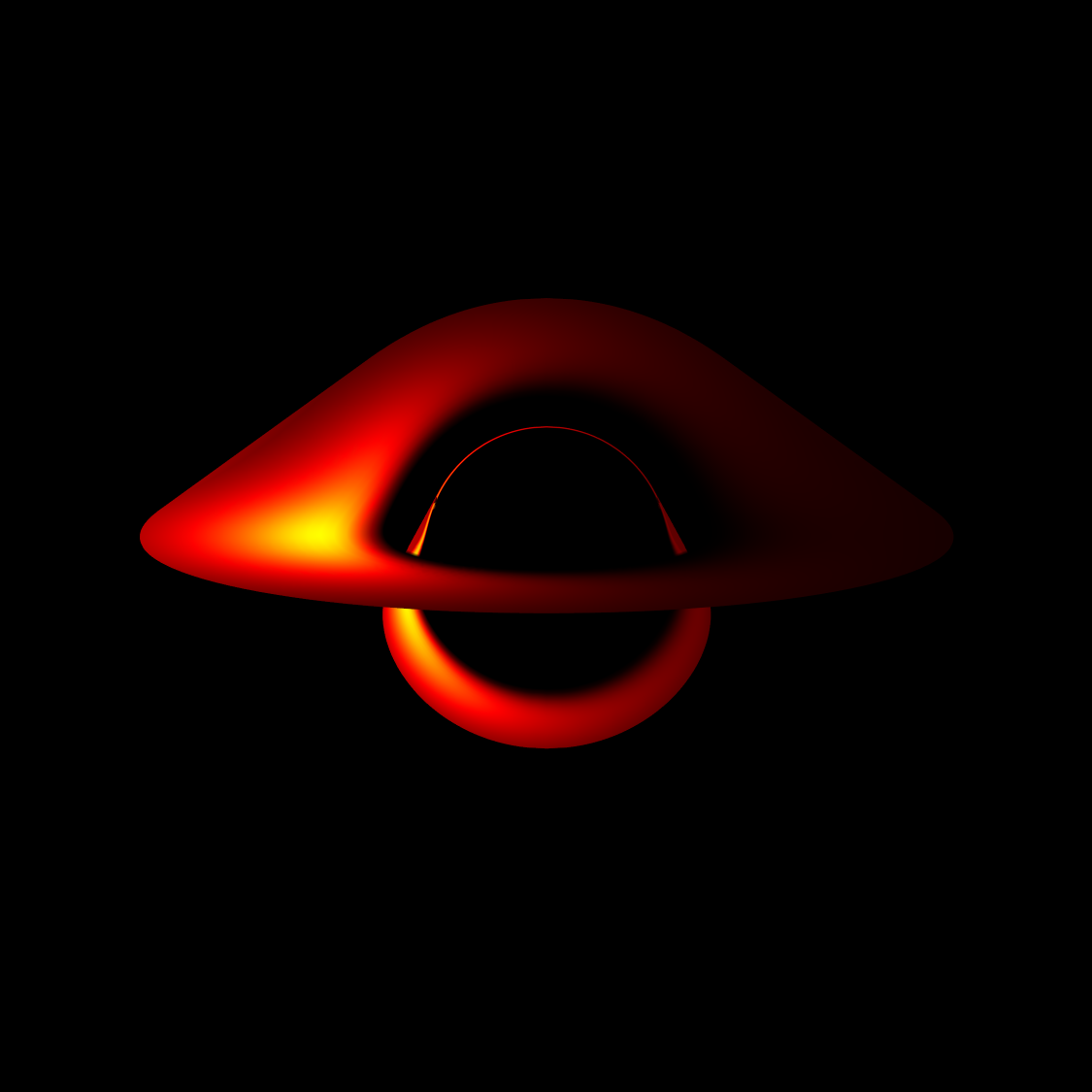}
					\put(0,103){\color{black}\large $r_{\mathrm{s}}=0.5,\rho_{\mathrm{s}}=1,\theta=80^{\circ}$} 
					\put(-10,48){\color{black} Y}
					\put(48,-10){\color{black} X}
				\end{overpic}
			\end{minipage}
			&
			\begin{minipage}[t]{0.3\textwidth}
				\centering
				\begin{overpic}[width=0.75\textwidth]{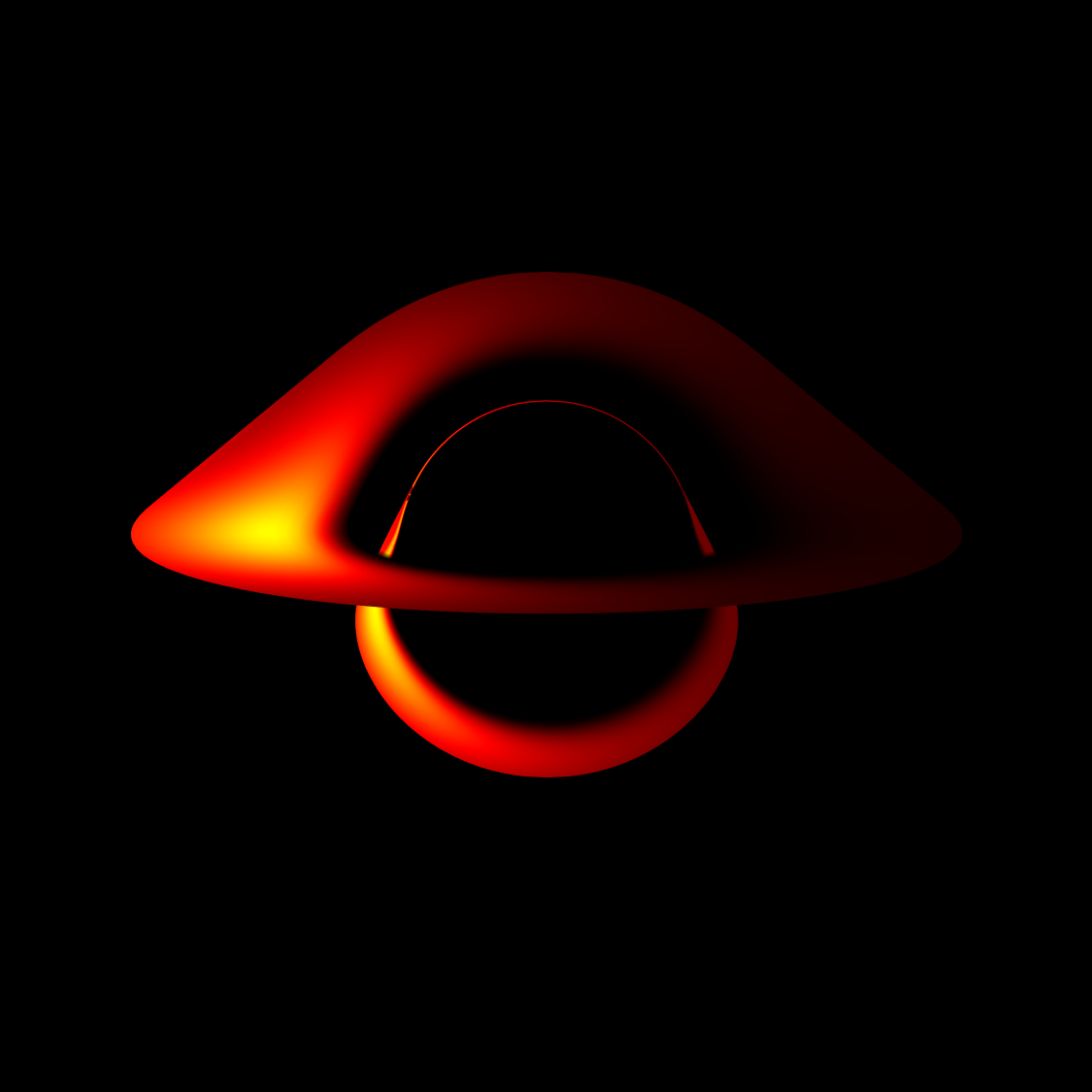}
					\put(0,103){\color{black}\large $r_{\mathrm{s}}=0.5,\rho_{\mathrm{s}}=1.5,\theta=80^{\circ}$}
					\put(-10,48){\color{black} Y}
					\put(48,-10){\color{black} X}
				\end{overpic}
			\end{minipage}
		\end{tabular}
		\caption{The complete apparent images of a thin accretion disk for different values of the parameter $\rho_{\mathrm{s}}$ and the inclination angle, with $r_{\mathrm{s}}=0.5$.}
		\label{xuanran2}
	\end{figure*}

	The observed flux $F_{\rm obs}$ at a location in the image plane differs from the locally emitted flux due to gravitational redshift. Accounting for this effect yields the following relation:
	\begin{equation}
		F_{\mathrm{obs}}=\frac{F(r)}{(1+z)^{4}}.\label{Fobs}
	\end{equation}
	where the total redshift factor $z$—accounting for both gravitational and Doppler contributions—can be written as \cite{Luminet:1979nyg}
	\begin{equation}
		1+z=\frac{1+\Omega b\sin\theta\cos\alpha}{\sqrt{-g_{tt}-g_{\phi\phi}\Omega^{2}}}.\label{hongyi}
	\end{equation}
	Combining Eqs. (\ref{flux}), (\ref{Fobs}), and (\ref{hongyi}), we obtain the observed flux as:
	\begin{equation}
		F_{\mathrm{obs}}=\frac{-\frac{\dot{M}\Omega_{,r}}{4\pi\sqrt{-g/g_{\theta \theta}}(E-\Omega L)^{2}}\int^{r}_{r_{\mathrm{ISCO}}}(E-\Omega L)L_{,r}\mathrm{d}r}{\left(\frac{1+\Omega b\sin\theta\cos\alpha}{\sqrt{-g_{tt}-g_{\phi\phi}\Omega^{2}}}\right)^{4}}.\label{Fobsz}
	\end{equation}
	The resulting observed flux maps of the accretion disk, shown in Figs.~\ref{xuanran1} and~\ref{xuanran2}, reveal how a Schwarzschild BH embedded in a Hernquist DM halo appears under different halo parameters. A key determinant of image morphology is the observer's inclination angle. Fig.~\ref{xuanran1} presents the results for fixed $\rho_{\mathrm{s}} = 0.4$, with the rows indicating distinct inclinations and the columns corresponding to increasing $r_{\mathrm{s}} = 0.3$,~ $0.6$,~ $0.9$. In contrast, Fig.~\ref{xuanran2} fixes $r_{\mathrm{s}} = 0.5$ and varies $\rho_{\mathrm{s}} = 0.5$,~ $1.0$,~ $1.5$ across columns. Higher inclinations induce pronounced flux asymmetry due to Doppler boosting, with the emission enhanced on the approaching side of the disk. Remarkably, despite variations in the DM distribution, the qualitative structure of the images remains consistent across both figures, highlighting the robustness of the lensing and beaming response to inclination.
	
	\section{Conclusion} 
	
	\label{section7}
	We have systematically investigated the influence of a Hernquist-type DM halo parameterized by the core radius $r_{\mathrm{s}}$ and the central density $\rho_{\mathrm{s}}$ on both the geodesic structure and the multi-messenger observables of a Schwarzschild BH surrounded by a geometrically thin, optically thick accretion disk. Our analysis reveals that the gravitational contribution of the DM halo significantly modifies the background spacetime, even though the central object remains non-rotating and uncharged.
	
	By examining the effective potential for timelike geodesics, we find that the radii of key circular orbits, specifically the ISCO and the MBO, increase monotonically with both $r_{\mathrm{s}}$ and $\rho_{\mathrm{s}}$. This outward shift arises because the additional mass from the DM halo enhances the overall gravitational pull at intermediate radii, effectively deepening the potential well and requiring the particles to orbit farther out to maintain stability.
	
	To translate these dynamical effects into observable predictions, we computed the high-resolution images of the accretion disk, including both direct- and secondary-lensing contributions, across a range of observer inclination angles. The resulting visual morphology exhibits notable dependence on the DM halo parameters: stronger halos (larger $r_{\mathrm{s}}$ or $\rho_{\mathrm{s}}$) produce broader but less concentrated emission rings, with diminished photon capture near the shadow boundary. Moreover, the associated thermal flux distribution derived under the standard Novikov-Thorne model shows a clear trend: as the DM halo becomes more massive or extended, the peak temperature of the disk decreases, and the total integrated luminosity drops significantly. In extreme cases, the disk can become substantially cooler and dimmer, potentially mimicking the observational appearance of lower-mass BHs if the DM environment is unaccounted for.
	
	Complementing these electromagnetic signatures, we also examined the impact of the DM halo on the GW emission from the periodic timelike orbits in the strong-field regime. Using a semi-analytical approach based on the geodesic integration and quadrupole-formula waveform construction, we demonstrated that the presence of the halo alters the characteristic zoom-whirl dynamics of such orbits. Specifically, the increased orbital periods and modified radial excursions induced by the DM potential lead to measurable phase delays and amplitude modulations in the emitted waveforms compared to the vacuum Schwarzschild case. These deviations grow with $r_{\mathrm{s}}$ and $\rho_{\mathrm{s}}$, suggesting that future space-based interferometers like LISA could, in principle, detect the imprint of DM through precision GW spectroscopy of EMRIs.
	
	These findings underscore the importance of incorporating realistic galactic-scale matter distributions, particularly DM halos, into the models of BH accretion, imaging, and GW emission. Future high-resolution observations by instruments such as the EHT, next-generation X-ray spectrometers, and space-based GW detectors may collectively provide complementary constraints on the properties of surrounding DM. While degeneracies with other effects (e.g., BH spin, magnetic fields, or accretion physics) remain a challenge, our results suggest that joint multi-wavelength, multi-inclination, and multi-messenger studies could help disentangle the role of DM in shaping both BH shadows and the GW signatures.
	
	\clearpage
	\begin{acknowledgments}
		This study is supported in part by National Natural Science Foundation of China (Grant
		No. 12333008).
	\end{acknowledgments}


\begin{thebibliography}{99}
		\bibitem{EventHorizonTelescope:2019dse}
		K.~Akiyama et al. (Event Horizon Telescope),
		{First M87 Event Horizon Telescope Results. I. The Shadow of the Supermassive Black Hole},
		\href{http://dx.doi.org/10.3847/2041-8213/ab0ec7}{Astrophys. J. Lett. {\bfseries 875}, L1 (2019)},
		[\href{http://arxiv.org/abs/1906.11238}{arXiv:1906.11238 [astro-ph.GA]}].
		
		\bibitem{EventHorizonTelescope:2021bee}
		K.~Akiyama et al. (Event Horizon Telescope),
		{First M87 Event Horizon Telescope Results. VII. Polarization of the Ring},
		\href{http://dx.doi.org/10.3847/2041-8213/abe71d}{Astrophys. J. Lett. {\bfseries 910}, L12 (2021)},
		[\href{http://arxiv.org/abs/2105.01169}{arXiv:2105.01169 [astro-ph.HE]}].
		
		\bibitem{EventHorizonTelescope:2022wkp}
		K.~Akiyama et al. (Event Horizon Telescope),
		{First Sagittarius A* Event Horizon Telescope Results. I. The Shadow of the Supermassive Black Hole in the Center of the Milky Way},
		\href{http://dx.doi.org/10.3847/2041-8213/ac6674}{Astrophys. J. Lett. {\bfseries 930}, L12 (2022)},
		[\href{http://arxiv.org/abs/2311.08680}{arXiv:2311.08680 [astro-ph.HE]}].
		
		\bibitem{LIGOScientific:2016aoc}
		B.~P.~Abbott et al. (LIGO Scientific, Virgo),
		{Observation of Gravitational Waves from a Binary Black Hole Merger},
		\href{http://dx.doi.org/10.1103/PhysRevLett.116.061102}{Phys. Rev. Lett. {\bfseries 116}, 061102 (2016)},
		[\href{http://arxiv.org/abs/1602.03837}{arXiv:1602.03837 [gr-qc]}].
		
		\bibitem{KAGRA:2021vkt}
		R.~Abbott et al. (LIGO Scientific, Virgo, KAGRA),
		{GWTC-3: Compact Binary Coalescences Observed by LIGO and Virgo during the Second Part of the Third Observing Run},
		\href{http://dx.doi.org/10.1103/PhysRevX.13.041039}{Phys. Rev. X {\bfseries 13}, 041039 (2023)},
		[\href{http://arxiv.org/abs/2111.03606}{arXiv:2111.03606 [gr-qc]}].
		
		\bibitem{GRAVITY:2020lpa}
		M.~Baub\"ock et al. (GRAVITY),
		{Modeling the orbital motion of Sgr A*{\textquoteright}s near-infrared flares},
		\href{http://dx.doi.org/10.1051/0004-6361/201937233}{Astron. Astrophys. {\bfseries 635}, A143 (2020)},
		[\href{http://arxiv.org/abs/2002.08374}{arXiv:2002.08374 [astro-ph.HE]}].
		
		\bibitem{Navarro:1995iw}
		J.~F.~Navarro, C.~S.~Frenk and S.~D.~M.~White,
		{The Structure of cold dark matter halos},
		\href{http://dx.doi.org/10.1086/177173}{Astrophys. J. {\bfseries 462}, 563--575 (1996)},
		[\href{http://arxiv.org/abs/astro-ph/9508025}{arXiv:astro-ph/9508025}].
		
		\bibitem{Burkert:1995yz}
		A.~Burkert,
		{The Structure of dark matter halos in dwarf galaxies},
		\href{http://dx.doi.org/10.1086/309560}{Astrophys. J. Lett. {\bfseries 447}, L25 (1995)},
		[\href{http://arxiv.org/abs/astro-ph/9504041}{arXiv:astro-ph/9504041}].
		
		\bibitem{Moore:1994yx}
		B.~Moore,
		{Evidence against dissipationless dark matter from observations of galaxy haloes},
		\href{http://dx.doi.org/10.1038/370629a0}{Nature {\bfseries 370}, 629 (1994)}.
		
		\bibitem{Dehnen:1993uh}
		W.~Dehnen,
		{A Family of Potential-Density Pairs for Spherical Galaxies and Bulges},
		\href{http://dx.doi.org/10.1093/mnras/265.2.250}{Mon. Not. Roy. Astron. Soc. {\bfseries 265}, 250 (1993)}.
		
		\bibitem{LIGOScientific:2016sjg}
		B.~P.~Abbott et al. (LIGO Scientific, Virgo),
		{GW151226: Observation of Gravitational Waves from a 22-Solar-Mass Binary Black Hole Coalescence},
		\href{http://dx.doi.org/10.1103/PhysRevLett.116.241103}{Phys. Rev. Lett. {\bfseries 116}, 241103 (2016)},
		[\href{http://arxiv.org/abs/1606.04855}{arXiv:1606.04855 [gr-qc]}].
		
		\bibitem{EventHorizonTelescope:2019uob}
		K.~Akiyama et al. (Event Horizon Telescope),
		{First M87 Event Horizon Telescope Results. II. Array and Instrumentation},
		\href{http://dx.doi.org/10.3847/2041-8213/ab0c96}{Astrophys. J. Lett. {\bfseries 875}, L2 (2019)},
		[\href{http://arxiv.org/abs/1906.11239}{arXiv:1906.11239 [astro-ph.IM]}].
		
		\bibitem{Glampedakis:2005hs}
		K.~Glampedakis,
		{Extreme mass ratio inspirals: LISA's unique probe of black hole gravity},
		\href{http://dx.doi.org/10.1088/0264-9381/22/15/004}{Class. Quant. Grav. {\bfseries 22}, S605--S659 (2005)},
		[\href{http://arxiv.org/abs/gr-qc/0509024}{arXiv:gr-qc/0509024}].
		
		\bibitem{Levin:2008mq}
		J.~Levin and G.~Perez-Giz,
		{A Periodic Table for Black Hole Orbits},
		\href{http://dx.doi.org/10.1103/PhysRevD.77.103005}{Phys. Rev. D {\bfseries 77}, 103005 (2008)},
		[\href{http://arxiv.org/abs/0802.0459}{arXiv:0802.0459 [gr-qc]}].
		
		\bibitem{Levin:2008ci}
		J.~Levin and B.~Grossman,
		{Dynamics of Black Hole Pairs. I. Periodic Tables},
		\href{http://dx.doi.org/10.1103/PhysRevD.79.043016}{Phys. Rev. D {\bfseries 79}, 043016 (2009)},
		[\href{http://arxiv.org/abs/0809.3838}{arXiv:0809.3838 [gr-qc]}].
		
		\bibitem{Grossman:2008yk}
		R.~Grossman and J.~Levin,
		{Dynamics of Black Hole Pairs II: Spherical Orbits and the Homoclinic Limit of Zoom-Whirliness},
		\href{http://dx.doi.org/10.1103/PhysRevD.79.043017}{Phys. Rev. D {\bfseries 79}, 043017 (2009)},
		[\href{http://arxiv.org/abs/0811.3798}{arXiv:0811.3798 [gr-qc]}].
		
		\bibitem{Poisson:1993vp}
		E.~Poisson,
		{Gravitational radiation from a particle in circular orbit around a black hole. 1: Analytical results for the nonrotating case},
		\href{http://dx.doi.org/10.1103/PhysRevD.47.1497}{Phys. Rev. D {\bfseries 47}, 1497--1510 (1993)}.
		
		\bibitem{Cutler:1993vq}
		C.~Cutler, E.~Poisson, G.~J.~Sussman and L.~S.~Finn,
		{Gravitational radiation from a particle in circular orbit around a black hole. 2: Numerical results for the nonrotating case},
		\href{http://dx.doi.org/10.1103/PhysRevD.47.1511}{Phys. Rev. D {\bfseries 47}, 1511--1518 (1993)}.
		
		\bibitem{Apostolatos:1993nu}
		T.~Apostolatos, D.~Kennefick, E.~Poisson and A.~Ori,
		{Gravitational radiation from a particle in circular orbit around a black hole. 3: Stability of circular orbits under radiation reaction},
		\href{http://dx.doi.org/10.1103/PhysRevD.47.5376}{Phys. Rev. D {\bfseries 47}, 5376--5388 (1993)}.
		
		\bibitem{Dai:2023cft}
		N.~Dai, Y.~Gong, Y.~Zhao and T.~Jiang,
		{Extreme mass ratio inspirals in galaxies with dark matter halos},
		\href{http://dx.doi.org/10.1103/PhysRevD.110.084080}{Phys. Rev. D {\bfseries 110}, 084080 (2024)},
		[\href{http://arxiv.org/abs/2301.05088}{arXiv:2301.05088 [gr-qc]}].
		
		\bibitem{GRAVITY:2020gka}
		R.~Abuter et al. (GRAVITY),
		{Detection of the Schwarzschild precession in the orbit of the star S2 near the Galactic centre massive black hole},
		\href{http://dx.doi.org/10.1051/0004-6361/202037813}{Astron. Astrophys. {\bfseries 636}, L5 (2020)},
		[\href{http://arxiv.org/abs/2004.07187}{arXiv:2004.07187 [astro-ph.GA]}].
		
		\bibitem{Cardoso:2014sna}
		V.~Cardoso, L.~C.~B.~Crispino, C.~F.~B.~Macedo, H.~Okawa and P.~Pani,
		{Light rings as observational evidence for event horizons: long-lived modes, ergoregions and nonlinear instabilities of ultracompact objects},
		\href{http://dx.doi.org/10.1103/PhysRevD.90.044069}{Phys. Rev. D {\bfseries 90}, 044069 (2014)},
		[\href{http://arxiv.org/abs/1406.5510}{arXiv:1406.5510 [gr-qc]}].
		
		\bibitem{Cunha:2022gde}
		P.~V.~P.~Cunha, C.~Herdeiro, E.~Radu and N.~Sanchis-Gual,
		{Exotic Compact Objects and the Fate of the Light-Ring Instability},
		\href{http://dx.doi.org/10.1103/PhysRevLett.130.061401}{Phys. Rev. Lett. {\bfseries 130}, 061401 (2023)},
		[\href{http://arxiv.org/abs/2207.13713}{arXiv:2207.13713 [gr-qc]}].
		
		\bibitem{Guo:2024cts}
		G.~Guo, P.~Wang and Y.-P.~Zhang,
		{Nonlinear stability of black holes with a stable light ring},
		\href{http://dx.doi.org/10.1103/xlsl-8dtq}{Phys. Rev. D {\bfseries 112}, 084023 (2025)},
		[\href{http://arxiv.org/abs/2403.02089}{arXiv:2403.02089 [gr-qc]}].
		
		\bibitem{Mummery:2022ana}
		A.~Mummery and S.~Balbus,
		{Inspirals from the Innermost Stable Circular Orbit of Kerr Black Holes: Exact Solutions and Universal Radial Flow},
		\href{http://dx.doi.org/10.1103/PhysRevLett.129.161101}{Phys. Rev. Lett. {\bfseries 129}, 161101 (2022)},
		[\href{http://arxiv.org/abs/2209.03579}{arXiv:2209.03579 [gr-qc]}].
		
		\bibitem{Fujita:2009bp}
		R.~Fujita and W.~Hikida,
		{Analytical solutions of bound timelike geodesic orbits in Kerr spacetime},
		\href{http://dx.doi.org/10.1088/0264-9381/26/13/135002}{Class. Quant. Grav. {\bfseries 26}, 135002 (2009)},
		[\href{http://arxiv.org/abs/0906.1420}{arXiv:0906.1420 [gr-qc]}].
		
		\bibitem{Healy:2009zm}
		J.~Healy, J.~Levin and D.~Shoemaker,
		{Zoom-Whirl Orbits in Black Hole Binaries},
		\href{http://dx.doi.org/10.1103/PhysRevLett.103.131101}{Phys. Rev. Lett. {\bfseries 103}, 131101 (2009)},
		[\href{http://arxiv.org/abs/0907.0671}{arXiv:0907.0671 [gr-qc]}].
		
		\bibitem{Wei:2019zdf}
		S.-W.~Wei, J.~Yang and Y.-X.~Liu,
		{Geodesics and periodic orbits in Kehagias-Sfetsos black holes in deformed Hor\v{r}ava-Lifshitz gravity},
		\href{http://dx.doi.org/10.1103/PhysRevD.99.104016}{Phys. Rev. D {\bfseries 99}, 104016 (2019)},
		[\href{http://arxiv.org/abs/1904.03129}{arXiv:1904.03129 [gr-qc]}].
		
		\bibitem{Azreg-Ainou:2020bfl}
		M.~Azreg-A\"{\i}nou, Z.~Chen, B.~Deng, M.~Jamil, T.~Zhu, Q.~Wu and Y.-K.~Lim,
		{Orbital mechanics and quasiperiodic oscillation resonances of black holes in Einstein-\AE ther theory},
		\href{http://dx.doi.org/10.1103/PhysRevD.102.044028}{Phys. Rev. D {\bfseries 102}, 044028 (2020)},
		[\href{http://arxiv.org/abs/2004.02602}{arXiv:2004.02602 [gr-qc]}].
		
		\bibitem{Deng:2020yfm}
		X.-M.~Deng,
		{Geodesics and periodic orbits around quantum-corrected black holes},
		\href{http://dx.doi.org/10.1016/j.dark.2020.100629}{Phys. Dark Univ. {\bfseries 30}, 100629 (2020)}.
		
		\bibitem{Wang:2022tfo}
		R.~Wang, F.~Gao and H.~Chen,
		{Periodic orbits around a static spherically symmetric black hole surrounded by quintessence},
		\href{http://dx.doi.org/10.1016/j.aop.2022.169167}{Annals Phys. {\bfseries 447}, 169167 (2022)}.
		
		\bibitem{QiQi:2024dwc}
		Q.~Qi, X.-M.~Kuang, Y.-Z.~Li and Y.~Sang,
		{Timelike bound orbits and pericenter precession around black hole with conformally coupled scalar hair},
		\href{http://dx.doi.org/10.1140/epjc/s10052-024-12989-y}{Eur. Phys. J. C {\bfseries 84}, 645 (2024)},
		[\href{http://arxiv.org/abs/2407.01958}{arXiv:2407.01958 [gr-qc]}].
		
		\bibitem{Alloqulov:2025ucf}
		M.~Alloqulov, T.~Xamidov, S.~Shaymatov and B.~Ahmedov,
		{Gravitational waveforms from periodic orbits around a Schwarzschild black hole embedded in a Dehnen-type dark matter halo},
		\href{http://dx.doi.org/10.1140/epjc/s10052-025-14529-8}{Eur. Phys. J. C {\bfseries 85}, 798 (2025)},
		[\href{http://arxiv.org/abs/2504.05236}{arXiv:2504.05236 [gr-qc]}].
		
		\bibitem{Wang:2025hla}
		C.-H.~Wang, X.-C.~Meng, Y.-P.~Zhang, T.~Zhu and S.-W.~Wei,
		{Equatorial periodic orbits and gravitational waveforms in a black hole free of Cauchy horizon},
		\href{http://dx.doi.org/10.1088/1475-7516/2025/07/021}{JCAP {\bfseries 07}, 021 (2025)},
		[\href{http://arxiv.org/abs/2502.08994}{arXiv:2502.08994 [gr-qc]}].
		
		\bibitem{Luminet:1979nyg}
		J.-P.~Luminet,
		{Image of a spherical black hole with thin accretion disk},
		Astron. Astrophys. {\bfseries 75}, 228--235 (1979).
		
		\bibitem{Hou:2022eev}
		Y.~Hou, Z.~Zhang, H.~Yan, M.~Guo and B.~Chen,
		{Image of a Kerr-Melvin black hole with a thin accretion disk},
		\href{http://dx.doi.org/10.1103/PhysRevD.106.064058}{Phys. Rev. D {\bfseries 106}, 064058 (2022)},
		[\href{http://arxiv.org/abs/2206.13744}{arXiv:2206.13744 [gr-qc]}].
		
		\bibitem{Zhang:2024lsf}
		Z.~Zhang, Y.~Hou, M.~Guo and B.~Chen,
		{Imaging thick accretion disks and jets surrounding black holes},
		\href{http://dx.doi.org/10.1088/1475-7516/2024/05/032}{JCAP {\bfseries 05}, 032 (2024)},
		[\href{http://arxiv.org/abs/2401.14794}{arXiv:2401.14794 [astro-ph.HE]}].
		
		\bibitem{Gyulchev:2019tvk}
		G.~Gyulchev, P.~Nedkova, T.~Vetsov and S.~Yazadjiev,
		{Image of the Janis-Newman-Winicour naked singularity with a thin accretion disk},
		\href{http://dx.doi.org/10.1103/PhysRevD.100.024055}{Phys. Rev. D {\bfseries 100}, 024055 (2019)},
		[\href{http://arxiv.org/abs/1905.05273}{arXiv:1905.05273 [gr-qc]}].
		
		\bibitem{Shaikh:2019hbm}
		R.~Shaikh and P.~S.~Joshi,
		{Can we distinguish black holes from naked singularities by the images of their accretion disks?},
		\href{http://dx.doi.org/10.1088/1475-7516/2019/10/064}{JCAP {\bfseries 10}, 064 (2019)},
		[\href{http://arxiv.org/abs/1909.10322}{arXiv:1909.10322 [gr-qc]}].
		
		\bibitem{Bambi:2019tjh}
		C.~Bambi, K.~Freese, S.~Vagnozzi and L.~Visinelli,
		{Testing the rotational nature of the supermassive object M87* from the circularity and size of its first image},
		\href{http://dx.doi.org/10.1103/PhysRevD.100.044057}{Phys. Rev. D {\bfseries 100}, 044057 (2019)},
		[\href{http://arxiv.org/abs/1904.12983}{arXiv:1904.12983 [gr-qc]}].
		
		\bibitem{Johannsen:2016uoh}
		T.~Johannsen,
		{Testing the No-Hair Theorem with Observations of Black Holes in the Electromagnetic Spectrum},
		\href{http://dx.doi.org/10.1088/0264-9381/33/12/124001}{Class. Quant. Grav. {\bfseries 33}, 124001 (2016)},
		[\href{http://arxiv.org/abs/1602.07694}{arXiv:1602.07694 [astro-ph.HE]}].
		
		\bibitem{Gates:2020sdh}
		D.~E.~A.~Gates, S.~Hadar and A.~Lupsasca,
		{Maximum observable blueshift from circular equatorial Kerr orbiters},
		\href{http://dx.doi.org/10.1103/PhysRevD.102.104041}{Phys. Rev. D {\bfseries 102}, 104041 (2020)},
		[\href{http://arxiv.org/abs/2009.03310}{arXiv:2009.03310 [gr-qc]}].
		
		\bibitem{Okyay:2021nnh}
		M.~Okyay and A.~\"Ovg\"un,
		{Nonlinear electrodynamics effects on the black hole shadow, deflection angle, quasinormal modes and greybody factors},
		\href{http://dx.doi.org/10.1088/1475-7516/2022/01/009}{JCAP {\bfseries 01}, 009 (2022)},
		[\href{http://arxiv.org/abs/2108.07766}{arXiv:2108.07766 [gr-qc]}].
		
		\bibitem{Liu:2024brf}
		A.~Liu, T.-Y.~He, M.~Liu, Z.-W.~Han and R.-J.~Yang,
		{Possible signatures of higher dimension in thin accretion disk around brane world black hole},
		[\href{http://arxiv.org/abs/2404.14131}{arXiv:2404.14131 [gr-qc]}].
		
		
		\bibitem{Feng:2025ljz}
		H.~Feng, Z.~Cai, H.-P.~Yan, R.-J.~Yang and J.-J.~Zhang,
		{Images of shadow and thin accretion disk around Bardeen black hole surrounded by perfect fluid dark matter},
		[\href{http://arxiv.org/abs/2512.00824}{arXiv:2512.00824 [astro-ph.HE]}].
		
		\bibitem{Yin:2025coq}
		J.-J.~Yin, T.-Y.~He, M.~Liu, H.-M.~Fan, B.-H.~Chen, Z.-W.~Han and R.-J.~Yang,
		{Observational properties of black hole in quantum fluctuation modified gravity},
		[\href{http://arxiv.org/abs/2503.00488}{arXiv:2503.00488 [gr-qc]}].
		
		
		\bibitem{Feng:2024iqj}
		H.~Feng, R.-J.~Yang and W.-Q.~Chen,
		{Thin accretion disk and shadow of Kerr–Sen black hole in Einstein–Maxwell–dilaton–axion gravity},
		[\href{http://arxiv.org/abs/2403.18541}{arXiv:2403.18541 [gr-qc]}].
		
		
		\bibitem{He:2022lrc}
		T.-Y.~He, Z.~Cai and R.-J.~Yang,
		{Thin accretion disks around a black hole in Einstein-Aether-scalar theory},
		[\href{http://arxiv.org/abs/2208.03723}{arXiv:2208.03723 [gr-qc]}].
		
		
		
		\bibitem{Huang:2023ilm}
		Y.-X.~Huang, S.~Guo, Y.-H.~Cui, Q.-Q.~Jiang and K.~Lin,
		{Influence of accretion disk on the optical appearance of the Kazakov-Solodukhin black hole},
		\href{http://dx.doi.org/10.1103/PhysRevD.107.123009}{Phys. Rev. D {\bfseries 107}, 123009 (2023)},
		[\href{http://arxiv.org/abs/2311.00302}{arXiv:2311.00302 [gr-qc]}].
		
		\bibitem{Cai:2025pan}
		Z.~Cai, Z.~Ban, L.~Wang, H.~Feng and Z.-W.~Long,
		{Shadow and thin accretion disk around Ay\'on-Beato--Garc\'{\i}a black hole coupled with cloud of strings},
		\href{http://dx.doi.org/10.1016/j.dark.2025.102169}{Phys. Dark Univ. {\bfseries 50}, 102169 (2025)},
		[\href{http://arxiv.org/abs/2506.22744}{arXiv:2506.22744 [gr-qc]}].
		
		\bibitem{Cai:2025rst}
		Z.~Cai, Z.~Ban, L.~Wang, H.~Feng and Z.-W.~Long,
		{Thin accretion disk around Schwarzschild-like black hole in bumblebee gravity},
		\href{http://dx.doi.org/10.1088/1475-7516/2025/10/101}{JCAP {\bfseries 10}, 101 (2025)},
		[\href{http://arxiv.org/abs/2503.08424}{arXiv:2503.08424 [gr-qc]}].
		
		\bibitem{Guo:2023grt}
		S.~Guo, Y.-X.~Huang, Y.-H.~Cui, Y.~Han, Q.-Q.~Jiang, E.-W.~Liang and K.~Lin,
		{Unveiling the unconventional optical signatures of regular black holes within accretion disk},
		\href{http://dx.doi.org/10.1140/epjc/s10052-023-12208-0}{Eur. Phys. J. C {\bfseries 83}, 1059 (2023)},
		[\href{http://arxiv.org/abs/2310.20523}{arXiv:2310.20523 [gr-qc]}].
		
		\bibitem{Liu:2021lvk}
		C.~Liu, L.~Tang and J.~Jing,
		{Image of the Schwarzschild black hole pierced by a cosmic string with a thin accretion disk},
		\href{http://dx.doi.org/10.1142/S0218271822500419}{Int. J. Mod. Phys. D {\bfseries 31}, 2250041 (2022)},
		[\href{http://arxiv.org/abs/2109.01867}{arXiv:2109.01867 [gr-qc]}].
		
		\bibitem{Guo:2022rql}
		S.~Guo, G.-R.~Li and E.-W.~Liang,
		{Observable characteristics of the charged black hole surrounded by thin disk accretion in Rastall gravity},
		\href{http://dx.doi.org/10.1088/1361-6382/ac6fa8}{Class. Quant. Grav. {\bfseries 39}, 135004 (2022)},
		[\href{http://arxiv.org/abs/2205.11241}{arXiv:2205.11241 [astro-ph.HE]}].
		
		\bibitem{Jha:2025xjf}
		S.~K.~Jha,
		{Thermodynamics, weak gravitational lensing, and parameter estimation of a Schwarzschild black hole immersed in Hernquist dark matter halo},
		\href{http://dx.doi.org/10.1088/1475-7516/2025/06/033}{JCAP {\bfseries 06}, 033 (2025)},
		[\href{http://arxiv.org/abs/2503.19938}{arXiv:2503.19938 [gr-qc]}].
		
		\bibitem{Yang:2024lmj}
		S.~Yang, Y.-P.~Zhang, T.~Zhu, L.~Zhao and Y.-X.~Liu,
		{Gravitational waveforms from periodic orbits around a quantum-corrected black hole},
		\href{http://dx.doi.org/10.1088/1475-7516/2025/01/091}{JCAP {\bfseries 01}, 091 (2025)},
		[\href{http://arxiv.org/abs/2407.00283}{arXiv:2407.00283 [gr-qc]}].
		
		\bibitem{Shabbir:2025kqh}
		O.~Shabbir, M.~Jamil and M.~Azreg-A\"{\i}nou,
		{Periodic orbits and their gravitational wave radiations around the Schwarzschild-MOG black hole},
		\href{http://dx.doi.org/10.1016/j.dark.2025.101816}{Phys. Dark Univ. {\bfseries 47}, 101816 (2025)},
		[\href{http://arxiv.org/abs/2501.04367}{arXiv:2501.04367 [gr-qc]}].
		
		\bibitem{Jiang:2024cpe}
		H.~Jiang, M.~Alloqulov, Q.~Wu, S.~Shaymatov and T.~Zhu,
		{Periodic orbits and plasma effects on gravitational weak lensing by self-dual black hole in loop quantum gravity},
		\href{http://dx.doi.org/10.1016/j.dark.2024.101627}{Phys. Dark Univ. {\bfseries 46}, 101627 (2024)}.
		
		\bibitem{Babak:2006uv}
		S.~Babak, H.~Fang, J.~R.~Gair, K.~Glampedakis and S.~A.~Hughes,
		{'Kludge' gravitational waveforms for a test-body orbiting a Kerr black hole},
		\href{http://dx.doi.org/10.1103/PhysRevD.75.024005}{Phys. Rev. D {\bfseries 75}, 024005 (2007)},
		[\href{http://arxiv.org/abs/gr-qc/0607007}{arXiv:gr-qc/0607007}].
		[Erratum: Phys.\ Rev.\ D {\bfseries 77}, 04990 (2008)].
		
		\bibitem{Maselli:2021men}
		A.~Maselli, N.~Franchini, L.~Gualtieri, T.~P.~Sotiriou, S.~Barsanti and P.~Pani,
		{Detecting fundamental fields with LISA observations of gravitational waves from extreme mass-ratio inspirals},
		\href{http://dx.doi.org/10.1038/s41550-021-01589-5}{Nature Astron. {\bfseries 6}, 464--470 (2022)},
		[\href{http://arxiv.org/abs/2106.11325}{arXiv:2106.11325 [gr-qc]}].
		
		\bibitem{Liang:2022gdk}
		D.~Liang, R.~Xu, Z.-F.~Mai and L.~Shao,
		{Probing vector hair of black holes with extreme-mass-ratio inspirals},
		\href{http://dx.doi.org/10.1103/PhysRevD.107.044053}{Phys. Rev. D {\bfseries 107}, 044053 (2023)},
		[\href{http://arxiv.org/abs/2212.09346}{arXiv:2212.09346 [gr-qc]}].
		
		\bibitem{Will:2016sgx}
		C.~M.~Will,
		{Gravity: Newtonian, Post-Newtonian, and General Relativistic},
		in {\em Gravity: Where Do We Stand?}, edited by R.~Peron, M.~Colpi, V.~Gorini and U.~Moschella
		(Springer, 2016), pp.~9--72,
		\href{http://dx.doi.org/10.1007/978-3-319-20224-2_2}{}.
		
		\bibitem{You:2024uql}
		L.~You, R.-B.~Wang, S.-J.~Ma, J.-B.~Deng and X.-R.~Hu,
		{Optical properties of Euler-Heisenberg black hole in the Cold Dark Matter Halo},
		[\href{http://arxiv.org/abs/2403.12840}{arXiv:2403.12840 [gr-qc]}].
		
		\bibitem{Page:1974he}
		D.~N.~Page and K.~S.~Thorne,
		{Disk-Accretion onto a Black Hole. Time-Averaged Structure of Accretion Disk},
		\href{http://dx.doi.org/10.1086/152990}{Astrophys. J. {\bfseries 191}, 499--506 (1974)}.
		
		\bibitem{Collodel:2021gxu}
		L.~G.~Collodel, D.~D.~Doneva and S.~S.~Yazadjiev,
		{Circular Orbit Structure and Thin Accretion Disks around Kerr Black Holes with Scalar Hair},
		[\href{http://arxiv.org/abs/2101.05073}{arXiv:2101.05073 [astro-ph.HE]}].
	\end{thebibliography}
\end{document}